\documentclass[reprint,superscriptaddress,pra,nofootinbib,preprintnumbers]{revtex4-2}

\usepackage[utf8]{inputenc}
\usepackage{amsmath}
\usepackage{amsfonts}
\usepackage{xcolor}
\usepackage{graphicx}
\usepackage{dsfont}
\usepackage{physics}
\usepackage{placeins}
\usepackage{amssymb}
\usepackage{mathtools}
\usepackage{tikz}
\usepackage[normalem]{ulem} 
\usepackage[caption=false]{subfig}
\usepackage[colorlinks=true, urlcolor=blue,citecolor=blue,anchorcolor=blue]{hyperref}
\usetikzlibrary{quantikz}

\providecommand{\CO}{\mathcal{O}}
\providecommand{\CW}{\mathcal{W}}
\providecommand{\uone}{${\rm U}(1)$\,}
\providecommand{\sutwo}{${\rm SU}(2)$\,}
\providecommand{\suthree}{${\rm SU}(3)$\,}
\providecommand{\suN}{${\rm SU}(N)$\,}

\begin{document}
\title{Efficient quantum implementation of 2+1 U(1) lattice gauge theories \protect\\ with Gauss law constraints}

\author{Christopher Kane}
\email{cfkane@arizona.edu}
\affiliation{Department of Physics, University of Arizona, Tucson, AZ 85719, USA}

\author{Dorota M. Grabowska}
\email{grabow@uw.edu}
\affiliation{InQubator for Quantum Simulation (IQuS), Department of Physics, University of Washington, Seattle, WA 98195}
\affiliation{Theoretical Physics Department, CERN, 1211 Geneva 23, Switzerland}

\author{Benjamin Nachman}
\email{bpnachman@lbl.gov}
\affiliation{Physics Division, Lawrence Berkeley National Laboratory, Berkeley, CA 94720, USA}

\author{Christian W. Bauer}
\email{cwbauer@lbl.gov}
\affiliation{Physics Division, Lawrence Berkeley National Laboratory, Berkeley, CA 94720, USA}

\preprint{CERN-TH-2022-195}
\date{\today}

\begin{abstract}
The study of real-time evolution of lattice quantum field theories using classical computers is known to scale exponentially with the number of lattice sites. Due to a fundamentally different computational strategy, quantum computers hold the promise of allowing for detailed studies of these dynamics from first principles. However, much like with classical computations, it is important that quantum algorithms do not have a cost that scales exponentially with the volume. Recently, it was shown how to break the exponential scaling of a naive implementation of a \uone gauge theory in two spatial dimensions through an operator redefinition. In this work, we describe modifications to how operators must be sampled in the new operator basis to keep digitization errors small. We compare the precision of the energies and plaquette expectation value between the two operator bases and find they are comparable. Additionally, we provide an explicit circuit construction for the Suzuki-Trotter implementation of the theory using the Walsh function formalism. The gate count scaling is studied as a function of the lattice volume, for both exact circuits and approximate circuits where rotation gates with small arguments have been dropped. We study the errors from finite Suzuki-Trotter time-step, circuit approximation, and quantum noise in a calculation of an explicit observable using IBMQ superconducting qubit hardware. We find the gate count scaling for the approximate circuits can be further reduced by up to a power of the volume without introducing larger errors.

\end{abstract}

\maketitle
{ \hypersetup{linkcolor=black}
\tableofcontents}

\section{Introduction}
With a fundamentally different computational strategy than their classical counterparts, quantum computers hold the promise of allowing first-principles calculations of real-time evolution in quantum systems. Of particular interest are a class of quantum field theories, known as gauge theories, which appear in many areas of physics, including condensed matter physics, materials physics, and particle physics. Specifically, the Standard Model of Particle Physics, which encapsulates the majority of our understanding of the fundamental nature of the universe, is formulated in terms of three gauge theories, namely ${\rm U}(1)$, ${\rm SU}(2)$, and ${\rm SU}(3)$. The process of developing a quantum circuit for simulation of gauge theories is highly non-trivial, and further developments are imperative to keep up with rapid developments in hardware.

One of the necessary steps is to discretize the theory by putting it on a finite lattice. Many choices are possible, and recent work formulating the gauge theories using improved Hamiltonians which have reduced discretization errors has been done in Ref.~\cite{Carena:2022kpg}. 

A second necessary step when implementing gauge theories onto a quantum computer is the development of Hamiltonian formulations of their interactions. Gauge theories naturally contain gauge redundancies, which can lead to a significant increase in qubit cost, due to an enlarged Hilbert space. Formulations have been developed that impose gauge invariance both while still including these unphysical states \cite{Halimeh_2020, Halimeh:2020ecg, Halimeh:2021lnv, Halimeh:2021vzf, Lamm_Scott_Yukari_2020, Tran_2021, PhysRevLett.109.125302, Banerjee_2012, https://doi.org/10.48550/arxiv.2012.08620, PhysRevX.3.041018, PhysRevA.90.042305, Stannigel_2014, Stryker_2019}, and removing them \textit{a priori} \cite{Kaplan:2018vnj, Haase:2020kaj, Bauer_2021}. Reviews of the many different approaches, both analogue and digital, can be found in Refs.~\cite{Wiese_2013, Zohar_2015, Dalmonte_2016, Aidelsburger_2021, Ba_uls_2020, Zohar_2021, Klco_2022, Bauer:2022hpo}.

Finally, one must digitize the theory to make the Hilbert space finite. It is important that the chosen digitization scheme matches the undigitized theory to a sufficient accuracy and has quantifiable errors. 

Much work has been done developing various methods, including loop string hadron formulations \cite{Anishetty_Mathur_Raychowdhury_2009, Raychowdhury_Stryker_2020, Raychowdhury_Stryker_SU2_2020, Anishetty_Mathur_Raychowdhury_SU3_2009, doi:10.1063/1.3464267, doi:10.1063/1.3660195, Raychowdhury_thesis}, discrete subgroups and group space decimation \cite{Alexandru_2019, PhysRevD.91.054506, PhysRevA.95.023604, PhysRevD.102.114513, Alexandru:2021jpm, Gustafson:2022xdt}, mesh digitization \cite{PhysRevA.99.062341}, light-front formulations \cite{Kreshchuk_2022, e23050597}, orbifold lattice methods \cite{Kaplan_2003, Buser_2021}, and magnetic and dual formulations \cite{Kaplan:2018vnj, Haase:2020kaj, Bender_2020, Unmuth-Yockey:2018xak, Bauer_2021}. Different experimental realizations can be found in Refs.~\cite{Muschik_2017, Schweizer_2019, Kokail_2019, Yang_2020, doi:10.1126/science.aaz5312, Riechert_2022}. 

Another important choice is which algorithm is used to perform the time evolution. A number of methods exist, including Suzuki-Trotter methods \cite{Childs_2021}, QDRIFT \cite{Campbell_2019}, qubitization \cite{Low_2017, Low_2019}, implementing a truncated Taylor series \cite{Berry_2015}, quantum walks \cite{10.5555/2231036.2231040}, and variational methods \cite{Yao_2021}.

Previous studies quantifying gate count requirements have looked at time evolution in effective quantum electrodynamics (QED) \cite{Stetina:2020abi} and $1+1$ dimensional QED \cite{Nguyen:2021hyk, Shaw_2020}. There has also been work on minimizing the gate count for variational methods, as applied to a $2+1$ \uone\ gauge theory \cite{Paulson:2020zjd}. Detailed gate counts required for implementing 1+1 \suN gauge theories were presented in  Ref.~\cite{https://doi.org/10.48550/arxiv.2207.01731}, and an implementation of 1+1 \suthree was done in Ref.~\cite{Farrell:2022vyh}. In Ref.~\cite{Murairi:2022zdg}, a general procedure for simulating gauge theories was presented and applied to an \sutwo gauge theory. A general study of gate counts required for implementing the Kogut-Susskind formulation \cite{PhysRevD.11.395} of ${\rm U}(1)$, ${\rm SU}(2)$, and ${\rm SU}(3)$ gauge theories in $d$ dimensions was done in Ref.~\cite{https://doi.org/10.48550/arxiv.2107.12769}.

In this work, we construct explicit quantum circuits for Suzuki-Trotter time evolution of a formulation of a \uone\ gauge theory in two spatial dimensions developed in Ref. \cite{Bauer_2021}. We choose to focus on Suzuki-Trotter time-evolution for several reasons. The first is that it has been shown in a $1+1$ \uone gauge theory that Suzuki-Trotter methods scale linearly in the electric field cutoff compared to qubitization, truncated Taylor series methods, and QDRIFT, which all scale quadratically \cite{Shaw_2020}. The second is that Suzuki-Trotter methods require no ancillary qubits, potentially making them better suited for noisy intermediate-scale quantum (NISQ) \cite{Preskill_2018} calculations.

As explained in Sec.~\ref{sec:resourceStudy}, this formulation requires circuits to efficiently implement the Fourier transform associated with a \uone gauge theory, as well as diagonal matrices. The Fourier transform for a \uone theory is the usual quantum Fourier transform (QFT) and can be done efficiently \cite{nielsen_chuang_2010}. To implement the diagonal matrices, we use the Walsh function formalism \cite{Welch_2014}, which is an efficient quantum algorithm for implementing diagonal matrices without using acillary qubits. This method involves approximating the quantum circuit by dropping single qubit rotation gates with small arguments, and then removing CNOT gates using circuit identities. This method is therefore relevant for NISQ calculations. More importantly, because the number of T-gates required to implement a fault-tolerant rotation gate increases with precision \cite{Bocharov_2015}, the Walsh formalism for dropping gates with small angles is important to limit the number of T-gates required in a fault-tolerant calculation.

It was shown in Ref.~\cite{https://doi.org/10.48550/arxiv.2208.03333} that enforcing magnetic Gauss' law introduces a term that naively requires a number of gates that scales exponentially in the volume. While the Walsh function formalism can in principle be used to break this exponential \textit{quantum} cost, one caveat of this method is that, for a particular class of matrices, there is a \textit{classical} computing cost, associated with constructing the desired efficient quantum circuit, that scales exponentially in the volume. We will demonstrate that, for a particular term in the magnetic Hamiltonian, this classical cost will be prohibitively expensive for realistic lattice sizes.

In Ref.~\cite{https://doi.org/10.48550/arxiv.2208.03333}, it was shown that the exponential volume scaling in the gate count can be reduced to polynomial by performing a carefully chosen operator basis change. More importantly, this operator basis change also breaks the exponential volume scaling in the classical computational cost associated with constructing the quantum circuit. Note that this gate count reduction is for an exact implementation, and that using the Walsh function formalism to approximate the circuit will further reduce the required gate count. In this work we will implement this approach explicitly, which will allow us to study the required quantum resources in more detail.

This paper is organized as follows. In Sec.~\ref{ssec:originalBasisReview} we provide a brief overview of the \uone\ formulation constructed in Ref.~\cite{Bauer_2021}. In Sec.~\ref{ssec:weavedBasisDigitization} we then review the operator change of basis method presented in Ref.~\cite{https://doi.org/10.48550/arxiv.2208.03333} and discuss modifications required to minimized digitization errors. From there, in Sec.~\ref{ssec:PrecisionComparison}, we perform a numerical study of the digitization errors for both the original basis and weaved basis. In Sec.~\ref{sec:WalshIntro} we review the Walsh function formalism and how it is used to construct diagonal matrices. We then discuss in detail the scaling of the number of gates, as well as the classical computing overhead, required to implement the time evolution of a 2+1 dimensional compact \uone\ gauge theory in the original basis. In Sec.~\ref{ssec:ChangeOfBasisGateCount} we show that in the new operator basis both the quantum gate count and the associated classical computing overhead scale polynomially with the volume. We then implement the quantum circuits using the Walsh function formalism and show how the gate counts can be further reduced by approximating the circuit. After some numerical studies, including a simulation on an IBMQ quantum devices, we present our conclusions.

\section{U(1) formulations used in this paper}
\label{sec:U1Review}
In this section we start by reviewing the formulation of a compact \uone\ gauge theory with a Hilbert space that has been constrained to satisfy Gauss' law. We also review the representation of the magnetic and electric operators introduced in Ref.~\cite{Bauer_2021}, which can be used at all values of the coupling. Next, we review the operator basis change method in Ref.~\cite{https://doi.org/10.48550/arxiv.2208.03333} and discuss the changes in the digitization scheme necessary to keep an efficient representation in this new operator basis. We conclude by performing numerical tests comparing the precision of the weaved basis to the original basis, and find they are comparable.

\subsection{Standard rotor formulation}
\label{ssec:originalBasisReview}
The Hamiltonian considered in Ref.~\cite{Bauer_2021} is formulated in terms of electric rotor and magnetic plaquette operators, given by $\hat{R}(x)$ and $\hat{B}(x)$, respectively. These operators satisfy
\begin{align}
    \left[ \hat B(x), \hat R(y) \right] = i \,  \delta^3(x-y)
    \,.
\end{align}
The discretized version of the theory we consider introduces a periodic lattice of $N_x$ and $N_y$ evenly spaced lattice points in the $\hat x$ and $\hat y$ dimensions with a lattice spacing $a$. The lattice version of the continuum Hamiltonian is defined in terms of operators, $\hat{R}_p$ and $\hat{B}_p$, with the index $p$ denoting a specific plaquette in the lattice volume. 
The Hamiltonian can be written in terms of an electric and magnetic component
\begin{align}
\hat{H} = \hat{H}_E + \hat{H}_B\,,
\end{align}
with the electric Hamiltonian given by
\begin{align}
\hat{H}_E = \frac{g^2}{2a}\sum_{p=1}^{N_x \cdot N_y}
     (\vec{\nabla} \times \hat R_p)^2 
     \,.
\end{align}
The magnetic Hamiltonian can be written in one of two ways. The first is the `non-compact' formulation, given by
\begin{align}
    \hat{H}_B^{\rm (NC)} = \frac{1}{2a\,g^2 }\sum_{p=1}^{N_x \cdot N_y}
    ( \hat B_p)^2
    \,,
\label{eq:NonCompactH}
\end{align}
while the second formulation, called the `compact' formulation, is given by
\begin{align}
    \hat{H}_B^{\rm (C)} =  \frac{1}{a\, g^2}\sum_{p=1}^{N_x \cdot N_y}
    (1-\cos \hat B_p).
    \label{eq:compactHB_before_constraint}
\end{align}
Note that the true continuum limit of these two formulations is not necessarily the same and in fact, the continuum limit of the compact formulation is not unique or universal. However, the compact theory shares several interesting features with more complicated gauge theories, which is why its study is illuminating in its own right. Second, in non-Abelian gauge theories the gauge field is necessarily compact, providing another reason why a compact \uone\ theory is often studied in detail. 

We will call the bases where the operators $\hat{H}_E$ and $\hat{H}_B$ are diagonal the electric and magnetic basis, respectively. Furthermore, we denote operators in the electric and magnetic basis with superscripts $(e)$ and $(m)$, respectively. Because $\hat{R}_p$ and $\hat{B}_p$ are conjugate operators, the electric and magnetic basis are related by a Fourier transformation.

The magnetic Gauss' law gives rise to one additional constraint, namely
\begin{align}
\label{eq:constraint}
    \hat R_{N_x \cdot N_y} = 0\,, \qquad \hat B_{N_x \cdot N_y} = -\sum_{p=1}^{N_x \cdot N_y - 1} \hat{B}_p
    \,.
\end{align}
The number of independent plaquettes is therefore $N_p \equiv N_x N_y - 1$. Taking this constraint into account, the compact magnetic Hamiltonian in Eq.~\eqref{eq:compactHB_before_constraint} becomes (up to an overall constant) 
\begin{align}
    \hat{H}_B^{\rm (C)} =  -\frac{1}{a\, g^2}\left[\sum_{p=1}^{N_p} \cos \hat B_p + \cos \left(\sum_{p=1}^{N_p} \hat{B}_p \right) \right].
    \label{eq:compactHB_after_constraint}
\end{align}
Notice that the $\cos (\sum_{p} \hat{B}_p)$ term couples the entire lattice together. As explained in Ref.~\cite{https://doi.org/10.48550/arxiv.2208.03333}, this fact, combined with the non-polynomial functional form, is the source of the exponential volume scaling in the circuit depth. We show in Sec.~\ref{ssec:resourceOriginalBasis} that this term is also the source of the classical computing overhead that scales exponentially in the lattice volume when building the time evolution operator. While it may be possible to perform time evolution without enforcing magnetic Gauss' law \textit{a priori}, we focus in this work on implementations that do enforce all gauge constraints.

Various representations for the operators $\hat{R}_p$ and $\hat{B}_p$ are possible, and different choices result in differing levels of efficiency (in both the number of qubits and the number of gates required), as well as faithfulness in the representation of the low-lying eigenstates. Most representations are typically optimized for either weak or strong couplings, where either the magnetic or electric Hamiltonian dominate. In Ref.~\cite{Bauer_2021} a representation was developed that is precise at both weak and strong coupling that works in the magnetic basis. That representation depends on a maximum value $b_{\rm max}$ for the magnetic field, which is chosen to optimize the representation and depends on the number of digitized states (for details, see Ref.~\cite{Bauer_2021}). Note that the standard choice of $b_{\text{max}}=\pi$ recovers the full range of possible magnetic field values in the compact theory. However, as explained in Ref.~\cite{Bauer_2021} smaller values of $b_{\text{max}}$ should be used to obtain an efficient representation at small coupling. The procedure for choosing $b_\text{max}$ is explained later in this section.

The basis states of this representation are labeled by a set of integers $0 \leq l_p \leq N-1$
for each plaquette, where 
\begin{align}
    N = 2^{n_q}
    \,,
\end{align}
and $n_q$ is the number of qubits used to represent each lattice site. The dimension of the Hilbert space is therefore
\begin{align}
    {\rm dim}_H = 2^{n_q N_p}
    \,.
\end{align}
The magnetic field operator in the magnetic basis is diagonal and defined as
\begin{align}
    \hat{B}^{(m)}_p \ket{b^{(p)}_l} = \left(-b_\text{max}+ l_p \, \delta b\right)\ket{b^{(p)}_l},
\end{align}
where $\delta b = 2b_\text{max}/N$. As previously mentioned, because $\hat{R}_p$ is conjugate to $\hat{B}_p$, the rotor operator in the magnetic basis can be written as
\begin{align}
    \hat{R}^{(m)}_p &= {\rm \widehat{FT}}^{-1} \hat{R}^{(e)}_p {\rm \widehat{FT}}, 
    \\
    \hat{R}^{(e)}_p \ket{r^{(p)}_l} &= \left(-r_\text{max}+l_p\, \delta r \right) \ket{r^{(p)}_l}\nonumber 
    \,,
    \label{eq:sampleR}
\end{align}
where ${\rm \widehat{FT}}$ denotes the usual quantum Fourier transform (QFT) and
\begin{align}
    r_\text{max}=\frac{\pi N}{2b_{\rm max}}\,, \qquad \delta r = \frac{\pi}{b_{\rm max} }
    \, ;
\end{align} 
note that the choice of eigenvalues of $\hat{B}_p$ coincide with a choice of twisted boundary conditions. 
This representation does not rely on writing either the electric or magnetic operators as raising or lowering operators as most representations use. We set $a=1$ for the remainder of the paper.

Before discussing the weaved basis, it will be useful to review the prescription for choosing $b^{(i)}_\text{max}$ for the plaquette $i$ in the original basis, as presented in Ref.~\cite{Bauer_2021}. For the non-compact formulation, the optimal choice was shown to be~\cite{Bauer_2021}
\begin{align}
    b^{(i), \text{NC}}_\text{max} &= g \frac{2^{n_q}}{2} \sqrt{\frac{\beta_R^{(i)}}{\beta_B^{(i)}}} \sqrt{\frac{\sqrt{8} \pi}{2^{n_q}}}
    \,,
    \label{eq:bMaxNC}
\end{align}
 where $b_\text{max}^{\text{NC}}$ indicates the operator $\hat{B}_i$ in the non-compact theory is sampled in the range $[-b_\text{max}^\text{NC},\, b_\text{max}^\text{NC}]$. The variables $\beta_R^{(i)}$ and $\beta_B^{(i)}$ are found by matching the non-compact magnetic Hamiltonian to a Hamiltonian, inspired by the quantum harmonic oscillator, of the form
\begin{align}
H_\text{QHO} = \frac{g^2}{2} (\beta_R^{(i)})^2 \hat{R}^2_i + \frac{1}{2g^2} (\beta_B^{(i)})^2 \hat{B}^2_i
\,,
\end{align}
and ignoring the cross-terms. As the operator $\hat{B}_i$ is allowed to take on any value in $\mathbb{R}$, this optimal value works well for all values of the gauge couplings. Note that this result is found by replacing $2l+1$ in the results of Ref.~\cite{Bauer_2021} with $2^{n_q}$.

For the compact formulation, the choice of $b_\text{max}$ is more complicated. For small values of $g$, the wavefunction in the magnetic basis is sharply peaked around $b = 0$ and so choosing a width according to the non-compact theory will sample the wavefunction efficiently in the region where it has support; another way to understand this result is that the compact Hamiltonian reduces to the non-compact Hamiltonian in the limit of weak coupling (though this has to be treated carefully). As the gauge coupling is increased, the region where the wavefunction has support increases. Eventually, the wavefunction has support over the entire interval $[-\pi, \pi]$ and $b_\text{max}$ cannot become any larger. The upper limit of the value of $b_\text{max}^{(i)}$ is chosen such that the Fourier transform of $(1-\cos \hat{B}_i)$ is the discrete version of a second order derivative with periodic boundary conditions, which can also be written in terms of raising operators, lowering operators and the identity matrix. This choice of $b_\text{max}$ in the large $g$ limit is motivated by the formulation of Ref.~\cite{Bauer_2021} needing to match onto the Kogut-Susskind formulation \cite{PhysRevD.11.395} at large coupling. Therefore, for the compact formulation we use
\begin{align}
    b^{(i), \text{C}}_\text{max} &= \text{min} \Big(b^{(i), \text{NC}}_\text{max}, \pi\Big),
\end{align}
where $b^{(i), \text{NC}}_\text{max}$ is given in  Eq.~\eqref{eq:bMaxNC}.

This prescription for $b^{(i), \text{C}}_\text{max}$ in the large $g$ limit assumes that the coefficient of a magnetic field operator $\hat{B}_i$ is unity anywhere it appears in the compact magnetic Hamiltonian; for any operator basis where this is not true, it must be modified. We discuss the modified procedure in the next section.

\subsection{Weaved basis}
\label{ssec:weavedBasisDigitization}
The method presented in Ref.~\cite{https://doi.org/10.48550/arxiv.2208.03333} that breaks the previously mentioned exponential volume scaling in the gate count involves performing an operator change of basis of the form
\begin{align}
\hat{B}_p &\rightarrow \mathcal{W}_{p p'} \hat{B}_{p'} \nonumber \\
\hat{R}_p &\rightarrow \mathcal{W}_{p p'} \hat{R}_{p'},
\label{eq:opBasisChange}
\end{align}
where $\CW$ is a carefully chosen orthogonal matrix. We will refer to the operator redefinition as the weaved basis. Specifically, it was shown that, for any value of $N_p$, there exists a choice of $\CW$ that reduces the scaling from exponential to polynomial in $N_p$ \cite{https://doi.org/10.48550/arxiv.2208.03333}.
We will not review the weaved basis here, since all required details can be found in the original paper. 

In this weaved basis, if one chooses to represent the $\hat{B}_p$ and $\hat{R}_p$ operators as in Ref.~\cite{Bauer_2021}, the procedure for choosing $b_{\rm max}$ must be changed in order to keep digitization errors comparable to the original basis. While this was mentioned briefly in Ref.~\cite{https://doi.org/10.48550/arxiv.2208.03333}, we will provide more details in the following paragraphs.

As already mentioned in Sec.~\ref{ssec:originalBasisReview}, this prescription for $b^{(i), \text{C}}_\text{max}$ in the large $g$ limit assumes that the coefficient of a magnetic field operator $\hat{B}_i$ is unity anywhere it appears in the compact magnetic Hamiltonian. In the weaved basis, this will in general no longer be true, with some operators having coefficients smaller than one. Consequently, even if $b_{\rm max}^{(i),C} = \pi$, an operator $\hat{B}_i$ inside a given cosine will not get sampled between the full range of $[-\pi, \pi]$. To fix this problem, we propose scaling the upper limit for each $b_{\rm max}^{(i),C}$ by the smallest coefficient of the operator $\hat{B}_i$ anywhere it appears in the Hamiltonian. As a simple example demonstrating this method, we work through the $N_p=3$ case. The rotation matrix used is
\begin{align}
    \label{eq:weavedNp3}
    \CW = \frac{1}{\sqrt{6}} \left(
\begin{array}{ccc}
 \sqrt{2} & -2 & 0 \\
 \sqrt{2} & 1 & -\sqrt{3} \\
 \sqrt{2} & 1 & \sqrt{3} \\
\end{array}
\right)
    \,,
\end{align}
which leads to the following Hamiltonian
\begin{align}
    \hat{H}_B^{C, \text{w}} &= -\frac{1}{g^2} \Bigg( \cos\left[\sqrt{3} \hat{B}_1\right] + \cos\left[\frac{\hat{B}_1 - \sqrt{2} \hat{B}_2}{\sqrt{3}}\right] 
    \nonumber\\
    & \quad + \cos\left[\frac{\sqrt{2} \hat{B}_1+\hat{B}_2 - \sqrt{3} \hat{B}_3}{\sqrt{6}}\right] \nonumber \\
    &\quad +\cos\left[\frac{\sqrt{2} \hat{B}_1+\hat{B}_2 + \sqrt{3} \hat{B}_3}{\sqrt{6}}\right] \Bigg)
    \,.
    \label{eq:HExamp}
\end{align}
As already mentioned, in order to ensure that the each $\hat{B}_i$ gets sampled between $[-\pi, \pi]$ in all of the cosine terms, we must scale the upper limit for each $b_\text{max}^{(i),C}$. In this example one would choose $b_\text{max}^{(i),C}$ values according to
\begin{align}
    b^{(1), \text{C}}_\text{max} &= \text{min} \Big(b^{(1), \text{NC}}_\text{max}, \sqrt{3} \pi\Big) \nonumber
    \\
    b^{(2), \text{C}}_\text{max} &= \text{min} \Big(b^{(2), \text{NC}}_\text{max}, \sqrt{6} \pi \Big) \nonumber
    \\
    b^{(3), \text{C}}_\text{max} &= \text{min} \Big(b^{(3), \text{NC}}_\text{max}, \sqrt{2} \pi \Big)\,.
\label{eq:scaled_bmax}
\end{align}
With this choice, the arguments of certain terms will exceed $\pi$; for example, $\sqrt{2/3}$ times the upper limit for $b_\text{max}^{(2),C}$ will be greater than $\pi$. 
Because the wavefunction is periodic, these values still provide information about the wavefunction, but they likely sample $\hat{B}_i$ at non-optimal values, which can lead to larger digitization errors for a given number of qubits per lattice site. Note that even though one is only performing an operator basis change in this case, the numerical results can still differ due to the presence of the cutoff $b_{\rm max}$. We study the digitization effects of both basis in the next section.

\subsection{Precision comparison}
\label{ssec:PrecisionComparison}
In this section we compare the precision of the weaved basis to the original basis for the $2\times 2$ site lattice with $N_p=3$ for three quantities, namely the energies of the non-compact Hamiltonian, the energies of the compact Hamiltonian, and the expectation value of the plaquette.

\begin{figure}[t]
    \centering
    \includegraphics[width=0.45\textwidth]{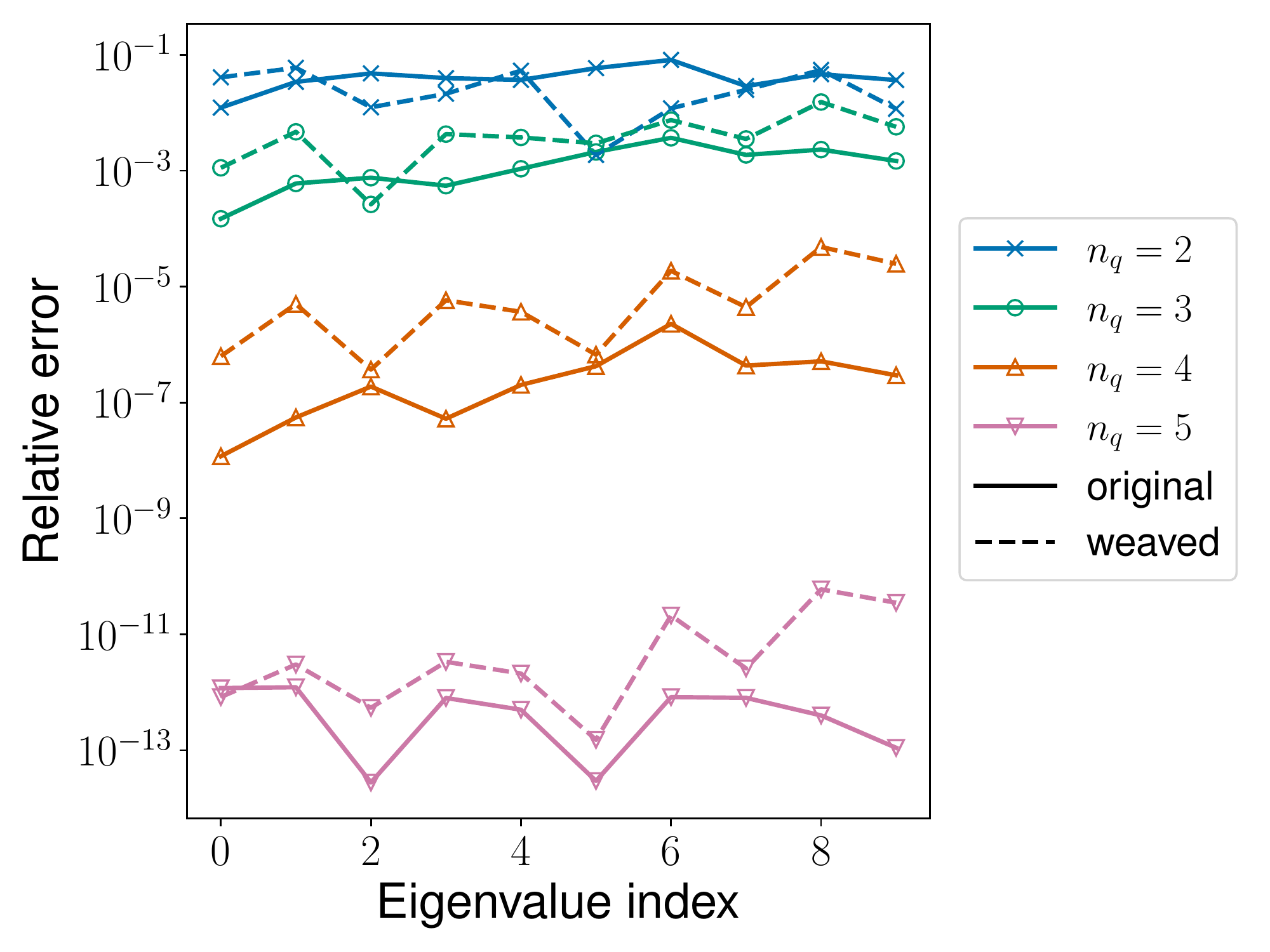}
    \caption{The relative error between the lowest 10 exactly known eigenvalues in the non-compact theory to the numerically calculated eigenvalues in the digitized theory using $N_p=3$. The different colored lines indicate different values of $n_q$. The solid (dashed) lines indicate the original (weaved) basis.}
    \label{fig:error_nc_evals}
\end{figure}

We start by comparing the precision of the digitized energies of the two bases in the non-compact case for $N_p=3$. Specifically, we compare the digitized energies to the analytically known energies of the non-compact Hamiltonian. Note that the eigenvalues in the non-compact formulation are independent of the coupling $g$. Figure \ref{fig:error_nc_evals} shows the relative error on the lowest 10 non-compact eigenvalues for both the original basis and weaved basis for different values of $n_q$. The weaved matrix used is given in Eq.~\eqref{eq:weavedNp3}. We notice that the errors are comparable, and for $n_q=3,4,5$ the original basis is generally slightly more precise. One possibility for this behavior is the previously mentioned fact that our digitization scheme in the weaved basis likely samples the magnetic field operators at non-optimal values. However, for $n_q\geq 3$ the difference in precision between the weaved and original basis will not be relevant unless a precision of $0.1\%$ is required for the quantum computation.

\begin{figure}[t]
    \centering
    \includegraphics[width=0.45\textwidth]{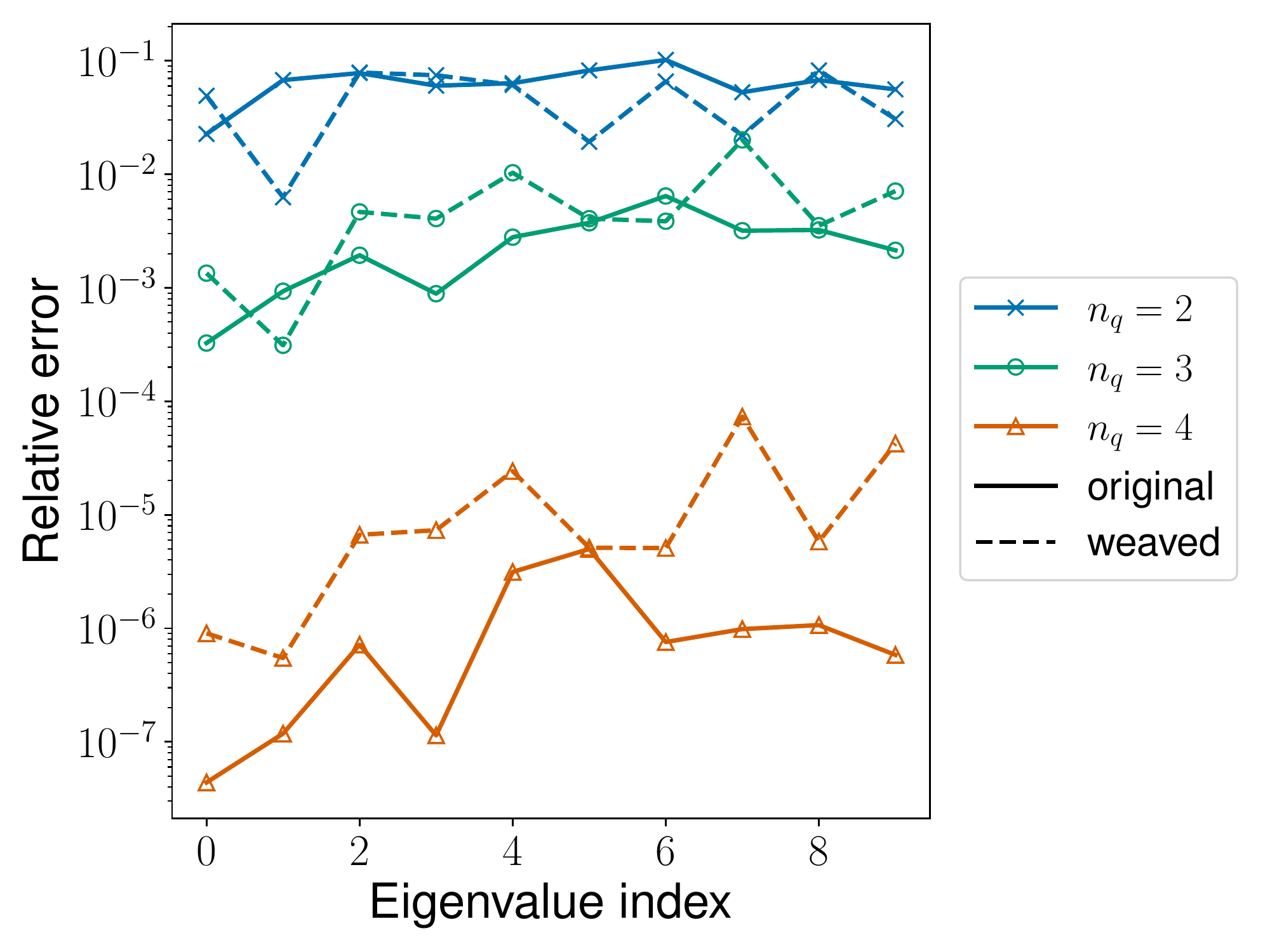}
    \caption{The relative error in the compact theory between the ``exact'' eigenvalues and numerically calculated eigenvalues in the digitized theory for $N_p=3$ and $g=0.2$. The different colored lines indicate different values of $n_q$. The solid (dashed) lines indicate the original (weaved) basis. The ``exact'' eigenvalues were taken to be the original basis with $n_q=5$.}
    \label{fig:dE_compact}
\end{figure}

Next, we compare the energies of the two bases in the compact formulation using a gauge coupling of $g=0.2$. Because the eigenvalues are not known analytically in the compact case, we choose the original basis eigenvalues with $n_q=5$ as the ``exact'' values to compare against. Figure \ref{fig:dE_compact} shows the relative error of the lowest 10 eigenvalues for both bases. For $n_q=2$, the weaved basis is generally slightly more precise. For $n_q=3$ the original and weaved basis have similar precision and $n_q=4$ the original basis is more precise (however, both basis have precision below 1 part in $10^4$). As with the non-compact case, for $n_q\geq 3$ the differences in precision between the two bases will not be relevant unless a precision of $\sim 0.1\%$ or greater is required.  This difference in precision will not be important for calculations done using NISQ hardware due to noise in the quantum device. Even if one uses fault-tolerant quantum computers, many physically relevant calculations will not require a sub-percent precision to provide phenomenologically useful results and so the difference in precision will not be important for these cases either.

Lastly, we compare the expectation value of the plaquette in the compact theory, defined as 
\begin{align}
    \langle \square \rangle = 1+\frac{g^2}{N_p+1} \bra{\Psi_0} \hat{H}_B^{(C), (\text{w})} \ket{\Psi_0}
    \,, 
\end{align}
where $\ket{\Psi_0}$ is the ground state of the full Hamiltonian and $N_p=3$. To understand the expected behavior for extreme values of $g$, it is sufficient to set $b_\text{max}=\pi$ and work in the original basis (assuming $n_q$ can be taken large to remove digitization effects). In this case, at small $g$, the Hamiltonian tends towards a harmonic oscillator Hamiltonian and so the ground state is a Gaussian centered at $b_{l_p}^{(p)}=0$ with a width $\sim g$. As $g$ approaches zero, the contribution from $b_{l_p}^{(p)}=0$ dominates. Looking at Eq.~\eqref{eq:compactHB_after_constraint} and plugging in $b_{l_p}^{(p)}=0$, we see that $\bra{\Psi_0}\hat{H}_B^{(C)}\ket{\Psi_0} = -(N_p+1)/g^2$ which implies that the plaquette expectation value is zero at small $g$. At large $g$, the electric Hamiltonian dominates and $\ket{\Psi_0}$ is the ground state of $\hat{H}_E$. If the rotors are sampled according to Eq.~\eqref{eq:sampleR}, the ground state is degenerate, and can be written as $|r^{(1)}_{l_1}, \dots r^{(N_p)}_{l_{N_p}} \rangle$. The canonical commutation relations between $\hat{R}_p$ and $\hat{B}_p$ imply that $e^{i \hat{B}_p}$ is a lowering operator on rotor eigenstates. Writing each $\cos \hat{B}_p = (e^{i \hat{B}_p}+e^{-i \hat{B}_p})/2$, we see that $\hat{H}_B^{(C)}$ applied to the electric ground state will be orthogonal to the electric ground state, and therefore the expectation value of the plaquette at large $g$ is one.

\begin{figure}[t]
    \centering
    \includegraphics[width=0.48\textwidth]{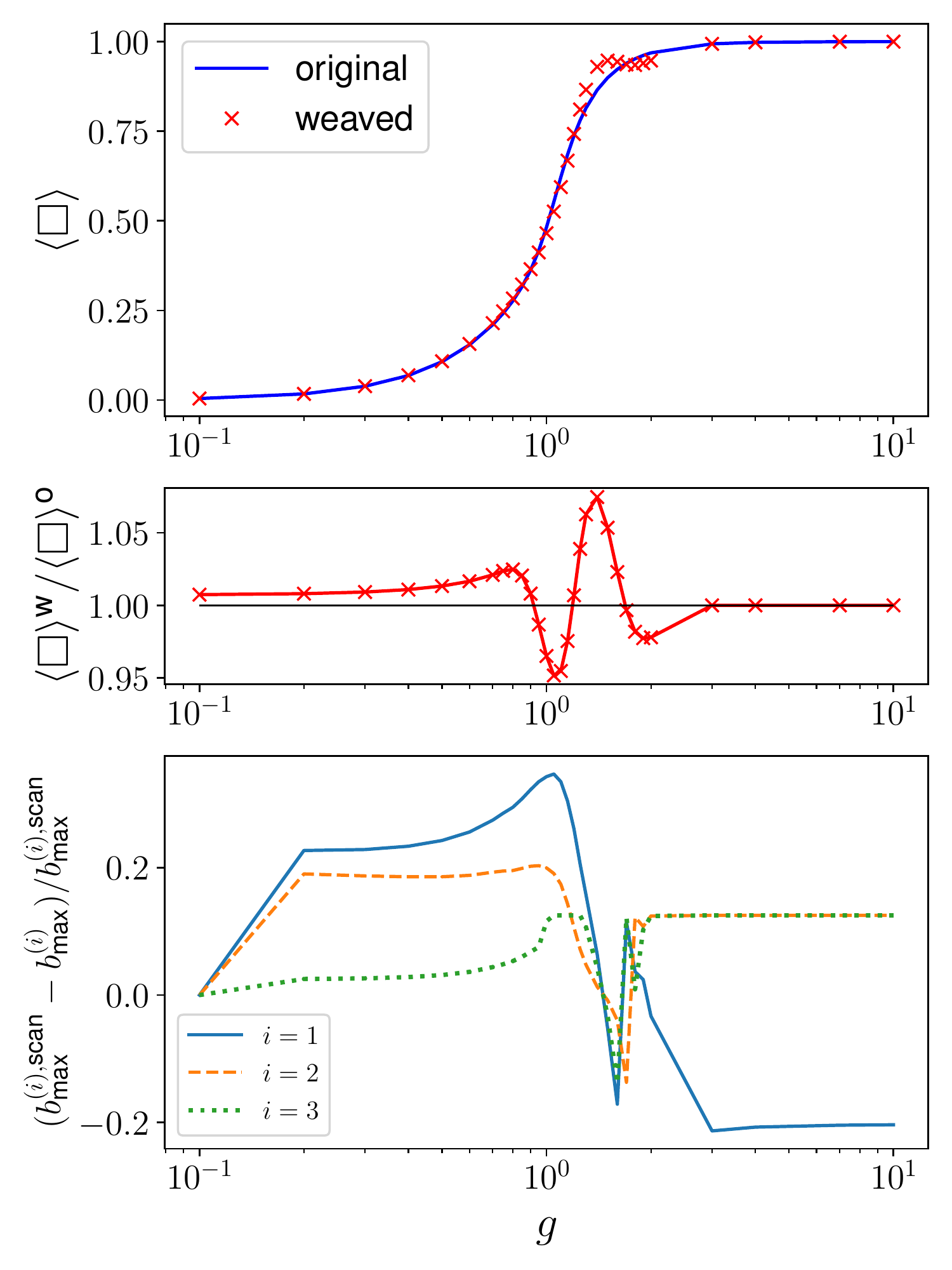}
    \caption{Top: Expectation value of the plaquette as a function of the coupling $g$ for $N_p=3$ and $n_q=3$. The solid blue line is the original basis, and the red cross marks are the weaved basis. Middle: Ratio of the expectation value of the plaquette for the weaved basis to the original basis as a function of the coupling $g$. Bottom: Relative difference of $b_\text{max}$ resulting from a scan which makes the weaved plaquette expectation value equal to the original plaquette expectation value and the value of $b_\text{max}$ used to calculate the plaquette expectation value in the weaved basis using Eq.~\ref{eq:scaled_bmax} as a function of $g$.}
    \label{fig:plaq_expval_weaved}
\end{figure}

The top plot in Fig.~\ref{fig:plaq_expval_weaved} shows the value of the plaquette calculated for both the original and weaved basis using $n_q=3$ which corresponds to sampling an operator 8 times per lattice site. In Ref.~\cite{Bauer_2021}, using $l=3$, the plaquette expectation value for the original basis was within 0.1\% of the value calculated using $l=6$ for the compact formulation \footnote{In Ref.~\cite{Bauer_2021}, the number of samplings for an operator was given by $2\ell+1$} and so we take the expectation value calculated using the original basis with $n_q = 3$ as the correct value. The middle plot in Fig.~\ref{fig:plaq_expval_weaved} shows the ratio of the plaquette expectation value in the weaved basis to the original basis. We see that the prescription works well for values all values of $g$ except $0.6 \lesssim g \lesssim 3.0$ where the largest deviation is 7.5\%.

By scanning for the optimal values of $b_\text{max}^{(i)}$ we were able to get the plaquette expectation value in the weaved basis to be equal to the original basis. A plot of the ratio of the values of $b_\text{max}^{(i)}$ calculated using Eq.~\eqref{eq:scaled_bmax} to the scanned values is shown in the bottom plot of Fig.~\ref{fig:plaq_expval_weaved}. The scanned values of $b_\text{max}$ are slightly different than the prescription we provide, with values varying more dramatically near $g\sim 1$. The values approach a constant because for large $g$, the values of $b_\text{max}$ reach their upper limit given in Eq.~\eqref{eq:scaled_bmax}. It is possible that one will have to perform some kind of scan to get the optimal values of $b_\text{max}$ in this basis when working in the coupling regime $g \sim 1$. A similar procedure for tuning the action parameters is required in modern lattice QCD methods when using effective actions, see e.g. Refs.~\cite{Aoki_2012, El_Khadra_1997,Aoki_2003, FermilabLattice:2010rur, Lin_2007, Aoki_2004}. Alternatively, one could use larger values of $n_q$ to reduce the digitization errors.

\section{Walsh function implementation of diagonal matrices}
\label{sec:WalshIntro}
As will be discussed in more detail in Sec.~\ref{sec:resourceStudy}, efficiently implementing time evolution via Susuki-Trotter methods of this formulation of a \uone gauge theory reduces to being able to efficiently implement diagonal matrices. In this section, we review the Walsh function formalism \cite{Welch_2014}, which is an efficient algorithm for constructing diagonal unitary operators without ancillary qubits. We first review how to construct an arbitrary $m$ qubit diagonal operator exactly, which in general requires $\CO(2^{m})$ gates. Then we review how this algorithm can be modified to implement arbitrary $m$ qubit diagonal matrices to a precision $\epsilon$ using $\text{poly}(m, 1/\epsilon)$ gates, where $\text{poly}(m, 1/\epsilon)$ is some polynomial function of $m$ and $1/\epsilon$. We conclude by discussing a class of matrices where the classical computing cost associated with constructing the efficient quantum circuit scales exponentially with the size of the circuit, and how this will be prohibitively expensive for large diagonal matrices. As will be discussed, the Walsh function implementation will result in circuits containing rotation matrices $R_z(\theta_i)$ sandwiched between CNOT gates, where the rotation angles are determined by the values of the diagonal matrix to be implemented. As we will see, for many matrices some of these rotation angles can turn out to be quite small. An exact implementation will include all of the rotations, while an approximation can be obtained where rotations with angles below a certain cutoff $\theta_{\rm min}$ are dropped. We will analyze the two different implementations in the following sections.

\subsection{Exact implementation (with $\theta_{\rm min} = 0$)}
\label{ssec:ExactWalsh}
Because arbitrary diagonal matrices acting on $n$ qubits in general have $2^{n}$ independent entries, we can expect that the construction of such a matrix on a quantum computer requires an exponential number of gates. Indeed, it was shown in  Ref.~\cite{Bullock_2003} that the asymptotically optimal gate count scaling for constructing arbitrary $n
$ qubit diagonal matrices is $\CO(2^{n})$. The Walsh function formalism provides an algorithm for implementing arbitrary ${n}$ qubit diagonal unitary matrices without ancillary qubits using $2^{{n}}-2$ CNOT gates and $2^{{n}}-1$ single qubit $R_z$ rotation gates \cite{Welch_2014}, which matches the asymptotically optimal scaling. In addition, because Walsh functions are analogous to the Fourier series, they share the beneficial properties with regards to systematically approximating functions to a desired precision. This property can be leveraged to efficiently implement $2^n \times 2^n$ diagonal unitary matrices to a precision $\epsilon$ using $\CO\left(\text{poly}(n, 1/\epsilon)\right)$ gates \cite{Welch_2014}. A more complete introduction to the Walsh function formalism can be found in Ref.~\cite{Welch_2014}. We also provide the necessary details for implementing the Walsh function formalism in Appendix~\ref{app:WalshImplementation}.

The task is to construct a $2^{n} \times 2^{n}$ diagonal unitary matrix $\hat{U} = e^{i \hat{H}}$. The $N = 2^n$ real phases are stored in the matrix $\hat{H}$. On a quantum computer, Walsh operators are diagonal operators with values $\pm 1$ on the diagonal and form a basis for real diagonal matrices. The $j$'th Paley-ordered Walsh operator for an ${n}$ qubit system is given by
\begin{align}
    \hat{w}_j = \bigotimes_{i=1}^{n} (\sigma^z_i)^{j_i}
    \,,
\end{align}
acting on a state representing a dyadic binary representation of the integer $k$
where, $j_i$ is the $i$'th bit of the binary expansion of the integer $j$, $\sigma^z_i$ is the Pauli matrix acting on the $i$'th qubit, and $j \in [0, N-1]$. These operators satisfy the orthogonality relation
\begin{align}
    \Tr[\hat{w}_j \hat{w}_k] = N \delta_{jk}
    \,.
\end{align}
The phases on the diagonal of $\hat{U}$ can be written as
\begin{align}
    \hat{H} = \sum_{j=0}^{N-1} a_j \hat{w}_j
    \,,
\end{align}
where the Walsh coefficients $a_j$ can be calculated by 
\begin{align}
    a_j = \frac{1}{N} \Tr[ \hat{H} \hat{w}_j ]
    \,.
\end{align}
Note that the Walsh coefficients can be calculated using the fast-Walsh-transform which requires $\CO(N \log_2 N)$ floating point operations \cite{Yarlagadda_Hershey}. 

\begin{figure}[t]
    \centering
    \begin{quantikz}[row sep={0.55cm,between origins}, column sep=0.4cm]
        \lstick{$q_1$} & \ctrl{3} & \qw & \qw & \qw & \ctrl{3} & \qw
        \\
        \lstick{$q_2$} & \qw & \qw & \qw & \qw & \qw & \qw
        \\
        \lstick{$q_3$} & \qw & \ctrl{1} & \qw & \ctrl{1} & \qw & \qw
        \\
        \lstick{$q_4$} & \targ{} & \targ{} & \gate{R_z(-2 a_{13})} & \targ{} & \targ{} & \qw 
    \end{quantikz}
    \caption{The quantum circuit corresponding to the operator $e^{i a_{13} \hat{w}_{13}} = \exp(i a_{13} \sigma_z \otimes \mathds{1} \otimes \sigma_z \otimes \sigma_z)$.}
    \label{fig:exp_of_walsh_example}
\end{figure}
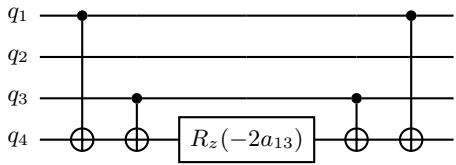

Many quantum algorithms (those involving time evolution for example) require the exponentiation of Walsh operators in the form $\exp(i a_j \hat{w}_j)$. They are constructed by placing an $R_z(\theta_j)$ gate between two sets of CNOT gates. The controls and targets of the CNOT gates are determined from the index $j$. The rotation angle of the $R_z(\theta_j)$ is given by $\theta_j = -2a_j$. As an example, the circuit for $\exp(i a_{13} w_{13})$ is shown in Fig.~\ref{fig:exp_of_walsh_example}. 
The diagonal unitary operator with the phases on the diagonal given by $\hat{H} = \sum_{i=1}^n a_i \hat{w}_i$ is implemented as a product of exponentials of Walsh operators, $\hat{U} = e^{i \hat{H}} = \prod_{j=0}^{N-1} e^{i a_j \hat{w}_j}$. 

Note that because the Walsh operators are diagonal, the terms in the product commute and one is free to choose an order that minimizes the number of CNOT gates in the circuit. This is achieved by ordering the Walsh operators according to their Gray binary ordering, \textit{i.e.} using sequency ordered Walsh operators \cite{Welch_2014}. For an $n$-qubit circuit, the optimal ordering requires $(2^{n}-1)$ $R_z$ gates and $(2^{n}-2)$ CNOT gates for a total of $(2^{n+1}-3)$ gates, which, as stated previously, is asymptotically optimal~\cite{Bullock_2003}. A nice worked-through example for constructing the optimal three qubit circuit can be found in Ref.~\cite{Welch_2014}. The resulting optimal three qubit circuit is shown in the top circuit of Fig.~\ref{fig:circ_simplification}. Larger circuits follow the same general pattern of alternating CNOT and $R_z$ gates.

When implementing diagonal unitary matrices where the phases on the diagonal are written as sums of different diagonal matrices, an extra step must be performed to produce the shortest depth circuit. For example, consider a Hamiltonian that contains the term
$\cos (\hat{B}_1 + \hat{B}_2)$ and another term $\cos (\hat{B}_2 + \hat{B}_3)$. 
Implementing each term independently requires $2\times\left( 2^{2n+1}-3\right)$ gates. However, the sum of these two terms can be written in terms of Walsh operators as 
\begin{align}
& \cos (\hat{B}_1 + \hat{B}_2) +\cos (\hat{B}_2 + \hat{B}_3) 
\nonumber\\
&= \sum_{i,j=0}^{2^{n}-1} a_{i} a_j \hat{w}_i \otimes \hat{w}_j \otimes \mathds{1} \nonumber\\
&\qquad  + \sum_{k,l=0}^{2^{n}-1} b_k b_l \mathds{1} \otimes \hat{w}_k \otimes \hat{w}_l.
\end{align}
Because $\hat{w}_0 = \mathds{1}$, the Walsh operators in the sums with $i=0$ and $l=0$ will be the same and we can group common Walsh operators as
\begin{align}
& \cos (\hat{B}_1 + \hat{B}_2) +\cos (\hat{B}_2 + \hat{B}_3)  \nonumber
\\
&= \sum_{m=0}^{2^{n}-1} (a_0 a_m + b_m b_0) \mathds{1} \otimes \hat{w}_m \otimes \mathds{1} \nonumber
\\
&\qquad + \sum_{i=1,j=0}^{2^{n}-1} a_{i} a_j \hat{w}_i \otimes \hat{w}_j \otimes \mathds{1} \nonumber
\\
&\qquad  + \sum_{k=0,l=1}^{2^{n}-1} b_k b_l \mathds{1} \otimes \hat{w}_k \otimes \hat{w}_l.
\end{align}
Instead of implementing the common Walsh operators separately, they can be implemented once as long as one uses a Walsh coefficient given by the sum of the individual Walsh coefficients. Taking into account that these two terms share $2^{n}$ Walsh operators will result in a gate count less than $2(2^{2n+1}-3)$.

It is known that certain classes of diagonal unitary matrices can be implemented in less than $(2^{n+1}-3)$ gates, see \textit{e.g.} Ref.~\cite{PhysRevA.99.052335}. 
To see how this simplification occurs using the Walsh function formalism, consider constructing the operator $e^{i \hat{x}}$ using $n_q$ qubits, where $\hat{x}$ is a diagonal matrix with evenly spaced entries on the diagonal. 
As pointed out in Ref.~\cite{PhysRevA.99.052335}, such an operator $\hat{x}$ can be written as $\hat{x} = \frac{x_\text{max}}{2^{n_q}-1}\sum_{j=1}^{n_q} 2^j \sigma^z_j$. 
We immediately see that this operator is a sum of only $n_q$ Walsh functions, as opposed to the $2^{n_q}$ required for an arbitrary diagonal unitary.
The complex exponential of this operator can therefore be implemented in $\mathcal{O}(n_q)$ gates. From this, one can also see that any product of two such operators, \textit{e.g.} $\hat{x}^2, \hat{x}_1 \otimes \hat{x}_2$, is a sum of $\mathcal{O}(n_q^2)$ Walsh functions and exponentiation requires $\mathcal{O}(n_q^2)$ gates. This simplification will be important for reducing the cost of implementing exponentials of both the electric Hamiltonian and the non-compact magnetic Hamiltonian, which are linear combinations of such bilinear terms. The scaling for a general arbitrary unitary can be found by writing the desired phases in terms of the operator $\hat{x}$. 

\subsection{Efficient implementation (with non-zero $\theta_{\rm min}$)}
\label{ssec:removingSmallWalsh}
We now review a modification to this algorithm, originally described in Ref.~\cite{Welch_2014}, that leads to an efficient algorithm for building arbitrary diagonal unitary matrices. It is intuitively clear that if some of the rotation angles $\theta_i=-2a_i$ get very small, they can be dropped from the circuit without introducing large errors. Doing so also reduces the number of CNOT gates required, making this an important method for NISQ era devices. Furthermore, for fault-tolerant devices, $R_z$ gates with high precision are expensive to implement, requiring $1.15 \log_2(1/\epsilon)$ \mbox{T-gates} \cite{Bocharov_2015}, where $\epsilon$ is the precision which the gate is implemented. Dropping rotation gates with angles less than some threshold $\theta_\text{min}$ is equivalent to rounding to some fixed point accuracy \cite{CK_Yuen} and therefore imposes a limit to the precision $\epsilon$. This in turn limits the number of \mbox{T-gates} required to implement a given fault-tolerant rotation gate, and therefoer this method is also important for fault-tolerant devices.

We follow Ref.~\cite{Welch_2014} and define the error between an exact time evolution operator $\hat{U}(t)=e^{i \hat{H} t}$ and an approximate time evolution operator $\hat{U}_\epsilon(t)=e^{i \hat{H}_\epsilon t}$ as $||\hat{U}_\epsilon(t)-\hat{U}(t)||$, where $||\hat{A}|| = {\text{max}}_{\ket{\psi}} |\hat{A} \ket{\psi}|$ is the spectral norm of operator $\hat{A}$, and $\ket{\psi}$ is any vector satisfying $\langle \psi | \psi \rangle =1$. 
The condition
\begin{align}
    ||\hat{U}_\epsilon(t)-\hat{U}(t)|| \leq \epsilon
\end{align}
is satisfied iff
\begin{align}
    |\hat{H}_\epsilon(x)-\hat{H}(x)| \leq \frac{\epsilon}{t}, \forall x
    \,.
\end{align}
In Ref.~\cite{Welch_2014}, it is shown that it is possible to approximate an arbitrary diagonal unitary operator represented by $n$ qubits with some error $\epsilon$ using $\mathcal{O}(\text{poly}(n, 1/\epsilon))$ gates and in that sense this method is ``efficient''. 
The approximate operator $\hat{U}_\epsilon(t)$ is constructed by keeping only Walsh operators with $R_z$ rotation angles with absolute value above some cutoff $\theta_\text{min}$, referred to as threshold sampling.
While other methods exist for choosing the partial Walsh series used to represent $\hat{H}_\epsilon(x)$, it was shown in Ref.~\cite{CK_Yuen} that threshold sampling generally produces the shortest length series for the smallest error, with the caveat that error analysis is difficult to perform. A different method was presented in Ref.~\cite{CK_Yuen} that, at the cost of a larger partial Walsh series, allows for easier error estimation. It will be interesting to compare these two methods, which we leave for future work.

Suppose one places cutoffs $\theta_\text{min}^{E}$ and $\theta_\text{min}^{B}$ on the smallest allowed rotation angle when implementing the unitary operators $e^{-i \hat{H}_E \delta t}$ and $e^{-i \hat{H}_B \delta t}$. As shown in Appendix~\ref{app:errorUpperBound}, assuming a first order Suzuki-Trotter method, the total error is bounded by
\begin{align}
\label{eq:Uerror}
    ||\hat{U}_\epsilon(t)-\hat{U}(t)|| \leq \alpha \, t \, \delta t + c_E \theta_\text{min}^{E} t + c_B \theta_\text{min}^{B} t
    \,,
\end{align}
where $\alpha = || [\hat{H}_E, \hat{H}_B]||$ \cite{Welch_2014} and $c_E$ and $c_B$ are constants that depend on the particular details of the approximate Walsh series used. Because threshold sampling is equivalent to rounding to some fixed point precision \cite{CK_Yuen}, this upper bound assumes the error from dropping each Walsh operator adds coherently. However, because the individual errors will not in general always add coherently, the actual error from using cutoffs $\theta_\text{min}^E$ and $\theta_\text{min}^B$ will likely be better than the upper bound in Eq.~\eqref{eq:Uerror}.

Note, however, that when using Suzuki-Trotter methods for time evolution, the desired precision $\epsilon$ is not equal to the cutoff $\theta_\text{min}$ placed on the rotation angles.
Since the Hamiltonians are scaled by $\delta t$, the Walsh coefficients are given by
\begin{align}
    a_j = \frac{\delta t}{2^n} \Tr[\hat{H} \hat{w}_j]
    \,,
\end{align}
and also scale with $\delta t$. Therefore, in order for the circuit depth to stay constant as one changes $\delta t$, the minimum angle has to scale with $\delta t$ as well. 
This also ensures that the overall error in the time evolution \eqref{eq:Uerror} scales with $\delta t$. 

Note that, when using higher order Suzuki-Trotter methods, one will have to choose $\theta_\text{min}$ to scale with higher powers of $\delta t$ to ensure the truncation error is on the same order of magnitude as the Suzuki-Trotter error. For example, using a second order Suzuki-Trotter method, one will have to choose $\theta_\text{min} \sim \delta t^2$ and the gate count will not be constant for different choices of $\delta t$. 

We follow the procedure given in Ref.~\cite{Welch_2014} for constructing the circuit of a general approximate time evolution operator $\hat{U}_\epsilon(t)$. As in the case of creating the exact circuit, Walsh operators are placed in sequency order, but because some have been dropped, the number of CNOT gates between each $R_z$ gate can in general be more than one. As this procedure will not in general produce the circuit with the minimal number of CNOT gates, circuit optimization techniques can be used to reduce the CNOT gate count; the $R_z$ gate count is fixed for a given $\theta_\text{min}$ and cannot be reduced with circuit optimization techniques. 
While these $\mathcal{O}(1)-{\cal O}(10)$ optimizations will be important in an actual calculation, they will not change the scaling arguments we present in the following section.

As in the exact Walsh function implementation reviewed in Sec.~\ref{ssec:ExactWalsh}, it is important to combine Walsh coefficients that are common between different terms in the Hamiltonian in order to reduce the circuit depth. Additionally, when using this approximation method, one must be sure to do so \textit{before} deciding to drop them from the circuit. This removes the possibility that $\textit{e.g.}$ two Walsh coefficients are independently below the cutoff $\theta_\text{min}$, but their sum is not. Dropping the two Walsh operators from the circuit would be equivalent to dropping a single Walsh operator with a Walsh coefficient above the chosen threshold $\theta_\text{min}$, possibly resulting in larger errors than the bound in Eq.~\eqref{eq:Uerror}.

\begin{figure*}[t]
    \centering
    \begin{quantikz}[row sep={0.55cm,between origins}, column sep=0.4cm]
        \lstick{$q_1$} & \gate{R_{z,1}} & \ctrl{1} & \qw & \ctrl{1} & \qw & \qw & \qw & \ctrl{2} & \qw & \qw & \qw & \ctrl{2} & \qw & \qw
        \\
        \lstick{$q_2$} & \qw & \targ{} & \gate{R_{z,3}} & \targ{} & \gate{R_{z,2}} & \ctrl{1} & \qw & \qw & \qw & \ctrl{1} & \qw & \qw & \qw & \qw
        \\
        \lstick{$q_3$} & \qw & \qw & \qw & \qw & \qw & \targ{} & \gate{R_{z,6}} & \targ{} & \gate{R_{z,7}} & \targ{} & \gate{R_{z,5}} & \targ{} & \gate{R_{z,4}} & \qw
    \end{quantikz}
    \\
    \Big\downarrow
    \\~~\\
    drop gates $R_{z,3}, R_{z,7}, R_{z,5}$
    \\~~\\
    \Big\downarrow
    \\
    \begin{quantikz}[row sep={0.55cm,between origins}, column sep=0.4cm]
        \lstick{$q_1$} & \gate{R_{z,1}} & \ctrl{1} & \ctrl{1} & \qw & \qw & \qw & \ctrl{2} & \qw & \ctrl{2} & \qw & \qw
        \\
        \lstick{$q_2$} & \qw & \targ{} & \targ{} & \gate{R_{z,2}} & \ctrl{1} & \qw & \qw & \ctrl{1} & \qw & \qw & \qw
        \\
        \lstick{$q_3$} & \qw & \qw & \qw & \qw & \targ{} & \gate{R_{z,6}} & \targ{} & \targ{}  & \targ{} & \gate{R_{z,4}} & \qw
    \end{quantikz}
    \\
    \Big\downarrow
    \\~~\\
    simplify CNOT gates
    \\~~\\
    \Big\downarrow
    \\
    \begin{quantikz}[row sep={0.55cm,between origins}, column sep=0.4cm]
        \lstick{$q_1$} & \gate{R_{z,1}} & \qw & \qw & \qw & \qw & \qw & \qw
        \\
        \lstick{$q_2$} & \qw & \gate{R_{z,2}} & \ctrl{1} & \qw & \ctrl{1} & \qw & \qw
        \\
        \lstick{$q_3$} & \qw & \qw & \targ{} & \gate{R_{z,6}} & \targ{} & \gate{R_{z,4}} & \qw
    \end{quantikz}
    \caption{The top circuit shows the implementation of an arbitrary diagonal unitary matrix using three qubits. The middle circuit is obtained from the top by dropping the gates $R_{z,3}, R_{z,7}, R_{z,5}$. The bottom circuit is obtained from the middle by removing CNOT gates using circuit identities. We use the notation $R_{z,j} \equiv R_z(-2 a_j)$.}
    \label{fig:circ_simplification}
\end{figure*}
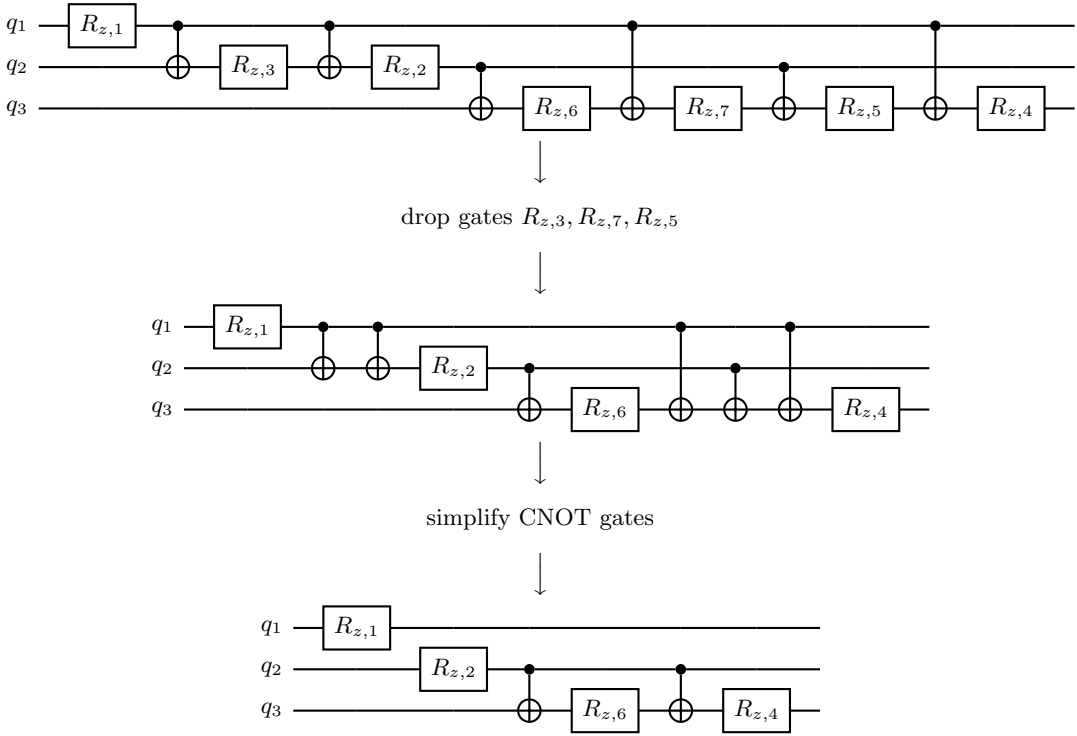

As a simple example of how the CNOT gate count can be reduced, consider the construction of a diagonal unitary operator using three qubits with Paley-ordered Walsh coefficients $(a_0, \dots a_7)$ and $R_z$ rotation angles $\theta_j=-2a_j$. The top circuit in Fig.~\ref{fig:circ_simplification} shows the exact implementation of the operator (the $a_0$ Walsh operator corresponds to a global phase and is not included). If we choose a cutoff $\theta_\text{min}$ such that the three angles $\theta_3, \theta_5, \theta_7$ are below the cutoff, we get the second circuit in Fig.~\ref{fig:circ_simplification}. After dropping the $R_z$ gates, four CNOT gates can be removed using circuit identities, resulting in the simpler circuit shown in the bottom of Fig.~\ref{fig:circ_simplification}.

\subsection{Class of matrices with classical exponential volume scaling overhead}
\label{ssec:expClassicalScaling}
One caveat with this method is that one must first calculate all of the Walsh coefficients in order to know if they are below the cutoff and therefore should be dropped. Using the fast Walsh transform, the number of floating point operations required to calculate $2^{n}$ Walsh coefficients scales as $\CO({n} 2^{n})$. In our case $n=n_q N_p$ and for realistic lattice sizes, the number of plaquettes will be ${N_p} \gtrsim 1000$ and the Walsh coefficients will be impossible to calculate classically for any value of $n_q$. One proposed method for breaking this classical exponential scaling is through the use of decision tree learning algorithms to find the best $k$ term Walsh series that approximates a function to a given precision \cite{decision_tree}. Note that this method is only valid for functions whose $L_1$ norm of their Walsh coefficients, \textit{i.e.} the sum of the absolute value of the Walsh coefficients, is not exponential in ${n}$. As we will discuss further in Sec.~\ref{ssec:resourceOriginalBasis}, the $L_1$ norm of the Walsh coefficients of the maximally coupled term in the compact magnetic Hamiltonian is observed to scale exponentially with the volume and therefore decision tree learning algorithms cannot be used to break this exponential classical cost. While this drawback will make simulations in the original operator basis prohibitively expensive, we study how imposing a lower limit on the Walsh coefficients affects the gate counts required in an implementation in Appendix~\ref{app:truncationOriginalBasis}.

We conclude by pointing out that, once the Walsh coefficients that will be included in the circuit are known, the classical resources required to create said circuit scales with the number of Walsh coefficients. Because any circuit implemented on realistic lattice sizes will only contain a polynomial number of Walsh operators, this classical cost is also polynomial in the volume.

\section{Resource study}
\label{sec:resourceStudy}
In this section, we focus on implementing time evolution via Suzuki-Trotter methods and study the gate counts needed to create the time evolution operators $\hat{U}^{(m)}_E(t) = \exp(i \hat{H}^{(m)}_E t)$ and $\hat{U}^{(m)}_B(t) = \exp(i \hat{H}^{(m)}_B t)$ separately. We work in the magnetic basis and therefore $\hat{H}^{(m)}_B$ is diagonal, and $\hat{H}^{(m)}_E$ is diagonalized by a QFT, which can be implemented efficiently on quantum computers. For this reason, we will construct $\hat{U}^{(m)}_E(t)$ by first rotating to the electric basis, constructing the diagonal operator $\hat{U}^{(e)}_E(t)$, and then rotating back. Because both $\hat{U}^{(e)}_E(t)$ and $\hat{U}^{(m)}_B(t)$ are diagonal, the task of time evolution reduces to the efficient construction of unitary diagonal matrices. 

We begin by discussing the gates required to implement time evolution in the original basis as a function of the volume. We show that, while the Walsh function formalism can in principle break the exponential volume gate count scaling, doing so requires an exponential volume classical computing cost, which will be prohibitively expensive for realistic lattice sizes. Then, we show that the weaved basis breaks the exponential volume scaling in both the quantum and classical cost. We construct the quantum circuits for implementing $\hat{U}^{(m)}_B(t)$ for different values of $N_p$ and study how imposing a cutoff $\theta_\text{min}$ reduces further the gate count. From there we study how the gate count scales as a function of $n_q$ for different values of the cutoff $\theta_\text{min}$. We then study of the gate count scaling as a function of the coupling $g$ and conclude by studying how the gate count scales with the cutoff $\theta_\text{min}$.

Before discussing exact gate counts, we point out that in a realistic calculation one would take a continuum limit. Because the lattice spacing $a$ and the value of the bare coupling $g$ are related through renormalization, understanding the dependence on the lattice spacing requires performing the calculation at many values of $g$. Methods for determining the lattice spacing $a$ for various gauge theories have been proposed in Refs.~\cite{Carena:2021ltu,Carena:2022hpz,Clemente:2022cka}.
Additionally, in order to understand the Suzuki-Trotter error, one would have to perform the calculation at different values of the Suzuki-Trotter step size $\delta t$ or using higher order methods. 
Because the Walsh coefficients will in general be different in these different calculations, the specific gate count will depend on the precise values chosen.

\subsection{Volume dependence of classical resources and quantum gate count in original basis}
\label{ssec:resourceOriginalBasis}

The electric Hamiltonian is a sum of products of two rotor operators, either $\hat{R}^2$ or $\hat{R}_i \otimes \hat{R}_j$, where $i$ and $j$ label neighboring plaquettes on the lattice.
These rotors in the electric basis are diagonal matrices with evenly spaced entries on the diagonal. 
As previously discussed in Sec.~\ref{ssec:ExactWalsh}, each of these products can be exponentiated in $\mathcal{O}(n_q^2)$ gates. 
In Ref.~\cite{Haase:2020kaj} it was shown that the number of terms in $\hat{H}_E$ is $\mathcal{O}(N_p)$. 
This, combined with the $\mathcal{O}(N_p n_q^2)$ gates required to perform a quantum Fourier transform at each lattice site, means that the electric Hamiltonian in the magnetic basis can be exponentiated in $\mathcal{O}(N_p n_q^2)$ gates, which is linear in the lattice volume; here and throughout this manuscript, we drop overall $\CO(1)$ pre-factors.

Similarly, the magnetic Hamiltonian in the non-compact theory, shown in Eq.~\eqref{eq:NonCompactH} and subject to the constraint \eqref{eq:constraint}, is a sum of $\mathcal{O}(N_p^2)$ two-body operators with linearly spaced entries on the diagonal. While enforcing magnetic Gauss' law introduces non-local interactions, the non-compact magnetic Hamiltonian can still be exponentiated using $\mathcal{O}(N_p^2 n_q^2)$ gates, which is quadratic in the lattice volume.

The compact magnetic Hamiltonian contains cosines of operators with evenly spaced values on the diagonal as shown in Eq.~\eqref{eq:compactHB_after_constraint}. There are two types of terms appearing, the cosine of a single $\hat{B}_i$ operator, and the cosine of the sum of all $\hat{B}_i$ operators. We will investigate the resources needed to implement these operators separately.

To find the gate count required to exponentiate $\cos \hat{B}_i$, we write the cosine using the Taylor series as
\begin{align}
    \cos(\hat{B}_i) = \sum_{k=0}^{\infty} \frac{(-1)^k }{(2k)!} (\hat{B}_i)^{2k}.
\end{align}
We discussed in Sec.~\ref{ssec:originalBasisReview} that $\hat{B}_i$ is a diagonal operator with evenly spaced entries on the diagonal. As previously discussed in Sec.~\ref{ssec:ExactWalsh}, each $(\hat{B_i})^{2k}$ term will require $\CO(2^{2k})$ terms to exponentiate (note that if $2^{2k} \geq 2^{n_q}$, the number of gates will be $\CO(2^{n_q})$). From this we see that each $\cos \hat{B}_i$ operator is composed of $2^{n_q}$ Walsh coefficients and therefore exponentiation of all $N_p$ terms requires $\mathcal{O}(N_p 2^{n_q})$ gates to implement. 

This is only exponential in $n_q$, but linear in $N_p$. Thus, if the desired precision of the calculation can be achieved while keeping $n_q$ relatively small, these terms will not be prohibitively expensive. The digitization errors for the non-compact formulation of a \uone gauge theory was studied in Ref.~\cite{Bauer_2021} for the $2 \times 2$ site lattice. It was found that the low-lying spectrum could be reproduced to per-mille level accuracy while only sampling the operators a relatively small number of times, corresponding to using $n_q=3$. In general, the convergence to the undigitized result in the weak-coupling limit is exponential in $n_q$ \cite{Macridin:2018oli,Macridin:2018gdw}.

We now turn our attention to the term which is the cosine of a sum of all $\hat{B}_i$ operators. Because this term couples the entire lattice together, the number of gates required to implement this term scales exponentially with the volume, which corresponds to the total number of qubits utilized in the simulation. When $b_\text{max}=\pi$, a symmetry in this term causes certain Walsh coefficients to be zero. We observe numerically the gate count scales roughly as $\CO(2^{N_p(n_q-1)})$. For $b_\text{max}<\pi$, this symmetry is broken and the gate count scales as $\CO(2^{N_p n_q})$. Because calculations will need to be done at different values of $g$, and therefore values of $b_\text{max}<\pi$, to take the continuum limit, we proceed assuming the worse scaling. Even if $n_q$ is small, this term will be impossible to implement exactly at lattice sizes needed for realistic calculations. For example, even for a very small lattices of $3 \times 3$, $4 \times 4$, and $5\times 5$, with only two qubits per plaquette, this naive implementation would require $\CO(10^4), \CO(10^9)$ and $\CO(10^{14})$ gates per Suzuki-Trotter step, respectively.

While the exponential volume scaling in the gate count can be broken using the truncation method described in Sec.~\ref{ssec:removingSmallWalsh}, the Walsh coefficients have to be calculated classically and so there is also an exponential classical computing cost. In particular, using the Fast-Walsh-Transform, this step requires 
$\CO(N_p n_q 2^{N_p n_q})$ classical floating point operations. As mentioned in Sec.~\ref{ssec:removingSmallWalsh}, implementing a decision tree learning algorithm could break this classical exponential cost, but only if the $L_1$ norm of the Walsh coefficients of the maximally coupled term is not exponential in the volume. In Appendix \ref{sec:L1_norm}, we show the $L_1$ norm of the Walsh coefficients of the maximally non-local term actually does scales exponentially in the volume of the system. This finding indicates that a completely different method is required to be able to implement the maximally non-local cosine term in the compact theory of \uone gauge theories using Susuki-Trotter methods. In the next section, we review the operator basis change method in Ref.~\cite{https://doi.org/10.48550/arxiv.2208.03333}, and show that it breaks the exponential volume scaling in both the quantum \textit{and} classical computational costs.

\subsection{Volume dependence of classical resources and quantum gate count in weaved basis}
\label{ssec:ChangeOfBasisGateCount}
In this section, we start by providing a more detailed review of the change of basis originally given in Ref.~\cite{https://doi.org/10.48550/arxiv.2208.03333} and how it breaks the exponential cost associated with simulating the maximally coupled term in the compact formulation of \uone gauge theories. We then study the gate counts of the magnetic Hamiltonian after performing the operator basis change using the Walsh function formalism. 

As discussed in Sec.~\ref{ssec:resourceOriginalBasis}, implementing the exponential of the maximally coupled component of the magnetic Hamiltonian, the cosine of the sum of all magnetic plaquettes, requires a circuit whose depth scales exponentially with the lattice volume.
On the other hand, implementing the cosine of a single magnetic plaquette is only exponential in the number of qubits per lattice site, which is typically manageable.

A redefinition that reduces the total number of magnetic field operators included in the maximally coupled term is possible, but at the cost of increasing the number of magnetic field operators in the previously completely local terms.
Choosing a redefinition which balances these two effects such that each term in the Hamiltonian has no more than $\CO(\log_2 N_p)$ magnetic field operators appearing in a single cosine would break the exponential scaling in the total circuit size. It was shown in Ref.~\cite{https://doi.org/10.48550/arxiv.2208.03333} that for any lattice volume, there is always a choice of $\mathcal{W}$ that achieves this, where $\mathcal{W}$ is the orthogonal matrix that rotates from the original to the weaved basis. In particular, it was shown in Ref.~\cite{https://doi.org/10.48550/arxiv.2208.03333} that for any $N_p$, it is always possible to choose a $\CW$ such that the number of gates needed to implement the compact magnetic Hamiltonian for this specific basis is
\begin{align}
\CO\left(N_p^{n_q} + N_p \left(\frac{N_p}{\log_2 N_p}\right)^{n_q}\right) \, ,
\label{eq:gate_scaling}
\end{align}
which scales polynomially with the volume where the exponent of the largest polynomial is given by $n_q$.

Note that this redefinition does introduce more two-body terms in $\hat{H}_E$, up to a maximum of $\CO(N_p^2)$ terms. However, as was discussed in Sec.~\ref{ssec:originalBasisReview}, each of these terms is possible to implement using $\CO(n_q^2)$ gates, leading to an upper limit of $\CO(n_q^2 N_p^2)$ gates required to implement $\hat{H}_E$ in the weaved basis. This quadratic volume scaling is more than made up for by the breaking of the previously exponential volume scaling. 

Note that using this weaved Hamiltonian would also reduce the resources necessary for implementing a variational approach to simulating this theory. In the rest of the manuscript, unless otherwise stated, all gate counts of weaved Hamiltonians were calculated using weaved matrices $\CW$ that minimize the gate count.

Before accepting this method as successfully transforming the exponential volume scaling to polynomial, it is important to also consider the classical computing cost required to calculate the necessary Walsh coefficients. Recall from Sec.~\ref{ssec:resourceOriginalBasis} that we deemed the truncation method inefficient since, while it resulted in a quantum circuit depth that scaled polynomially in the volume, there was a corresponding classical computational cost that scaled exponentially in the volume. For the weaved method, this is no longer true. As the dimensions of all the unitary matrices in the magnetic Hamiltonian are now $\CO(N_p^{n_q})$, the number of classical floating point operations required to calculate all the Walsh coefficients is $\CO(n_q N_p^{n_q} \log_2 N_p )$. Additionally, the method for generating the weaved matrices themselves and carrying out the change of operator basis was found to scale roughly linearly with the volume with a small coefficient due to the sparse nature of the weaved matrices \cite{https://doi.org/10.48550/arxiv.2208.03333}. 

To get a sense of the full scope of the gate count reduction from the operator redefinition, we have implemented this new basis using the Walsh function method. Recall that a single Suzuki-Trotter step for the compact magnetic Hamiltonian in the original basis using two qubits per lattice site required $\CO(10^4), \CO(10^9)$ and $\CO(10^{14})$ gates for a $3\times 3$, $4 \times 4$, and $5\times 5$ lattice, respectively. In the rotated basis, the same term for both the $3\times 3$ and $4\times 4$ lattices can be implemented using $\CO(10^2)$ gates, and $5\times 5$ lattice can be implemented using $\CO(10^3)$ gates.

\begin{figure*}[t]
\subfloat[\label{fig:Cx_vs_Np_weaved_dt}Walsh function truncation scale, $\theta_\text{min}$, chosen relative to $\delta t$.]{\includegraphics[width=0.48\textwidth]{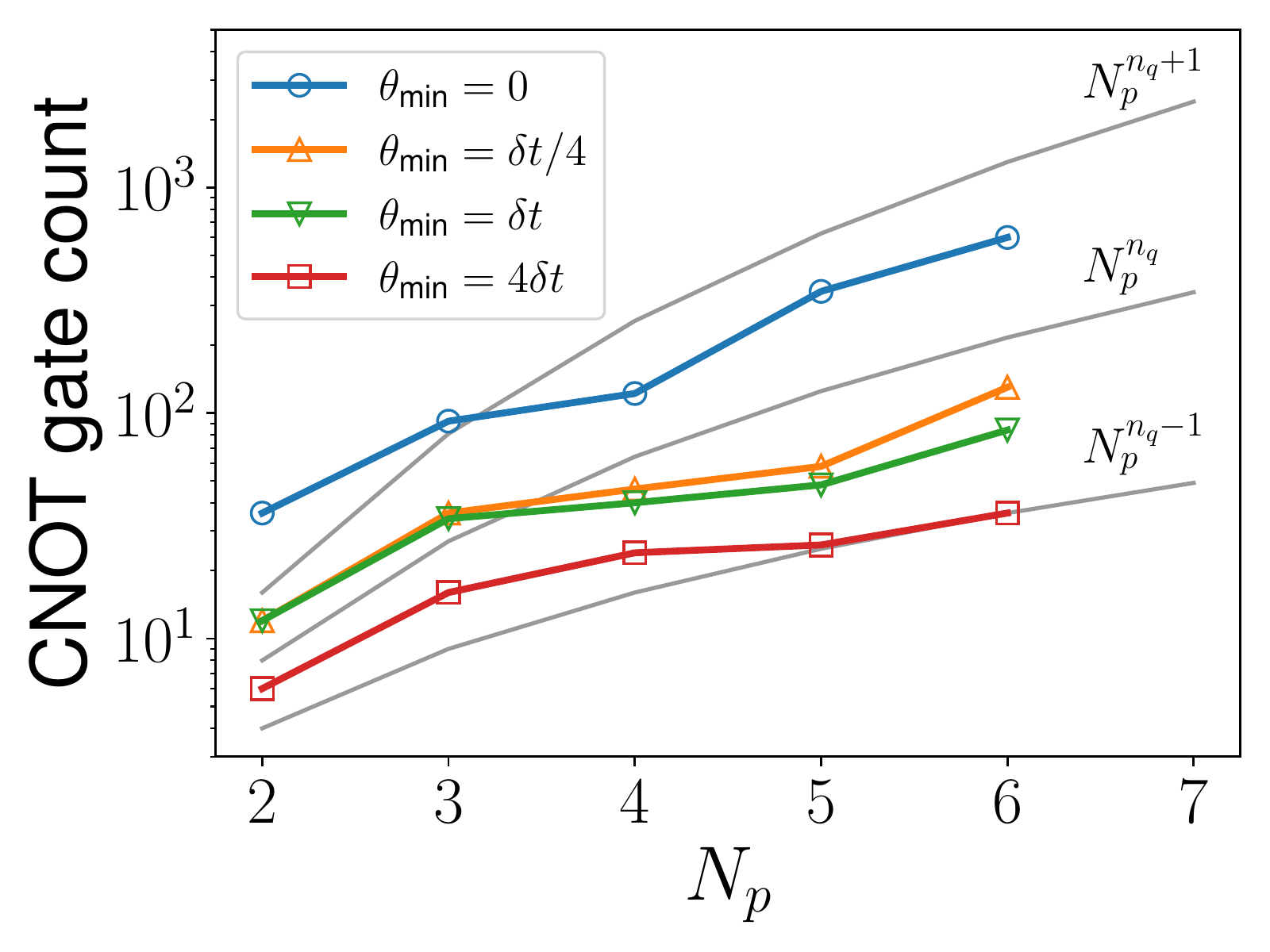}}
\hfill
\subfloat[\label{fig:Cx_vs_Np_weaved_dtsq}Walsh function truncation scale, $\theta_\text{min}$, chosen relative to $\delta t^2$, with $\delta t = 1/4$.]{\includegraphics[width=0.48\textwidth]{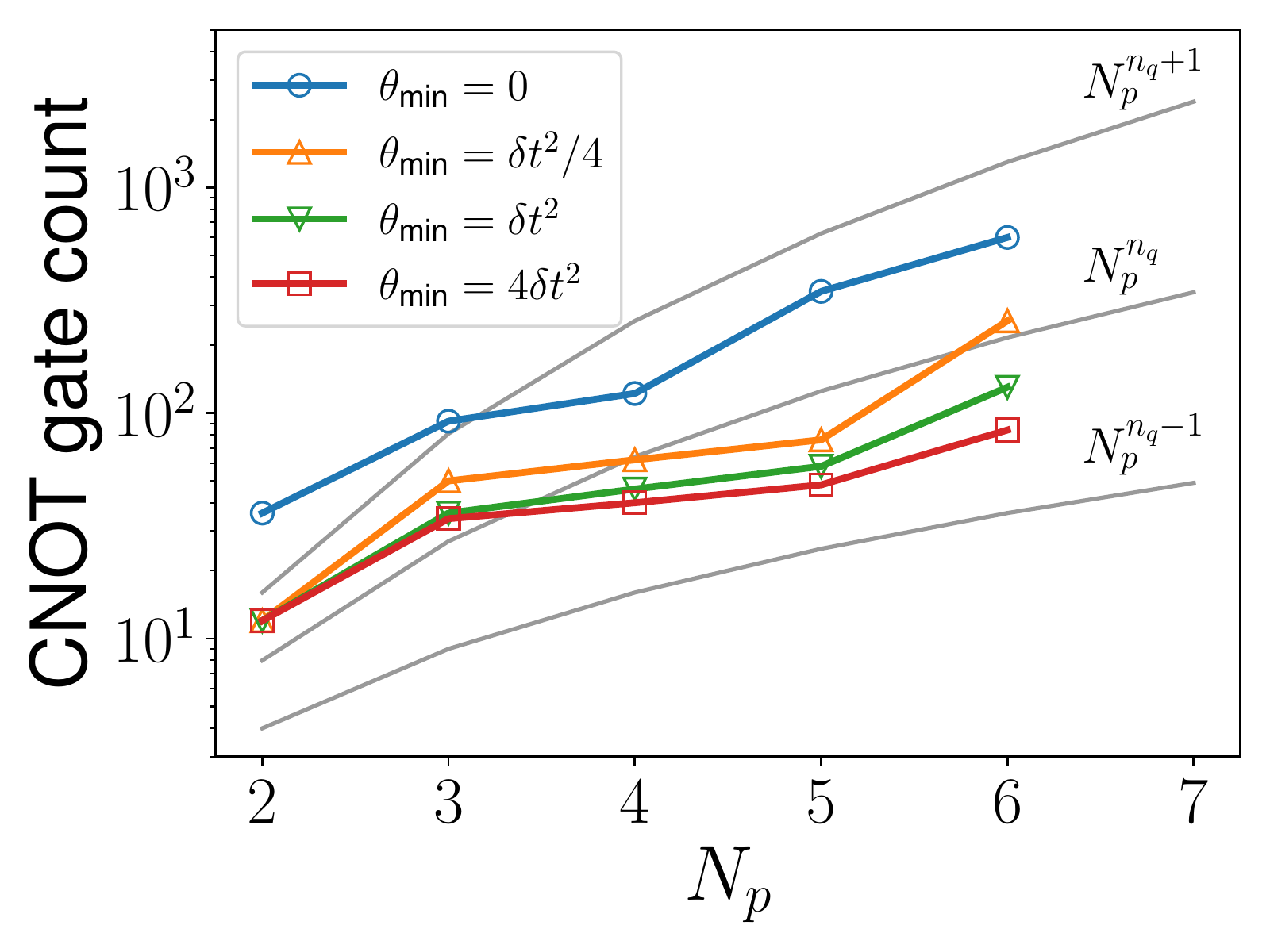}}
\caption{CNOT gate count to implement the complex exponential of the weaved compact magnetic Hamiltonian as a function of the number of plaquettes $N_p$ using $g = 0.1$ and $n_q=3$. The values of $b_\text{max}$ were chosen to match the original basis values. The gray lines show the functions $N_p^{n_q-1},  N_p^{n_q}$ and $N_p^{n_q+1}$ for reference.}
\end{figure*}

A truncated Walsh series can be used to further improve the gate requirements. We choose to study the gate count considering choosing $\theta_\text{min}$ assuming a $1^\text{st}$ and $2^\text{nd}$ order Susuki-Trotter implementation, which can be found in Figs.~\ref{fig:Cx_vs_Np_weaved_dt} and \ref{fig:Cx_vs_Np_weaved_dtsq}, respectively. Assuming a $1^\text{st}$ order method, we see that for values of $1/8 < \theta_\text{min}/\delta t < 4$ the scaling is between $\CO(N_p^{n_q-1})$ and $\CO(N_p^{n_q})$.
The scaling assuming a $2^\text{nd}$ order method for $1/2 \leq \theta_\text{min}/(\delta t)^2 < 4$ is also between $\CO(N_p^{n_q-1})$ and $\CO(N_p^{n_q})$, which indicates a $2^\text{nd}$ order method could offer significant gate count reductions by allowing one to use a larger Suzuki-Trotter step $\delta t$.

\subsection{Gate count dependence on qubits per lattice site in weaved basis}
\label{ssec:gateVsnqWeaved}
While the change of operator basis discussed in Sec.~\ref{ssec:ChangeOfBasisGateCount} breaks the exponential scaling with $N_p$, it does not break the exponential scaling with $n_q$ (even though the scaling has been reduced from $(2^{N_p})^{n_q}$ to $N_p^{n_q}$, the scaling is still exponential). This is perhaps expected because $2^{n_q}$ is the number of times the operators of the theory are sampled. Looking at this scaling from this point of view, we see that the gate count in the weaved basis actually scales better than linearly with the number of times we sample the operators, and therefore it may not a problem that the gate count scales exponentially with $n_q$ as well. On the other hand, because $n_q$ must take integer values, increasing $n_q$ by the smallest increment possible (one) will dramatically change the cost of the quantum computation. Therefore, it may be that this `exponential' scaling must also be overcome. 

\begin{figure*}[t]
\subfloat[\label{fig:Cx_vs_nq_weaved_1st_order} Walsh function truncation scale $\theta_\text{min}$, chosen relative to $\delta t$.]{\includegraphics[width=0.48\textwidth]{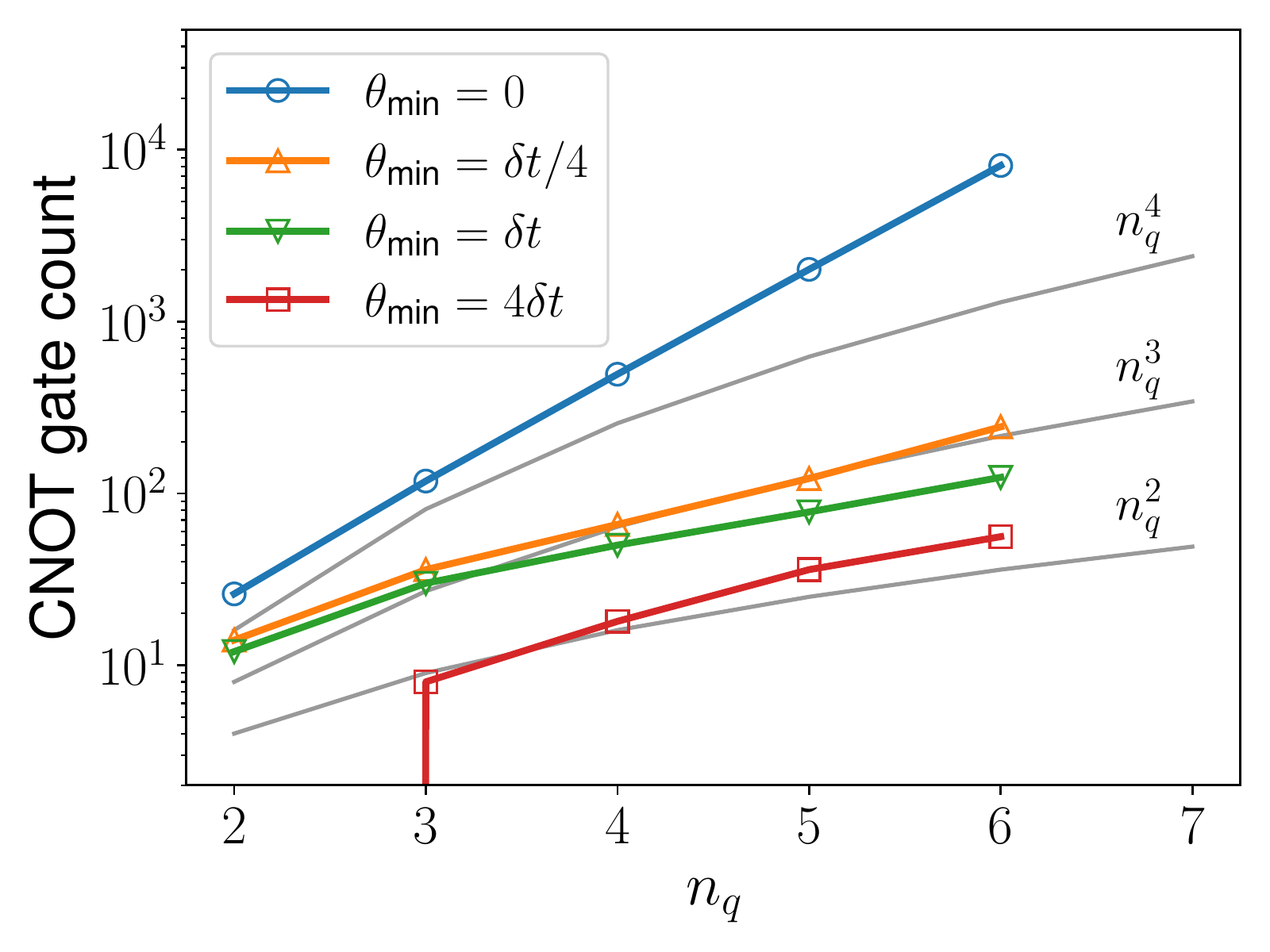}}
\hfill
\subfloat[\label{fig:Cx_vs_nq_weaved_2nd_order} Walsh function truncation scale $\theta_\text{min}$, chosen relative to $\delta t^2$, with $\delta t = 1/4$.]{\includegraphics[width=0.48\textwidth]{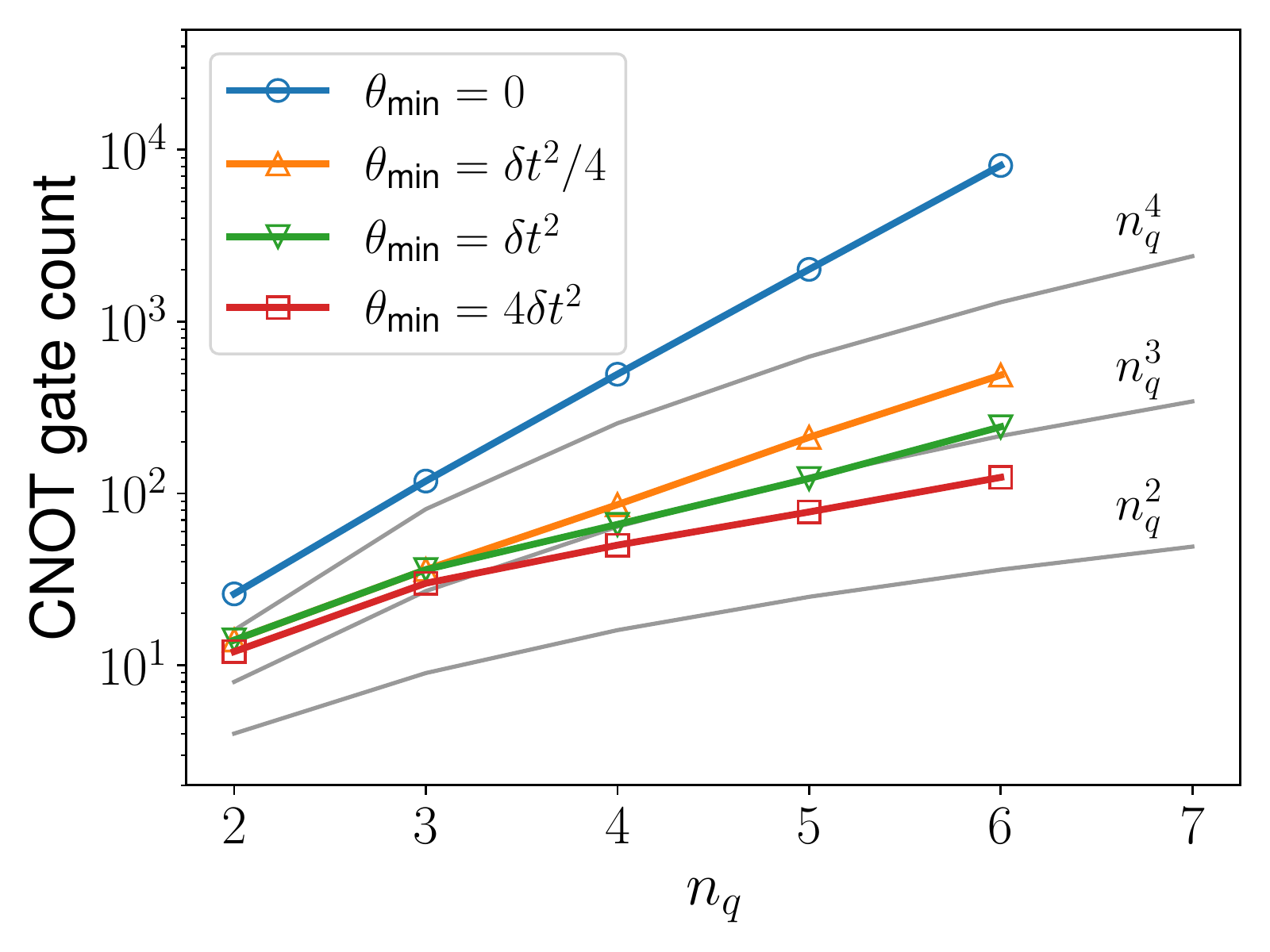}}
\caption{CNOT gate count to implement the complex exponential of the weaved compact magnetic Hamiltonian as a function of the number of qubits per plaquette $n_q$ for fixed $N_p=3$ with $g = 0.1$. The values of $b_\text{max}$ were chosen using the procedure in Sec.~\ref{ssec:weavedBasisDigitization}. The gray lines show the functions $n_q^2, n_q^3$, and $n_q^4$ for reference.}
\end{figure*}

In this section we study the scaling in $n_q$ using the truncation method for the weaved basis. The scaling for the original basis can be found in Appendix~\ref{app:truncationOriginalBasis}. The Hamiltonian we study is the weaved compact magnetic Hamiltonian with $N_p=3$ and $g=0.1$. Our results for several values of $\theta_\text{min}$ in a $1^\text{st}$ and $2^\text{nd}$ order Suzuki-Trotter method can be found in Figs.~\ref{fig:Cx_vs_nq_weaved_1st_order} and~\ref{fig:Cx_vs_nq_weaved_2nd_order}, respectively. Considering a $1^\text{st}$ order method, we see that for values $1/4 \leq \theta_\text{min}/\delta t \leq 4$, the scaling is between quadratic and cubic in $n_q$. Assuming a $2^\text{nd}$ order method, the scaling increases slightly. For values $1 \leq \theta_\text{min}/\delta t \leq 4$ the scaling is between $\CO(n_q^2)$ and $\CO(n_q^3)$ and for values $1/4 \leq \theta_\text{min}/\delta t < 1$ the scaling is between $\CO(n_q^3)$ and $\CO(n_q^4)$. In all cases, the scaling is polynomial, and the exponential scaling has been broken. In addition, as with the scaling with the number of plaquettes $N_p$, the scaling between considering $1^\text{st}$ and $2^\text{nd}$ order methods is comparable and it is possible that using a $2^\text{nd}$ order method could offer reduced gate counts.

\subsection{Gate count dependence on coupling strength}
\label{ssec:gatesVsCouplingWeaved}
\begin{figure}[t]
    \centering
    \includegraphics[width=0.48\textwidth]{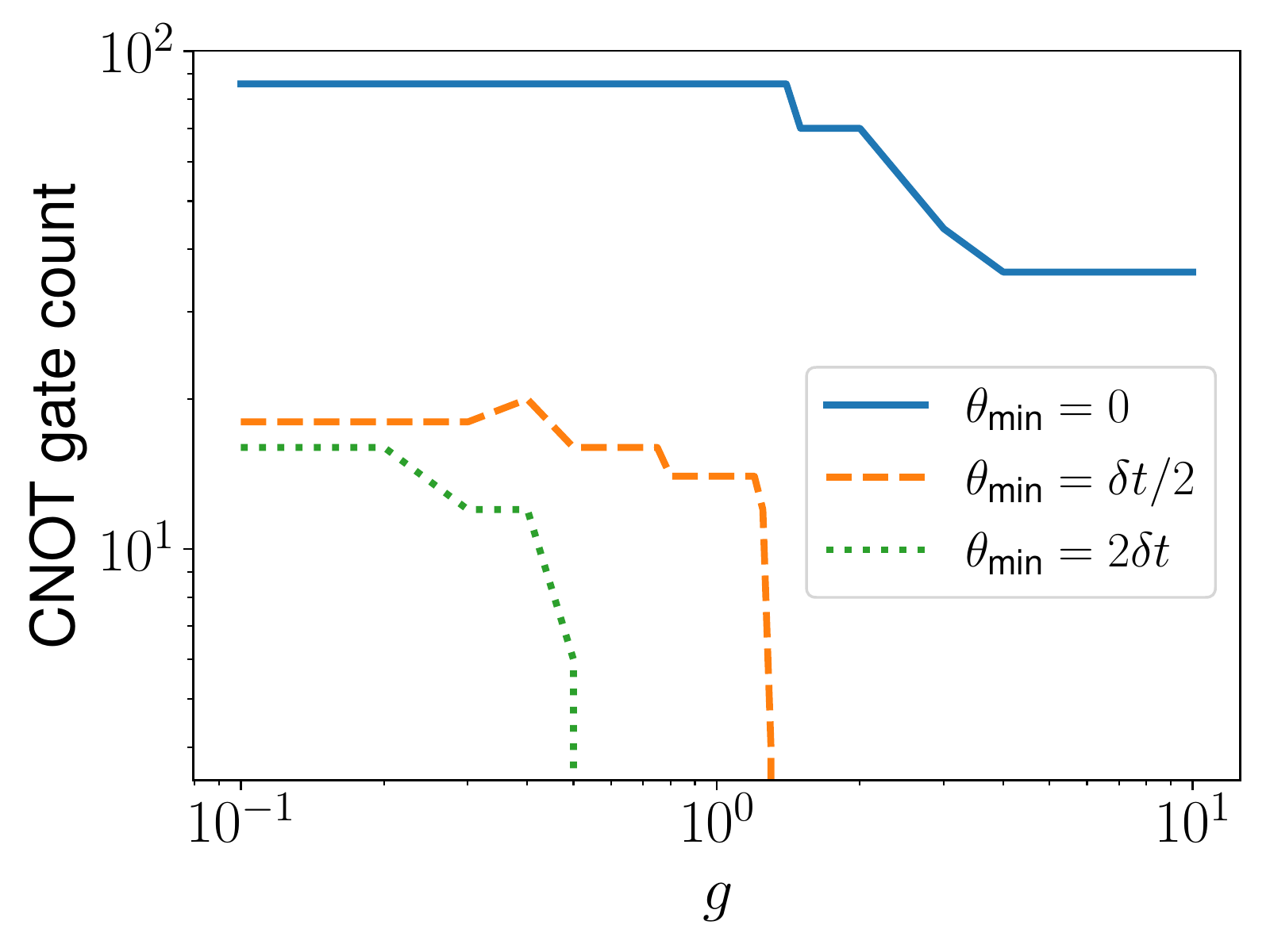}
    \caption{Number of CNOT gates required to implement $e^{i H_B \delta t}$ in the weaved basis as a function of the coupling $g$ for $N_p=5$ and $n_q=2$. The different lines show different values of the cutoff $\theta_\text{min}$.}
    \label{fig:cx_vs_g}
\end{figure}

Because the value of the lattice spacing depends on the choice of lattice coupling, any continuum extrapolation requires performing calculations for many values of the coupling $g$. 
Figure~\ref{fig:cx_vs_g} shows the number of CNOT gates required to implement $e^{i \delta t \hat{H}_B}$ in the weaved basis for $N_p=5$ and $n_q=2$ as a function of the coupling for different values of $\theta_\text{min}$. For small values of $g$ the gate count is constant. 
As $g$ increases, because $\hat{H}_B \sim 1/g^2$, the Walsh coefficients get smaller in magnitude and more gates are dropped from the circuit. Once all Walsh coefficients are below the cutoff $\theta_\text{min}$, the gate count drops to zero.
This implies that the magnetic Hamiltonian can be neglected given the desired precision, and in that case it is more efficient to work directly in the electric basis.  

Another interesting feature in Fig.~\ref{fig:cx_vs_g} is that the number CNOT gates for $\theta_\text{min}=0$ is not constant as $g$ varies. 
As explained in Sec. \ref{ssec:originalBasisReview}, for small $g$, the wavefunction in magnetic field space is highly localized and $b_\text{max}<\pi$. 
As $g$ increases, eventually $b_\text{max}$ reaches its upper limit, which restores a symmetry of the magnetic Hamiltonian causing some Walsh coefficients to be zero. Because there are three different values of $b_\text{max}$ each with different upper limits, there are three such drops in the gate count for $\theta_\text{min}=0$ as a function of $g$. 

\subsection{Gate count dependence on cutoff $\theta_{\rm min}$}
\label{ssec:CxvsThetaMin}
\begin{figure}[t]
    \centering
    \includegraphics[width=0.48\textwidth]{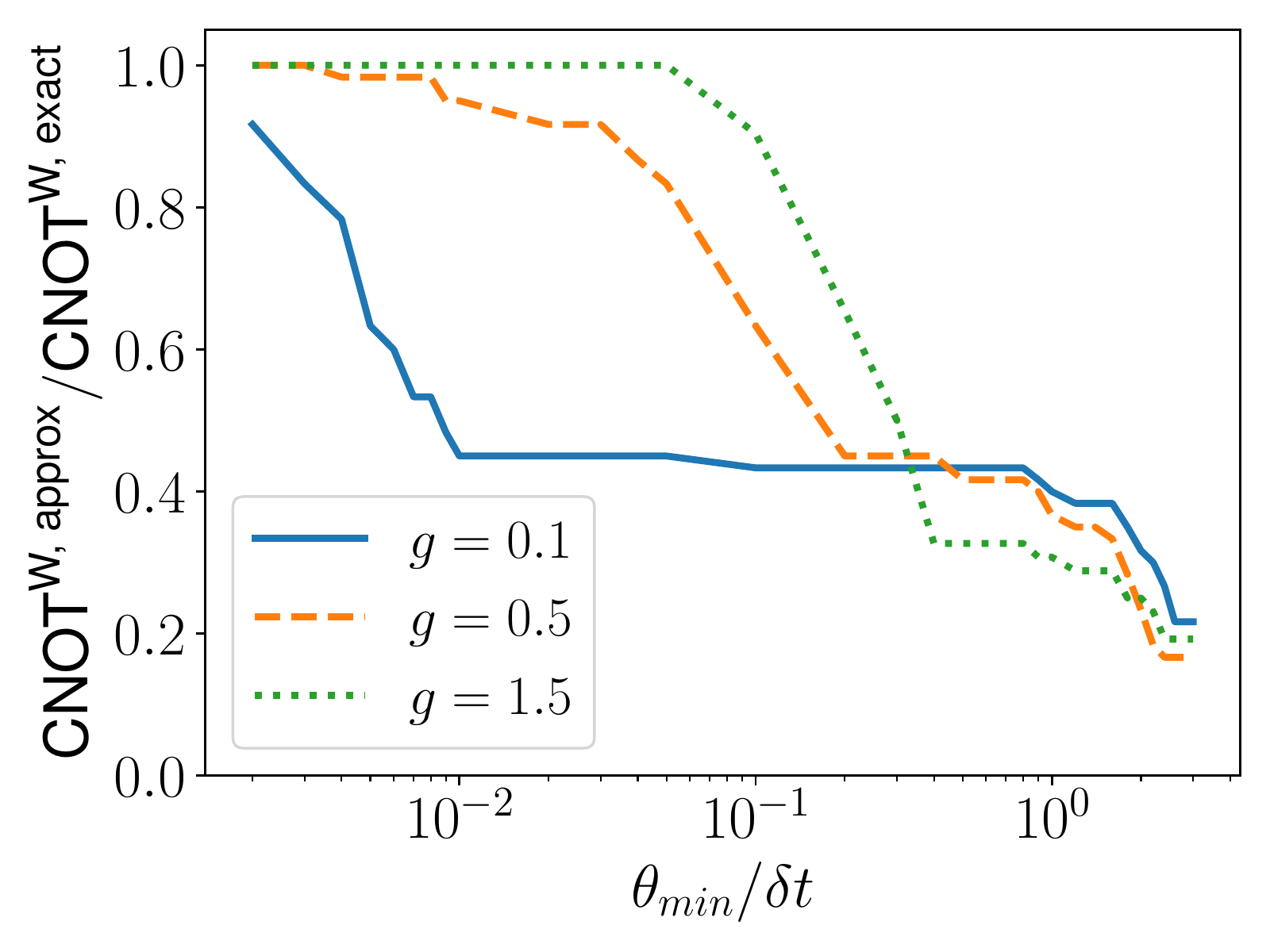}
    \caption{Ratio of the number of CNOT gates required, for a single Suzuki-Trotter step, between the truncated weaved operator basis and the exact untruncated weaved operator basis as a function of the cutoff relative to the Suzuki-Trotter step-size $\theta_\text{min} / \delta t$. CNOT gate counts were calculated for $N_p=5$ and $n_q=2$. The different lines show different values of the coupling $g$.}
    \label{fig:cx_vs_theta_min}
\end{figure}

Another important scaling to study is the number of gates as a function of the cutoff $\theta_\text{min}$. Figure~\ref{fig:cx_vs_theta_min} shows the ratio of the number of CNOT gates required to implement a single Suzuki-Trotter step between the truncated weaved basis and the exact untruncated weaved basis for $N_p=5$ and $n_q=2$. The quantum Fourier transform circuit is implemented exactly. The same $\theta_\text{min}$ was used for $\hat{H}_E^{(e)}$ and $\hat{H}_B^{(m)}$. We see that, for all values of $g$, the number of gates drops by a factor of two when $0.05 \lesssim \theta_\text{min}/\delta t \lesssim 0.5$. For values of $\theta_\text{min}/\delta t \gtrsim 0.5$, the number of CNOT gates can be reduced by a factor of four or more. The most dramatic reductions in gate count occur when choosing a cutoff $\theta_\text{min} \sim \delta t$. 

\section{Error analysis of explicit observable}
\label{sec:errorAnalysis}

\begin{figure*}[t]
\subfloat[\label{fig:errorAnalysisSimulator}]{\includegraphics[width=0.48\textwidth]{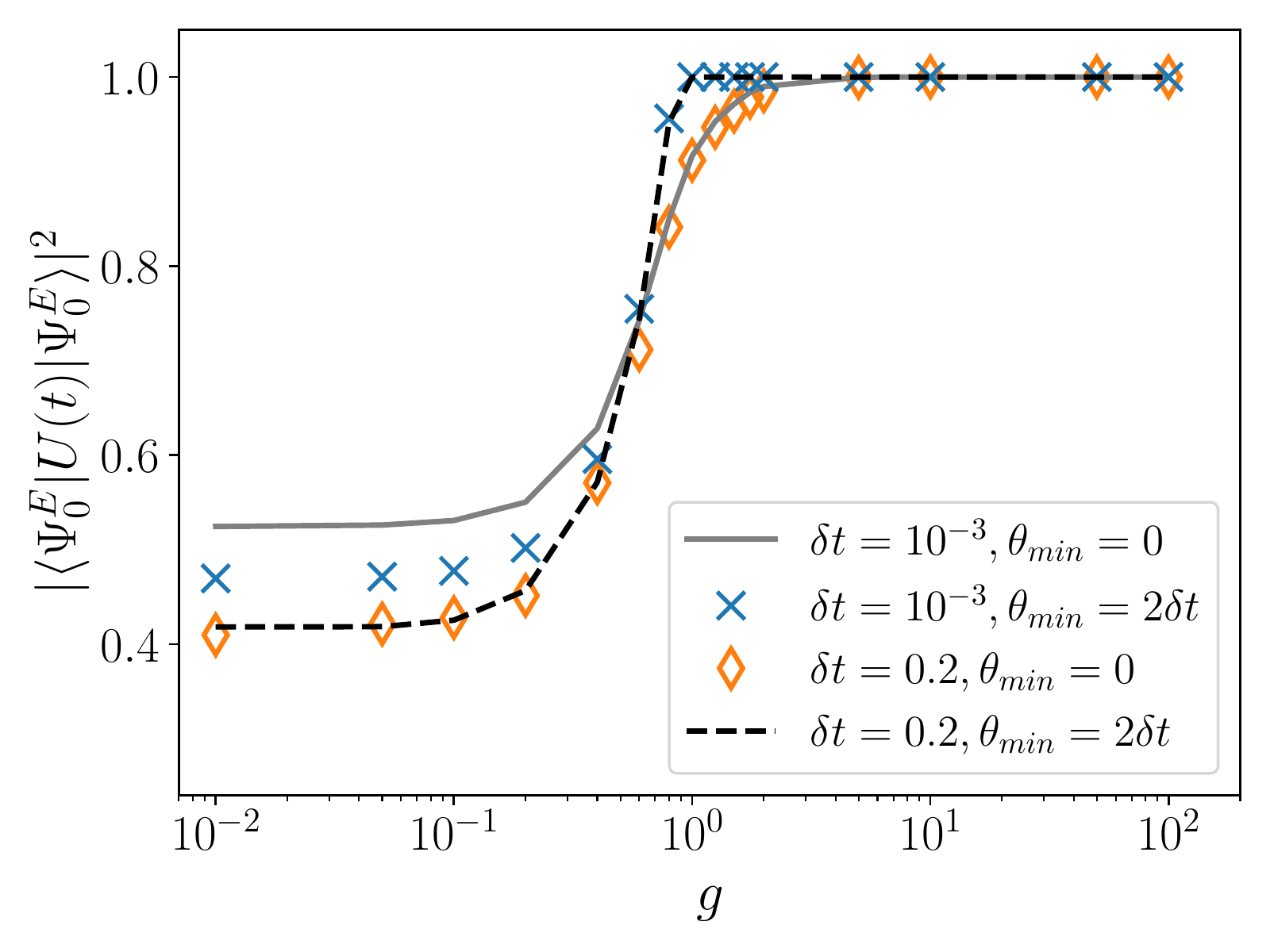}}
\hfill
\subfloat[\label{fig:errorAnalysisIBMQ}]{\includegraphics[width=0.48\textwidth]{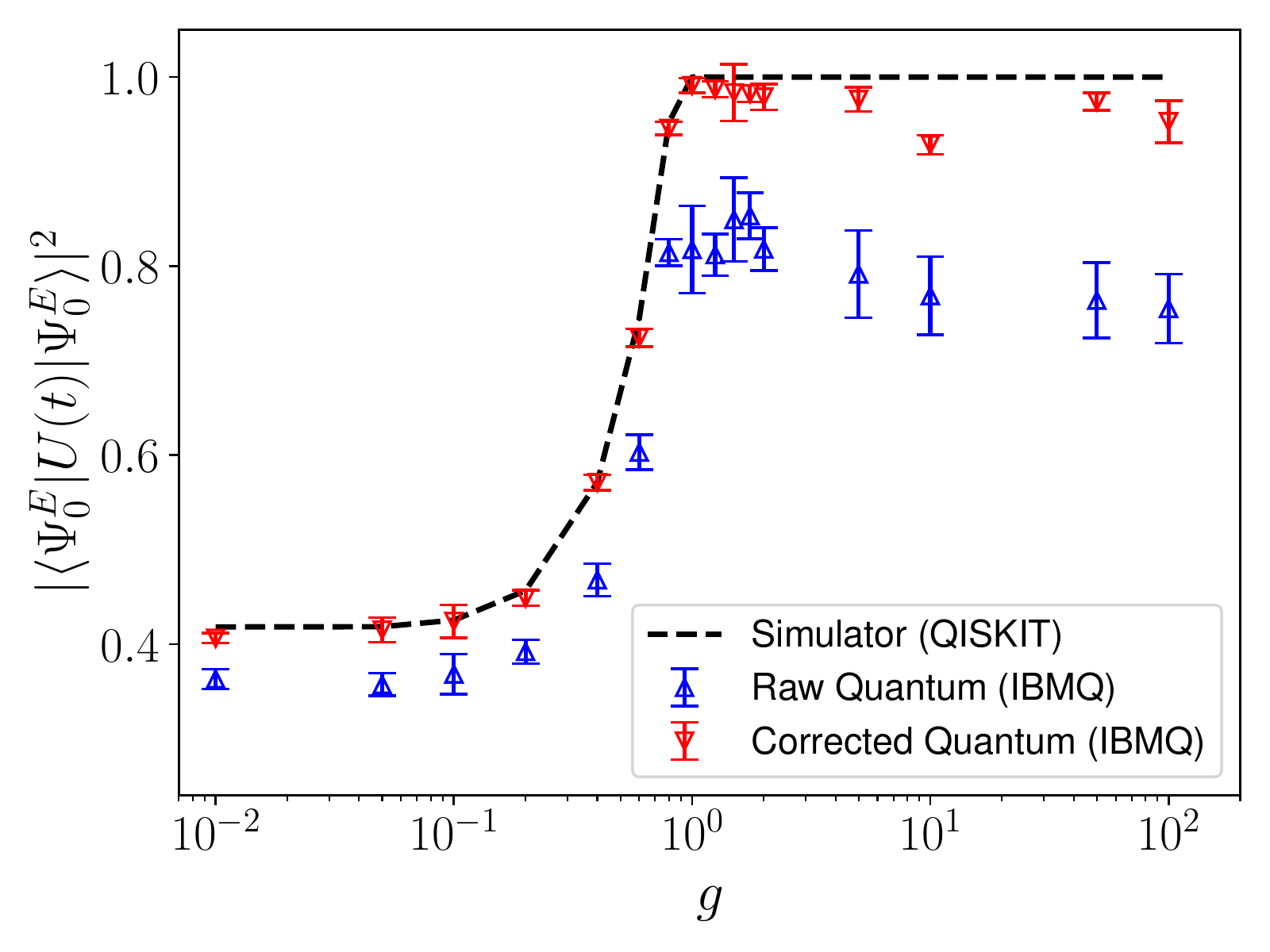}}
\caption{Value of $|\hat{U}(t)|^2 \equiv |\langle \Psi^E_0| \hat{U}(t) | \Psi^E_0 \rangle|^2$ calculated as a function of the coupling $g$ for $t=0.2$. Figure~\ref{fig:errorAnalysisSimulator} shows $|\hat{U}(t)|^2$ calculated using a simulator, where different lines and data points correspond to different combinations of Suzuki-Trotter step-sizes $\delta t$ and cutoffs $\theta_\text{min}$. In Fig.~\ref{fig:errorAnalysisIBMQ}, the blue (red) points show the raw (corrected) output of $|\hat{U}(t)|^2$ resulting from a quantum calculation using IBMQ superconducting qubit hardware with $\delta t=0.2$ and $\theta_\text{min}=2 \delta t$. The black dashed line in Fig.~\ref{fig:errorAnalysisIBMQ} is the same quantity calculated using a simulator, and is the same as the black dashed line in Fig.~\ref{fig:errorAnalysisSimulator}.}
\end{figure*}
In the gate count studies in Sec.~\ref{sec:resourceStudy}, we found that using a cutoff $\theta_\text{min}$ on the same order as the Suzuki-Trotter step size significantly reduces the gate count for a single Suzuki-Trotter step. This choice in turn requires a systematic study of the errors introduced by the truncation of the Walsh series to test how they compare with the Suzuki-Trotter error. In this section we perform a simple error analysis for a particular observable by calculating its expectation value on a noiseless quantum simulator. We then performing the same calculation on a (noisy) quantum computer and study how imperfect hardware affects the precision of the calculation.

The system we consider is a $2\times 3$ lattice with $N_p=5$ in the weaved basis. To keep the system size small enough to study using quantum computers, we choose $n_q=1$. The observable we choose to calculate is $|\langle \hat{U}(t) \rangle|^2 \equiv |\langle \Psi_0^E | \hat{U}(t) | \Psi_0^E \rangle|^2$ for some fixed value of $t$ as a function of the coupling $g$, where $\hat{U}(t)$ is the time evolution operator, $|\Psi_0^E \rangle$ is the ground state of the electric Hamiltonian, and $t=0.2$. Before comparing the various errors in the calculation, we argue how $|\langle \hat{U}(t) \rangle|^2$ scales with $g$, which is done by studying how $\hat{H}_E$ and $\hat{H}_B$ scale as a function of $g$. The electric Hamiltonian goes as $\hat{H}_E \sim g^2 \hat{R}^2$ and the rotor operator goes as $\hat{R} \sim r_\text{max}$. As discussed in Sec.~\ref{ssec:originalBasisReview}, the formulation we use chooses $r_\text{max} \sim 1/b_\text{max}$. Since at small values of $g$ one has $b_\text{max} \sim 1/g$, the magnitude of the electric Hamiltonian in that region is roughly independent of $g$ at small $g$. The compact magnetic Hamiltonian at small $g$ scales as $\hat{H}_B^{(C)} \sim \hat{B}^2/g^2$ and because $\hat{B} \sim g$, $\hat{H}_B^{(C)}$ is also roughly independent of $g$. Thus, at small $g$ the Hamiltonian becomes independent of $g$ and the observable tends to some constant between $0$ and $1$. At large $g$, $\hat{H}_B^{(C)} \sim 1/g^2$. We therefore expect the electric Hamiltonian to dominate in that region, such that the operator $\hat U(t)$ is only adding a phase to the state $|\Psi_E\rangle$. This finally implies that $\langle \hat{U}(t) \rangle$ will approach one at large $g$.

Figure~\ref{fig:errorAnalysisSimulator} shows $|\langle \hat{U}(t=0.2) \rangle|^2$ as a function of $g$, calculated using two values of the Suzuki-Trotter step size $\delta t = \{10^{-3}, 0.2\}$ and two values of the cutoff $\theta_\text{min} = \{0, 2\delta t\}$ using a simulator. The first observation is that all four combinations studied exhibit the expected qualitative behavior, starting at some intermediate value at low $g$ and approaching one as $g$ is increased. The features that are quantitatively different are the value at low $g$, and the steepness of the approach to one as $g$ is increased. The solid gray line was calculated using $\delta t=10^{-3}$ and with $\theta_\text{min}=0$, and is considered the ``exact'' curve for our comparison.

We first study the effects of using a using a non-zero cutoff $\theta_\text{min}$. The blue cross marks in Fig.~\ref{fig:errorAnalysisSimulator} were calculated using $\delta t=10^{-3}$ and $\theta_\text{min}=2\delta t$. We observe that value at small $g$ is lower by $\sim 10\%$, and the transition to one at large $g$ is steeper.

We now study the effects of using a larger Suzuki-Trotter step-size $\delta t$. The orange diamonds in Fig.~\ref{fig:errorAnalysisSimulator} were calculated using $\theta_\text{min}=0$ for a single Suzuki-Trotter step with $\delta t=0.2$. Looking at small $g$, the Suzuki-Trotter error results in a value that is lower by $\sim 20\%$, which is a larger difference than only using a non-zero cutoff $\theta_\text{min}$. As $g$ is increased, we observe that the behavior matches the exact curve more closely.

We are now in a position to study the effect of performing a calculation with both Suzuki-Trotter and finite cutoff errors. The dashed black curve in Fig.~\ref{fig:errorAnalysisSimulator} was calculated using a single Suzuki-Trotter step with $\delta t=0.2$ and using a cutoff $\theta_\text{min}=2\delta t$. Looking first at the small $g$ region, we see that the combined error is almost identical to only the Suzuki-Trotter error. In other words, imposing the cutoff $\theta_\text{min}=2\delta t$ on top of using a larger Suzuki-Trotter step size of $\delta t=0.2$ introduces no additional errors for small $g$. Similarly for $g\sim 1$, the combined error in the steepness is similar to only the cutoff error from a non-zero $\theta_\text{min}$. To summarize, our numerical results show that the error from using a larger Suzuki-Trotter step in addition to using a cutoff is only as bad as the worse of the two individual errors. 

The final component of our error analysis is to study how errors from performing the calculation on NISQ hardware change the result in addition to Suzuki-Trotter and finite cutoff errors. The dashed black line in Fig.~\ref{fig:errorAnalysisIBMQ} is the result of a noiseless quantum simulation implemented in \textsc{Qiskit}~\cite{gadi_aleksandrowicz_2019_2562111} using $\delta t = 0.2$ and $\theta_\text{min} = 2\delta t$ (this data is the same as the black dashed line in Fig.~\ref{fig:errorAnalysisSimulator}). 

The rest of the markers in Fig.~\ref{fig:errorAnalysisIBMQ} correspond to runs on IBMQ superconducting qubit quantum hardware using $\delta t = 0.2$ and $\theta_\text{min} = 2\delta t$. We use a number of quantum computers: \verb+ibm_hanoi+,  \verb+ibmq_montreal+,  \verb+ibmq_guadalupe+,  \verb+ibmq_jakarta+,  \verb+ibm_cairo+,  \verb+ibm_auckland+,  and \verb+ibm_washington+. These machines range from 7 qubits to 127 qubits, with most machines operating with 27 qubits and the Falcon processer. Readout and gate errors are about 1\%, with readout errors typically slightly higher than gate errors. The circuit requires 5 logical qubits and the routing onto physical qubits is performed using \textsc{Qiskit} (transpile, optimization level 3).  

As NISQ devices, the errors on these computers are significant, which leads the raw output (red up arrows in Fig.~\ref{fig:errorAnalysisIBMQ}) of the computer to deviate significantly from the expectations.  We apply rudimentary gate and readout error mitigation techniques.  Gate error mitigation is performed using zero noise extrapolation with the fixed identity insertion method using one identity insertion~\cite{PhysRevLett.120.210501,PhysRevLett.119.180509,He:2020udd}.  Readout error mitigation is performed using Lucy-Richardson deconvolution (sometimes called Iterative Bayesian Unfolding in high energy physics~\cite{DAgostini:1994fjx}) with 20 iterations~\cite{1974AJ.....79..745L,Richardson:72,1910.01969}.  With these corrections, blue down arrow data points in Fig.~\ref{fig:errorAnalysisIBMQ} track the expected dependence on the coupling through $g\sim 1$, after which there is a 10\% deviation from the fact that the correct answer is all contained in one state so any errors move the answer away from the correct result.  Furthermore, the entangling gate count increases from a few below $g\sim 1$ to $\mathcal{O}(10)$ for $g\gtrsim 1$. It will be interesting to explore larger system sizes in the future and to combine the results with additional error mitigation schemes to further improve the precision (see e.g. Ref.~\cite{PhysRevLett.127.270502,Endo_2021}).

\section{Summary and Conclusion}
In this work, we presented an explicit circuit implementation for Suzuki-Trotter time-evolution of a $2+1$ dimensional, gauge redundancy free, \uone gauge theory, with a gate count that scales polynomially in the lattice volume. 

We implemented the formulation given in Ref.~\cite{Bauer_2021}, which enforces Gauss' law at the level of the Hamiltonian. This, combined with the fact that we enforce magnetic Gauss' law \textit{a priori}, leads to a gauge-redundancy free formulation. The non-compact formulation, in which the Hamiltonian only contains bilinear terms, can be implemented exactly using the Walsh function formalism with a number of gates that scales quadratically in the volume. The compact formulation on the other hand requires a number of gates that scales exponentially with the volume, the source of which is the maximally coupled term in the magnetic Hamiltonian, the cosine of the sum of all plaquette terms. While it is possible to break this exponential scaling in the quantum circuit gate count by dropping Walsh operators with small arguments, it is first necessary to classically calculate exponentially many Walsh coefficients, which is unrealistic even for small lattice volumes.

As a solution to both the quantum and classical exponential volume scaling, we instead apply an operator basis change, put forth in Ref.~\cite{https://doi.org/10.48550/arxiv.2208.03333}, that reduces the maximum number of magnetic field operators in a given cosine in the magnetic Hamiltonian to be logarithmic in the volume. This in turn reduces the number of Walsh coefficients, and therefore the circuit depth, to scale polynomially with the lattice volume; additionally, the polynomial number of Walsh coefficients can be calculated in a reasonable amount of time, even for large lattice volumes. Therefore, this method breaks both the classical and quantum exponential scaling. This polynomial gate count scaling can be further reduced by approximating the circuit using the Walsh function formalism. 

In addition to performing gate count studies, we expanded on the work in Ref.~\cite{https://doi.org/10.48550/arxiv.2208.03333} by providing a modified procedure for choosing $b_\text{max}$ in the weaved basis required to maintain comparable precision to the original basis. We performed several numerical tests for the $2\times 2$ site lattice and found that, using this new procedure, the precision of the weaved basis was comparable to the original basis for equal qubits per lattice site in both the non-compact and compact formulations.

It is important to note that if only the change of basis procedure is implemented, there remains an exponential scaling with the number of qubits per plaquettes, $n_q$. We argue that the exponential scaling in $n_q$ is not too problematic as the number of times the operators are sampled, and thus the digitization errors, also scale exponentially with $n_q$. However, because simulations will have to be carried out at multiple values of $n_q$ to understand the digitization errors, breaking this exponential scaling may prove to be important. As was demonstrated in Sec.~\ref{ssec:gateVsnqWeaved}, if the change of basis procedure is combined with the truncation method, then both the exponential scaling with $N_p$ and $n_q$ is reduced to polynomial.

Additionally, we studied the gate count as a function of $g$. We found that for large $g$, the magnetic Hamiltonian in the weaved basis can be completely neglected for choices of $\theta_\text{min} \sim \delta t$, indicating that it might be more efficient to work in the electric basis for large $g$. Looking at the gate count as a function of $\theta_\text{min}$ indicated that significant gate count reductions occur when $\theta_\text{min} \sim \delta t$. We then carried out an error analysis of an explicit observable comparing the Suzuki-Trotter error and cutoff error, including performing the simulation on real quantum hardware to see how noise from imperfect hardware contributes to the overall error.

While we applied the Walsh function formalism to time-evolution of a particular formulation of a \uone gauge theory in $2+1$ dimensions, it can be applied to any problem where one needs to construct diagonal matrices efficiently. The fact that this algorithm does not require ancillary qubits, combined with the reduction in CNOT gate count, makes it useful for NISQ era computations. Additionally, it provides a formalism for dropping $R_z$ gates with small arguments, which will be useful for fault-tolerant computations. The Walsh function formalism therefore has potential applications to state preparation, higher dimensional generalizations of \uone gauge theories, and non-abelian gauge theories, using both NISQ and fault-tolerant devices.

\begin{acknowledgements}
The authors would like to thank Hank Lamm, Natalie Klco, Jesse Stryker and Tobias Osborne for useful discussions and comments on the manuscript. This work was supported by the U.S. Department of Energy (DOE), Office of Science under contract DE-AC02-05CH11231, through Quantum Information Science Enabled Discovery
(QuantISED) for High Energy Physics (KA2401032), by the Office of Advanced Scientific Computing Research (ASCR) through the Accelerated Research for Quantum Computing Program, by the U.S. Department of Energy grant DE-FG02-97ER-41014 (Farrell), and the U.S. Department of Energy, Office of Science, Office of Nuclear Physics and \href{https://iqus.uw.edu}{\color{black}}{InQubator for Quantum Simulation (IQuS)} under Award Number DOE (NP) Award DE-SC0020970. CK is supported by the DOE Computational Science Graduate Fellowship under award number DE-SC0020347. This work is also supported, in part, through the \href{https://phys.washington.edu}{\color{black}}{Department of Physics} and \href{https://www.artsci.washington. edu}{\color{black}}{the College of Arts and Sciences}  at the University of Washington. This research used resources of the Oak Ridge Leadership Computing Facility, which is a DOE Office of Science User Facility supported under Contract DE-AC05-00OR22725.
\end{acknowledgements}

\bibliographystyle{apsrev4-1}
\bibliography{references}

\begin{thebibliography}{98}%
\makeatletter
\providecommand \@ifxundefined [1]{%
 \@ifx{#1\undefined}
}%
\providecommand \@ifnum [1]{%
 \ifnum #1\expandafter \@firstoftwo
 \else \expandafter \@secondoftwo
 \fi
}%
\providecommand \@ifx [1]{%
 \ifx #1\expandafter \@firstoftwo
 \else \expandafter \@secondoftwo
 \fi
}%
\providecommand \natexlab [1]{#1}%
\providecommand \enquote  [1]{``#1''}%
\providecommand \bibnamefont  [1]{#1}%
\providecommand \bibfnamefont [1]{#1}%
\providecommand \citenamefont [1]{#1}%
\providecommand \href@noop [0]{\@secondoftwo}%
\providecommand \href [0]{\begingroup \@sanitize@url \@href}%
\providecommand \@href[1]{\@@startlink{#1}\@@href}%
\providecommand \@@href[1]{\endgroup#1\@@endlink}%
\providecommand \@sanitize@url [0]{\catcode `\\12\catcode `\$12\catcode
  `\&12\catcode `\#12\catcode `\^12\catcode `\_12\catcode `\%12\relax}%
\providecommand \@@startlink[1]{}%
\providecommand \@@endlink[0]{}%
\providecommand \url  [0]{\begingroup\@sanitize@url \@url }%
\providecommand \@url [1]{\endgroup\@href {#1}{\urlprefix }}%
\providecommand \urlprefix  [0]{URL }%
\providecommand \Eprint [0]{\href }%
\providecommand \doibase [0]{http://dx.doi.org/}%
\providecommand \selectlanguage [0]{\@gobble}%
\providecommand \bibinfo  [0]{\@secondoftwo}%
\providecommand \bibfield  [0]{\@secondoftwo}%
\providecommand \translation [1]{[#1]}%
\providecommand \BibitemOpen [0]{}%
\providecommand \bibitemStop [0]{}%
\providecommand \bibitemNoStop [0]{.\EOS\space}%
\providecommand \EOS [0]{\spacefactor3000\relax}%
\providecommand \BibitemShut  [1]{\csname bibitem#1\endcsname}%
\let\auto@bib@innerbib\@empty
\bibitem [{\citenamefont {Carena}\ \emph
  {et~al.}(2022{\natexlab{a}})\citenamefont {Carena}, \citenamefont {Lamm},
  \citenamefont {Li},\ and\ \citenamefont {Liu}}]{Carena:2022kpg}%
  \BibitemOpen
  \bibfield  {author} {\bibinfo {author} {\bibfnamefont {M.}~\bibnamefont
  {Carena}}, \bibinfo {author} {\bibfnamefont {H.}~\bibnamefont {Lamm}},
  \bibinfo {author} {\bibfnamefont {Y.-Y.}\ \bibnamefont {Li}}, \ and\ \bibinfo
  {author} {\bibfnamefont {W.}~\bibnamefont {Liu}},\ }\href@noop {} {\
  (\bibinfo {year} {2022}{\natexlab{a}})},\ \Eprint
  {http://arxiv.org/abs/2203.02823} {arXiv:2203.02823 [hep-lat]} \BibitemShut
  {NoStop}%
\bibitem [{\citenamefont {Halimeh}\ and\ \citenamefont
  {Hauke}(2020)}]{Halimeh_2020}%
  \BibitemOpen
  \bibfield  {author} {\bibinfo {author} {\bibfnamefont {J.~C.}\ \bibnamefont
  {Halimeh}}\ and\ \bibinfo {author} {\bibfnamefont {P.}~\bibnamefont
  {Hauke}},\ }\href {\doibase 10.1103/physrevlett.125.030503} {\bibfield
  {journal} {\bibinfo  {journal} {Physical Review Letters}\ }\textbf {\bibinfo
  {volume} {125}} (\bibinfo {year} {2020}),\
  10.1103/physrevlett.125.030503}\BibitemShut {NoStop}%
\bibitem [{\citenamefont {Halimeh}\ \emph
  {et~al.}(2021{\natexlab{a}})\citenamefont {Halimeh}, \citenamefont {Lang},
  \citenamefont {Mildenberger}, \citenamefont {Jiang},\ and\ \citenamefont
  {Hauke}}]{Halimeh:2020ecg}%
  \BibitemOpen
  \bibfield  {author} {\bibinfo {author} {\bibfnamefont {J.~C.}\ \bibnamefont
  {Halimeh}}, \bibinfo {author} {\bibfnamefont {H.}~\bibnamefont {Lang}},
  \bibinfo {author} {\bibfnamefont {J.}~\bibnamefont {Mildenberger}}, \bibinfo
  {author} {\bibfnamefont {Z.}~\bibnamefont {Jiang}}, \ and\ \bibinfo {author}
  {\bibfnamefont {P.}~\bibnamefont {Hauke}},\ }\href {\doibase
  10.1103/PRXQuantum.2.040311} {\bibfield  {journal} {\bibinfo  {journal} {PRX
  Quantum}\ }\textbf {\bibinfo {volume} {2}},\ \bibinfo {pages} {040311}
  (\bibinfo {year} {2021}{\natexlab{a}})},\ \Eprint
  {http://arxiv.org/abs/2007.00668} {arXiv:2007.00668 [quant-ph]} \BibitemShut
  {NoStop}%
\bibitem [{\citenamefont {Halimeh}\ \emph
  {et~al.}(2021{\natexlab{b}})\citenamefont {Halimeh}, \citenamefont {Homeier},
  \citenamefont {Schweizer}, \citenamefont {Aidelsburger}, \citenamefont
  {Hauke},\ and\ \citenamefont {Grusdt}}]{Halimeh:2021lnv}%
  \BibitemOpen
  \bibfield  {author} {\bibinfo {author} {\bibfnamefont {J.~C.}\ \bibnamefont
  {Halimeh}}, \bibinfo {author} {\bibfnamefont {L.}~\bibnamefont {Homeier}},
  \bibinfo {author} {\bibfnamefont {C.}~\bibnamefont {Schweizer}}, \bibinfo
  {author} {\bibfnamefont {M.}~\bibnamefont {Aidelsburger}}, \bibinfo {author}
  {\bibfnamefont {P.}~\bibnamefont {Hauke}}, \ and\ \bibinfo {author}
  {\bibfnamefont {F.}~\bibnamefont {Grusdt}},\ }\href@noop {} {\  (\bibinfo
  {year} {2021}{\natexlab{b}})},\ \Eprint {http://arxiv.org/abs/2108.02203}
  {arXiv:2108.02203 [cond-mat.quant-gas]} \BibitemShut {NoStop}%
\bibitem [{\citenamefont {Halimeh}\ \emph {et~al.}(2022)\citenamefont
  {Halimeh}, \citenamefont {Lang},\ and\ \citenamefont
  {Hauke}}]{Halimeh:2021vzf}%
  \BibitemOpen
  \bibfield  {author} {\bibinfo {author} {\bibfnamefont {J.~C.}\ \bibnamefont
  {Halimeh}}, \bibinfo {author} {\bibfnamefont {H.}~\bibnamefont {Lang}}, \
  and\ \bibinfo {author} {\bibfnamefont {P.}~\bibnamefont {Hauke}},\ }\href
  {\doibase 10.1088/1367-2630/ac5564} {\bibfield  {journal} {\bibinfo
  {journal} {New J. Phys.}\ }\textbf {\bibinfo {volume} {24}},\ \bibinfo
  {pages} {033015} (\bibinfo {year} {2022})},\ \Eprint
  {http://arxiv.org/abs/2106.09032} {arXiv:2106.09032 [cond-mat.quant-gas]}
  \BibitemShut {NoStop}%
\bibitem [{\citenamefont {Lamm}\ \emph {et~al.}(2020)\citenamefont {Lamm},
  \citenamefont {Lawrence},\ and\ \citenamefont
  {Yamauchi}}]{Lamm_Scott_Yukari_2020}%
  \BibitemOpen
  \bibfield  {author} {\bibinfo {author} {\bibfnamefont {H.}~\bibnamefont
  {Lamm}}, \bibinfo {author} {\bibfnamefont {S.}~\bibnamefont {Lawrence}}, \
  and\ \bibinfo {author} {\bibfnamefont {Y.}~\bibnamefont {Yamauchi}},\ }\href
  {\doibase 10.48550/ARXIV.2005.12688} {\enquote {\bibinfo {title} {Suppressing
  coherent gauge drift in quantum simulations},}\ } (\bibinfo {year}
  {2020})\BibitemShut {NoStop}%
\bibitem [{\citenamefont {Tran}\ \emph {et~al.}(2021)\citenamefont {Tran},
  \citenamefont {Su}, \citenamefont {Carney},\ and\ \citenamefont
  {Taylor}}]{Tran_2021}%
  \BibitemOpen
  \bibfield  {author} {\bibinfo {author} {\bibfnamefont {M.~C.}\ \bibnamefont
  {Tran}}, \bibinfo {author} {\bibfnamefont {Y.}~\bibnamefont {Su}}, \bibinfo
  {author} {\bibfnamefont {D.}~\bibnamefont {Carney}}, \ and\ \bibinfo {author}
  {\bibfnamefont {J.~M.}\ \bibnamefont {Taylor}},\ }\href {\doibase
  10.1103/prxquantum.2.010323} {\bibfield  {journal} {\bibinfo  {journal}
  {{PRX} Quantum}\ }\textbf {\bibinfo {volume} {2}} (\bibinfo {year} {2021}),\
  10.1103/prxquantum.2.010323}\BibitemShut {NoStop}%
\bibitem [{\citenamefont {Zohar}\ \emph {et~al.}(2012)\citenamefont {Zohar},
  \citenamefont {Cirac},\ and\ \citenamefont
  {Reznik}}]{PhysRevLett.109.125302}%
  \BibitemOpen
  \bibfield  {author} {\bibinfo {author} {\bibfnamefont {E.}~\bibnamefont
  {Zohar}}, \bibinfo {author} {\bibfnamefont {J.~I.}\ \bibnamefont {Cirac}}, \
  and\ \bibinfo {author} {\bibfnamefont {B.}~\bibnamefont {Reznik}},\ }\href
  {\doibase 10.1103/PhysRevLett.109.125302} {\bibfield  {journal} {\bibinfo
  {journal} {Phys. Rev. Lett.}\ }\textbf {\bibinfo {volume} {109}},\ \bibinfo
  {pages} {125302} (\bibinfo {year} {2012})}\BibitemShut {NoStop}%
\bibitem [{\citenamefont {Banerjee}\ \emph {et~al.}(2012)\citenamefont
  {Banerjee}, \citenamefont {Dalmonte}, \citenamefont {Müller}, \citenamefont
  {Rico}, \citenamefont {Stebler}, \citenamefont {Wiese},\ and\ \citenamefont
  {Zoller}}]{Banerjee_2012}%
  \BibitemOpen
  \bibfield  {author} {\bibinfo {author} {\bibfnamefont {D.}~\bibnamefont
  {Banerjee}}, \bibinfo {author} {\bibfnamefont {M.}~\bibnamefont {Dalmonte}},
  \bibinfo {author} {\bibfnamefont {M.}~\bibnamefont {Müller}}, \bibinfo
  {author} {\bibfnamefont {E.}~\bibnamefont {Rico}}, \bibinfo {author}
  {\bibfnamefont {P.}~\bibnamefont {Stebler}}, \bibinfo {author} {\bibfnamefont
  {U.-J.}\ \bibnamefont {Wiese}}, \ and\ \bibinfo {author} {\bibfnamefont
  {P.}~\bibnamefont {Zoller}},\ }\href {\doibase
  10.1103/physrevlett.109.175302} {\bibfield  {journal} {\bibinfo  {journal}
  {Physical Review Letters}\ }\textbf {\bibinfo {volume} {109}} (\bibinfo
  {year} {2012}),\ 10.1103/physrevlett.109.175302}\BibitemShut {NoStop}%
\bibitem [{\citenamefont {Kasper}\ \emph {et~al.}(2020)\citenamefont {Kasper},
  \citenamefont {Zache}, \citenamefont {Jendrzejewski}, \citenamefont
  {Lewenstein},\ and\ \citenamefont
  {Zohar}}]{https://doi.org/10.48550/arxiv.2012.08620}%
  \BibitemOpen
  \bibfield  {author} {\bibinfo {author} {\bibfnamefont {V.}~\bibnamefont
  {Kasper}}, \bibinfo {author} {\bibfnamefont {T.~V.}\ \bibnamefont {Zache}},
  \bibinfo {author} {\bibfnamefont {F.}~\bibnamefont {Jendrzejewski}}, \bibinfo
  {author} {\bibfnamefont {M.}~\bibnamefont {Lewenstein}}, \ and\ \bibinfo
  {author} {\bibfnamefont {E.}~\bibnamefont {Zohar}},\ }\href {\doibase
  10.48550/ARXIV.2012.08620} {\enquote {\bibinfo {title} {Non-abelian gauge
  invariance from dynamical decoupling},}\ } (\bibinfo {year}
  {2020})\BibitemShut {NoStop}%
\bibitem [{\citenamefont {Hauke}\ \emph {et~al.}(2013)\citenamefont {Hauke},
  \citenamefont {Marcos}, \citenamefont {Dalmonte},\ and\ \citenamefont
  {Zoller}}]{PhysRevX.3.041018}%
  \BibitemOpen
  \bibfield  {author} {\bibinfo {author} {\bibfnamefont {P.}~\bibnamefont
  {Hauke}}, \bibinfo {author} {\bibfnamefont {D.}~\bibnamefont {Marcos}},
  \bibinfo {author} {\bibfnamefont {M.}~\bibnamefont {Dalmonte}}, \ and\
  \bibinfo {author} {\bibfnamefont {P.}~\bibnamefont {Zoller}},\ }\href
  {\doibase 10.1103/PhysRevX.3.041018} {\bibfield  {journal} {\bibinfo
  {journal} {Phys. Rev. X}\ }\textbf {\bibinfo {volume} {3}},\ \bibinfo {pages}
  {041018} (\bibinfo {year} {2013})}\BibitemShut {NoStop}%
\bibitem [{\citenamefont {K\"uhn}\ \emph {et~al.}(2014)\citenamefont {K\"uhn},
  \citenamefont {Cirac},\ and\ \citenamefont {Ba\~nuls}}]{PhysRevA.90.042305}%
  \BibitemOpen
  \bibfield  {author} {\bibinfo {author} {\bibfnamefont {S.}~\bibnamefont
  {K\"uhn}}, \bibinfo {author} {\bibfnamefont {J.~I.}\ \bibnamefont {Cirac}}, \
  and\ \bibinfo {author} {\bibfnamefont {M.-C.}\ \bibnamefont {Ba\~nuls}},\
  }\href {\doibase 10.1103/PhysRevA.90.042305} {\bibfield  {journal} {\bibinfo
  {journal} {Phys. Rev. A}\ }\textbf {\bibinfo {volume} {90}},\ \bibinfo
  {pages} {042305} (\bibinfo {year} {2014})}\BibitemShut {NoStop}%
\bibitem [{\citenamefont {Stannigel}\ \emph {et~al.}(2014)\citenamefont
  {Stannigel}, \citenamefont {Hauke}, \citenamefont {Marcos}, \citenamefont
  {Hafezi}, \citenamefont {Diehl}, \citenamefont {Dalmonte},\ and\
  \citenamefont {Zoller}}]{Stannigel_2014}%
  \BibitemOpen
  \bibfield  {author} {\bibinfo {author} {\bibfnamefont {K.}~\bibnamefont
  {Stannigel}}, \bibinfo {author} {\bibfnamefont {P.}~\bibnamefont {Hauke}},
  \bibinfo {author} {\bibfnamefont {D.}~\bibnamefont {Marcos}}, \bibinfo
  {author} {\bibfnamefont {M.}~\bibnamefont {Hafezi}}, \bibinfo {author}
  {\bibfnamefont {S.}~\bibnamefont {Diehl}}, \bibinfo {author} {\bibfnamefont
  {M.}~\bibnamefont {Dalmonte}}, \ and\ \bibinfo {author} {\bibfnamefont
  {P.}~\bibnamefont {Zoller}},\ }\href {\doibase
  10.1103/physrevlett.112.120406} {\bibfield  {journal} {\bibinfo  {journal}
  {Physical Review Letters}\ }\textbf {\bibinfo {volume} {112}} (\bibinfo
  {year} {2014}),\ 10.1103/physrevlett.112.120406}\BibitemShut {NoStop}%
\bibitem [{\citenamefont {Stryker}(2019)}]{Stryker_2019}%
  \BibitemOpen
  \bibfield  {author} {\bibinfo {author} {\bibfnamefont {J.~R.}\ \bibnamefont
  {Stryker}},\ }\href {\doibase 10.1103/physreva.99.042301} {\bibfield
  {journal} {\bibinfo  {journal} {Physical Review A}\ }\textbf {\bibinfo
  {volume} {99}} (\bibinfo {year} {2019}),\
  10.1103/physreva.99.042301}\BibitemShut {NoStop}%
\bibitem [{\citenamefont {Kaplan}\ and\ \citenamefont
  {Stryker}(2020)}]{Kaplan:2018vnj}%
  \BibitemOpen
  \bibfield  {author} {\bibinfo {author} {\bibfnamefont {D.~B.}\ \bibnamefont
  {Kaplan}}\ and\ \bibinfo {author} {\bibfnamefont {J.~R.}\ \bibnamefont
  {Stryker}},\ }\href {\doibase 10.1103/PhysRevD.102.094515} {\bibfield
  {journal} {\bibinfo  {journal} {Phys. Rev. D}\ }\textbf {\bibinfo {volume}
  {102}},\ \bibinfo {pages} {094515} (\bibinfo {year} {2020})},\ \Eprint
  {http://arxiv.org/abs/1806.08797} {arXiv:1806.08797 [hep-lat]} \BibitemShut
  {NoStop}%
\bibitem [{\citenamefont {Haase}\ \emph {et~al.}(2021)\citenamefont {Haase},
  \citenamefont {Dellantonio}, \citenamefont {Celi}, \citenamefont {Paulson},
  \citenamefont {Kan}, \citenamefont {Jansen},\ and\ \citenamefont
  {Muschik}}]{Haase:2020kaj}%
  \BibitemOpen
  \bibfield  {author} {\bibinfo {author} {\bibfnamefont {J.~F.}\ \bibnamefont
  {Haase}}, \bibinfo {author} {\bibfnamefont {L.}~\bibnamefont {Dellantonio}},
  \bibinfo {author} {\bibfnamefont {A.}~\bibnamefont {Celi}}, \bibinfo {author}
  {\bibfnamefont {D.}~\bibnamefont {Paulson}}, \bibinfo {author} {\bibfnamefont
  {A.}~\bibnamefont {Kan}}, \bibinfo {author} {\bibfnamefont {K.}~\bibnamefont
  {Jansen}}, \ and\ \bibinfo {author} {\bibfnamefont {C.~A.}\ \bibnamefont
  {Muschik}},\ }\href {\doibase 10.22331/q-2021-02-04-393} {\bibfield
  {journal} {\bibinfo  {journal} {Quantum}\ }\textbf {\bibinfo {volume} {5}},\
  \bibinfo {pages} {393} (\bibinfo {year} {2021})},\ \Eprint
  {http://arxiv.org/abs/2006.14160} {arXiv:2006.14160 [quant-ph]} \BibitemShut
  {NoStop}%
\bibitem [{\citenamefont {Bauer}\ and\ \citenamefont
  {Grabowska}(2021)}]{Bauer_2021}%
  \BibitemOpen
  \bibfield  {author} {\bibinfo {author} {\bibfnamefont {C.~W.}\ \bibnamefont
  {Bauer}}\ and\ \bibinfo {author} {\bibfnamefont {D.~M.}\ \bibnamefont
  {Grabowska}},\ }\href {\doibase 10.48550/ARXIV.2111.08015} {\  (\bibinfo
  {year} {2021}),\ 10.48550/ARXIV.2111.08015}\BibitemShut {NoStop}%
\bibitem [{\citenamefont {Wiese}(2013)}]{Wiese_2013}%
  \BibitemOpen
  \bibfield  {author} {\bibinfo {author} {\bibfnamefont {U.-J.}\ \bibnamefont
  {Wiese}},\ }\href {\doibase 10.1002/andp.201300104} {\bibfield  {journal}
  {\bibinfo  {journal} {Annalen der Physik}\ }\textbf {\bibinfo {volume}
  {525}},\ \bibinfo {pages} {777} (\bibinfo {year} {2013})}\BibitemShut
  {NoStop}%
\bibitem [{\citenamefont {Zohar}\ \emph {et~al.}(2015)\citenamefont {Zohar},
  \citenamefont {Cirac},\ and\ \citenamefont {Reznik}}]{Zohar_2015}%
  \BibitemOpen
  \bibfield  {author} {\bibinfo {author} {\bibfnamefont {E.}~\bibnamefont
  {Zohar}}, \bibinfo {author} {\bibfnamefont {J.~I.}\ \bibnamefont {Cirac}}, \
  and\ \bibinfo {author} {\bibfnamefont {B.}~\bibnamefont {Reznik}},\ }\href
  {\doibase 10.1088/0034-4885/79/1/014401} {\bibfield  {journal} {\bibinfo
  {journal} {Reports on Progress in Physics}\ }\textbf {\bibinfo {volume}
  {79}},\ \bibinfo {pages} {014401} (\bibinfo {year} {2015})}\BibitemShut
  {NoStop}%
\bibitem [{\citenamefont {Dalmonte}\ and\ \citenamefont
  {Montangero}(2016)}]{Dalmonte_2016}%
  \BibitemOpen
  \bibfield  {author} {\bibinfo {author} {\bibfnamefont {M.}~\bibnamefont
  {Dalmonte}}\ and\ \bibinfo {author} {\bibfnamefont {S.}~\bibnamefont
  {Montangero}},\ }\href {\doibase 10.1080/00107514.2016.1151199} {\bibfield
  {journal} {\bibinfo  {journal} {Contemporary Physics}\ }\textbf {\bibinfo
  {volume} {57}},\ \bibinfo {pages} {388} (\bibinfo {year} {2016})}\BibitemShut
  {NoStop}%
\bibitem [{\citenamefont {Aidelsburger}\ \emph {et~al.}(2021)\citenamefont
  {Aidelsburger}, \citenamefont {Barbiero}, \citenamefont {Bermudez},
  \citenamefont {Chanda}, \citenamefont {Dauphin}, \citenamefont {Gonz{\'{a}
  }lez-Cuadra}, \citenamefont {Grzybowski}, \citenamefont {Hands},
  \citenamefont {Jendrzejewski}, \citenamefont {Jünemann}, \citenamefont
  {Juzeli{\={u}}nas}, \citenamefont {Kasper}, \citenamefont {Piga},
  \citenamefont {Ran}, \citenamefont {Rizzi}, \citenamefont {Sierra},
  \citenamefont {Tagliacozzo}, \citenamefont {Tirrito}, \citenamefont {Zache},
  \citenamefont {Zakrzewski}, \citenamefont {Zohar},\ and\ \citenamefont
  {Lewenstein}}]{Aidelsburger_2021}%
  \BibitemOpen
  \bibfield  {author} {\bibinfo {author} {\bibfnamefont {M.}~\bibnamefont
  {Aidelsburger}}, \bibinfo {author} {\bibfnamefont {L.}~\bibnamefont
  {Barbiero}}, \bibinfo {author} {\bibfnamefont {A.}~\bibnamefont {Bermudez}},
  \bibinfo {author} {\bibfnamefont {T.}~\bibnamefont {Chanda}}, \bibinfo
  {author} {\bibfnamefont {A.}~\bibnamefont {Dauphin}}, \bibinfo {author}
  {\bibfnamefont {D.}~\bibnamefont {Gonz{\'{a} }lez-Cuadra}}, \bibinfo {author}
  {\bibfnamefont {P.~R.}\ \bibnamefont {Grzybowski}}, \bibinfo {author}
  {\bibfnamefont {S.}~\bibnamefont {Hands}}, \bibinfo {author} {\bibfnamefont
  {F.}~\bibnamefont {Jendrzejewski}}, \bibinfo {author} {\bibfnamefont
  {J.}~\bibnamefont {Jünemann}}, \bibinfo {author} {\bibfnamefont
  {G.}~\bibnamefont {Juzeli{\={u}}nas}}, \bibinfo {author} {\bibfnamefont
  {V.}~\bibnamefont {Kasper}}, \bibinfo {author} {\bibfnamefont
  {A.}~\bibnamefont {Piga}}, \bibinfo {author} {\bibfnamefont {S.-J.}\
  \bibnamefont {Ran}}, \bibinfo {author} {\bibfnamefont {M.}~\bibnamefont
  {Rizzi}}, \bibinfo {author} {\bibfnamefont {G.}~\bibnamefont {Sierra}},
  \bibinfo {author} {\bibfnamefont {L.}~\bibnamefont {Tagliacozzo}}, \bibinfo
  {author} {\bibfnamefont {E.}~\bibnamefont {Tirrito}}, \bibinfo {author}
  {\bibfnamefont {T.~V.}\ \bibnamefont {Zache}}, \bibinfo {author}
  {\bibfnamefont {J.}~\bibnamefont {Zakrzewski}}, \bibinfo {author}
  {\bibfnamefont {E.}~\bibnamefont {Zohar}}, \ and\ \bibinfo {author}
  {\bibfnamefont {M.}~\bibnamefont {Lewenstein}},\ }\href {\doibase
  10.1098/rsta.2021.0064} {\bibfield  {journal} {\bibinfo  {journal}
  {Philosophical Transactions of the Royal Society A: Mathematical, Physical
  and Engineering Sciences}\ }\textbf {\bibinfo {volume} {380}} (\bibinfo
  {year} {2021}),\ 10.1098/rsta.2021.0064}\BibitemShut {NoStop}%
\bibitem [{\citenamefont {Ba{\~{n}}uls}\ \emph {et~al.}(2020)\citenamefont
  {Ba{\~{n}}uls}, \citenamefont {Blatt}, \citenamefont {Catani}, \citenamefont
  {Celi}, \citenamefont {Cirac}, \citenamefont {Dalmonte}, \citenamefont
  {Fallani}, \citenamefont {Jansen}, \citenamefont {Lewenstein}, \citenamefont
  {Montangero}, \citenamefont {Muschik}, \citenamefont {Reznik}, \citenamefont
  {Rico}, \citenamefont {Tagliacozzo}, \citenamefont {Acoleyen}, \citenamefont
  {Verstraete}, \citenamefont {Wiese}, \citenamefont {Wingate}, \citenamefont
  {Zakrzewski},\ and\ \citenamefont {Zoller}}]{Ba_uls_2020}%
  \BibitemOpen
  \bibfield  {author} {\bibinfo {author} {\bibfnamefont {M.~C.}\ \bibnamefont
  {Ba{\~{n}}uls}}, \bibinfo {author} {\bibfnamefont {R.}~\bibnamefont {Blatt}},
  \bibinfo {author} {\bibfnamefont {J.}~\bibnamefont {Catani}}, \bibinfo
  {author} {\bibfnamefont {A.}~\bibnamefont {Celi}}, \bibinfo {author}
  {\bibfnamefont {J.~I.}\ \bibnamefont {Cirac}}, \bibinfo {author}
  {\bibfnamefont {M.}~\bibnamefont {Dalmonte}}, \bibinfo {author}
  {\bibfnamefont {L.}~\bibnamefont {Fallani}}, \bibinfo {author} {\bibfnamefont
  {K.}~\bibnamefont {Jansen}}, \bibinfo {author} {\bibfnamefont
  {M.}~\bibnamefont {Lewenstein}}, \bibinfo {author} {\bibfnamefont
  {S.}~\bibnamefont {Montangero}}, \bibinfo {author} {\bibfnamefont {C.~A.}\
  \bibnamefont {Muschik}}, \bibinfo {author} {\bibfnamefont {B.}~\bibnamefont
  {Reznik}}, \bibinfo {author} {\bibfnamefont {E.}~\bibnamefont {Rico}},
  \bibinfo {author} {\bibfnamefont {L.}~\bibnamefont {Tagliacozzo}}, \bibinfo
  {author} {\bibfnamefont {K.~V.}\ \bibnamefont {Acoleyen}}, \bibinfo {author}
  {\bibfnamefont {F.}~\bibnamefont {Verstraete}}, \bibinfo {author}
  {\bibfnamefont {U.-J.}\ \bibnamefont {Wiese}}, \bibinfo {author}
  {\bibfnamefont {M.}~\bibnamefont {Wingate}}, \bibinfo {author} {\bibfnamefont
  {J.}~\bibnamefont {Zakrzewski}}, \ and\ \bibinfo {author} {\bibfnamefont
  {P.}~\bibnamefont {Zoller}},\ }\href {\doibase 10.1140/epjd/e2020-100571-8}
  {\bibfield  {journal} {\bibinfo  {journal} {The European Physical Journal D}\
  }\textbf {\bibinfo {volume} {74}} (\bibinfo {year} {2020}),\
  10.1140/epjd/e2020-100571-8}\BibitemShut {NoStop}%
\bibitem [{\citenamefont {Zohar}(2021)}]{Zohar_2021}%
  \BibitemOpen
  \bibfield  {author} {\bibinfo {author} {\bibfnamefont {E.}~\bibnamefont
  {Zohar}},\ }\href {\doibase 10.1098/rsta.2021.0069} {\bibfield  {journal}
  {\bibinfo  {journal} {Philosophical Transactions of the Royal Society A:
  Mathematical, Physical and Engineering Sciences}\ }\textbf {\bibinfo {volume}
  {380}} (\bibinfo {year} {2021}),\ 10.1098/rsta.2021.0069}\BibitemShut
  {NoStop}%
\bibitem [{\citenamefont {Klco}\ \emph {et~al.}(2022)\citenamefont {Klco},
  \citenamefont {Roggero},\ and\ \citenamefont {Savage}}]{Klco_2022}%
  \BibitemOpen
  \bibfield  {author} {\bibinfo {author} {\bibfnamefont {N.}~\bibnamefont
  {Klco}}, \bibinfo {author} {\bibfnamefont {A.}~\bibnamefont {Roggero}}, \
  and\ \bibinfo {author} {\bibfnamefont {M.~J.}\ \bibnamefont {Savage}},\
  }\href {\doibase 10.1088/1361-6633/ac58a4} {\bibfield  {journal} {\bibinfo
  {journal} {Reports on Progress in Physics}\ }\textbf {\bibinfo {volume}
  {85}},\ \bibinfo {pages} {064301} (\bibinfo {year} {2022})}\BibitemShut
  {NoStop}%
\bibitem [{\citenamefont {Bauer}\ \emph {et~al.}(2022)\citenamefont {Bauer}
  \emph {et~al.}}]{Bauer:2022hpo}%
  \BibitemOpen
  \bibfield  {author} {\bibinfo {author} {\bibfnamefont {C.~W.}\ \bibnamefont
  {Bauer}} \emph {et~al.},\ }\href@noop {} {\  (\bibinfo {year} {2022})},\
  \Eprint {http://arxiv.org/abs/2204.03381} {arXiv:2204.03381 [quant-ph]}
  \BibitemShut {NoStop}%
\bibitem [{\citenamefont {Anishetty}\ \emph
  {et~al.}(2009{\natexlab{a}})\citenamefont {Anishetty}, \citenamefont
  {Mathur},\ and\ \citenamefont
  {Raychowdhury}}]{Anishetty_Mathur_Raychowdhury_2009}%
  \BibitemOpen
  \bibfield  {author} {\bibinfo {author} {\bibfnamefont {R.}~\bibnamefont
  {Anishetty}}, \bibinfo {author} {\bibfnamefont {M.}~\bibnamefont {Mathur}}, \
  and\ \bibinfo {author} {\bibfnamefont {I.}~\bibnamefont {Raychowdhury}},\
  }\href {\doibase 10.1063/1.3122666} {\bibfield  {journal} {\bibinfo
  {journal} {J. Math. Phys.}\ }\textbf {\bibinfo {volume} {50}},\ \bibinfo
  {pages} {053503} (\bibinfo {year} {2009}{\natexlab{a}})},\ \Eprint
  {http://arxiv.org/abs/0901.0644} {arXiv:0901.0644 [math-ph]} \BibitemShut
  {NoStop}%
\bibitem [{\citenamefont {Raychowdhury}\ and\ \citenamefont
  {Stryker}(2020{\natexlab{a}})}]{Raychowdhury_Stryker_2020}%
  \BibitemOpen
  \bibfield  {author} {\bibinfo {author} {\bibfnamefont {I.}~\bibnamefont
  {Raychowdhury}}\ and\ \bibinfo {author} {\bibfnamefont {J.~R.}\ \bibnamefont
  {Stryker}},\ }\href {\doibase 10.1103/physrevresearch.2.033039} {\bibfield
  {journal} {\bibinfo  {journal} {Physical Review Research}\ }\textbf {\bibinfo
  {volume} {2}} (\bibinfo {year} {2020}{\natexlab{a}}),\
  10.1103/physrevresearch.2.033039}\BibitemShut {NoStop}%
\bibitem [{\citenamefont {Raychowdhury}\ and\ \citenamefont
  {Stryker}(2020{\natexlab{b}})}]{Raychowdhury_Stryker_SU2_2020}%
  \BibitemOpen
  \bibfield  {author} {\bibinfo {author} {\bibfnamefont {I.}~\bibnamefont
  {Raychowdhury}}\ and\ \bibinfo {author} {\bibfnamefont {J.~R.}\ \bibnamefont
  {Stryker}},\ }\href {\doibase 10.1103/physrevd.101.114502} {\bibfield
  {journal} {\bibinfo  {journal} {Physical Review D}\ }\textbf {\bibinfo
  {volume} {101}} (\bibinfo {year} {2020}{\natexlab{b}}),\
  10.1103/physrevd.101.114502}\BibitemShut {NoStop}%
\bibitem [{\citenamefont {Anishetty}\ \emph
  {et~al.}(2009{\natexlab{b}})\citenamefont {Anishetty}, \citenamefont
  {Mathur},\ and\ \citenamefont
  {Raychowdhury}}]{Anishetty_Mathur_Raychowdhury_SU3_2009}%
  \BibitemOpen
  \bibfield  {author} {\bibinfo {author} {\bibfnamefont {R.}~\bibnamefont
  {Anishetty}}, \bibinfo {author} {\bibfnamefont {M.}~\bibnamefont {Mathur}}, \
  and\ \bibinfo {author} {\bibfnamefont {I.}~\bibnamefont {Raychowdhury}},\
  }\href {\doibase 10.1063/1.3122666} {\bibfield  {journal} {\bibinfo
  {journal} {Journal of Mathematical Physics}\ }\textbf {\bibinfo {volume}
  {50}},\ \bibinfo {pages} {053503} (\bibinfo {year}
  {2009}{\natexlab{b}})}\BibitemShut {NoStop}%
\bibitem [{\citenamefont {Mathur}\ \emph {et~al.}(2010)\citenamefont {Mathur},
  \citenamefont {Raychowdhury},\ and\ \citenamefont
  {Anishetty}}]{doi:10.1063/1.3464267}%
  \BibitemOpen
  \bibfield  {author} {\bibinfo {author} {\bibfnamefont {M.}~\bibnamefont
  {Mathur}}, \bibinfo {author} {\bibfnamefont {I.}~\bibnamefont
  {Raychowdhury}}, \ and\ \bibinfo {author} {\bibfnamefont {R.}~\bibnamefont
  {Anishetty}},\ }\href {\doibase 10.1063/1.3464267} {\bibfield  {journal}
  {\bibinfo  {journal} {Journal of Mathematical Physics}\ }\textbf {\bibinfo
  {volume} {51}},\ \bibinfo {pages} {093504} (\bibinfo {year} {2010})},\
  \Eprint {http://arxiv.org/abs/https://doi.org/10.1063/1.3464267}
  {https://doi.org/10.1063/1.3464267} \BibitemShut {NoStop}%
\bibitem [{\citenamefont {Mathur}\ \emph {et~al.}(2011)\citenamefont {Mathur},
  \citenamefont {Raychowdhury},\ and\ \citenamefont
  {Sreeraj}}]{doi:10.1063/1.3660195}%
  \BibitemOpen
  \bibfield  {author} {\bibinfo {author} {\bibfnamefont {M.}~\bibnamefont
  {Mathur}}, \bibinfo {author} {\bibfnamefont {I.}~\bibnamefont
  {Raychowdhury}}, \ and\ \bibinfo {author} {\bibfnamefont {T.~P.}\
  \bibnamefont {Sreeraj}},\ }\href {\doibase 10.1063/1.3660195} {\bibfield
  {journal} {\bibinfo  {journal} {Journal of Mathematical Physics}\ }\textbf
  {\bibinfo {volume} {52}},\ \bibinfo {pages} {113505} (\bibinfo {year}
  {2011})},\ \Eprint {http://arxiv.org/abs/https://doi.org/10.1063/1.3660195}
  {https://doi.org/10.1063/1.3660195} \BibitemShut {NoStop}%
\bibitem [{\citenamefont {Raychowdhury}(2013)}]{Raychowdhury_thesis}%
  \BibitemOpen
  \bibfield  {author} {\bibinfo {author} {\bibfnamefont {I.}~\bibnamefont
  {Raychowdhury}},\ }\emph {\bibinfo {title} {Prepotential formulation of
  lattice gauge theories}},\ \href@noop {} {Ph.D. thesis},\ \bibinfo  {school}
  {University of Calcutta} (\bibinfo {year} {2013})\BibitemShut {NoStop}%
\bibitem [{\citenamefont {Alexandru}\ \emph {et~al.}(2019)\citenamefont
  {Alexandru}, \citenamefont {Bedaque}, \citenamefont {Harmalkar},
  \citenamefont {Lamm}, \citenamefont {Lawrence},\ and\ \citenamefont
  {and}}]{Alexandru_2019}%
  \BibitemOpen
  \bibfield  {author} {\bibinfo {author} {\bibfnamefont {A.}~\bibnamefont
  {Alexandru}}, \bibinfo {author} {\bibfnamefont {P.~F.}\ \bibnamefont
  {Bedaque}}, \bibinfo {author} {\bibfnamefont {S.}~\bibnamefont {Harmalkar}},
  \bibinfo {author} {\bibfnamefont {H.}~\bibnamefont {Lamm}}, \bibinfo {author}
  {\bibfnamefont {S.}~\bibnamefont {Lawrence}}, \ and\ \bibinfo {author}
  {\bibfnamefont {N.~C.~W.}\ \bibnamefont {and}},\ }\href {\doibase
  10.1103/physrevd.100.114501} {\bibfield  {journal} {\bibinfo  {journal}
  {Physical Review D}\ }\textbf {\bibinfo {volume} {100}} (\bibinfo {year}
  {2019}),\ 10.1103/physrevd.100.114501}\BibitemShut {NoStop}%
\bibitem [{\citenamefont {Zohar}\ and\ \citenamefont
  {Burrello}(2015)}]{PhysRevD.91.054506}%
  \BibitemOpen
  \bibfield  {author} {\bibinfo {author} {\bibfnamefont {E.}~\bibnamefont
  {Zohar}}\ and\ \bibinfo {author} {\bibfnamefont {M.}~\bibnamefont
  {Burrello}},\ }\href {\doibase 10.1103/PhysRevD.91.054506} {\bibfield
  {journal} {\bibinfo  {journal} {Phys. Rev. D}\ }\textbf {\bibinfo {volume}
  {91}},\ \bibinfo {pages} {054506} (\bibinfo {year} {2015})}\BibitemShut
  {NoStop}%
\bibitem [{\citenamefont {Zohar}\ \emph {et~al.}(2017)\citenamefont {Zohar},
  \citenamefont {Farace}, \citenamefont {Reznik},\ and\ \citenamefont
  {Cirac}}]{PhysRevA.95.023604}%
  \BibitemOpen
  \bibfield  {author} {\bibinfo {author} {\bibfnamefont {E.}~\bibnamefont
  {Zohar}}, \bibinfo {author} {\bibfnamefont {A.}~\bibnamefont {Farace}},
  \bibinfo {author} {\bibfnamefont {B.}~\bibnamefont {Reznik}}, \ and\ \bibinfo
  {author} {\bibfnamefont {J.~I.}\ \bibnamefont {Cirac}},\ }\href {\doibase
  10.1103/PhysRevA.95.023604} {\bibfield  {journal} {\bibinfo  {journal} {Phys.
  Rev. A}\ }\textbf {\bibinfo {volume} {95}},\ \bibinfo {pages} {023604}
  (\bibinfo {year} {2017})}\BibitemShut {NoStop}%
\bibitem [{\citenamefont {Ji}\ \emph {et~al.}(2020)\citenamefont {Ji},
  \citenamefont {Lamm},\ and\ \citenamefont {Zhu}}]{PhysRevD.102.114513}%
  \BibitemOpen
  \bibfield  {author} {\bibinfo {author} {\bibfnamefont {Y.}~\bibnamefont
  {Ji}}, \bibinfo {author} {\bibfnamefont {H.}~\bibnamefont {Lamm}}, \ and\
  \bibinfo {author} {\bibfnamefont {S.}~\bibnamefont {Zhu}} (\bibinfo
  {collaboration} {NuQS Collaboration}),\ }\href {\doibase
  10.1103/PhysRevD.102.114513} {\bibfield  {journal} {\bibinfo  {journal}
  {Phys. Rev. D}\ }\textbf {\bibinfo {volume} {102}},\ \bibinfo {pages}
  {114513} (\bibinfo {year} {2020})}\BibitemShut {NoStop}%
\bibitem [{\citenamefont {Alexandru}\ \emph {et~al.}(2022)\citenamefont
  {Alexandru}, \citenamefont {Bedaque}, \citenamefont {Brett},\ and\
  \citenamefont {Lamm}}]{Alexandru:2021jpm}%
  \BibitemOpen
  \bibfield  {author} {\bibinfo {author} {\bibfnamefont {A.}~\bibnamefont
  {Alexandru}}, \bibinfo {author} {\bibfnamefont {P.~F.}\ \bibnamefont
  {Bedaque}}, \bibinfo {author} {\bibfnamefont {R.}~\bibnamefont {Brett}}, \
  and\ \bibinfo {author} {\bibfnamefont {H.}~\bibnamefont {Lamm}},\ }\href
  {\doibase 10.1103/PhysRevD.105.114508} {\bibfield  {journal} {\bibinfo
  {journal} {Phys. Rev. D}\ }\textbf {\bibinfo {volume} {105}},\ \bibinfo
  {pages} {114508} (\bibinfo {year} {2022})},\ \Eprint
  {http://arxiv.org/abs/2112.08482} {arXiv:2112.08482 [hep-lat]} \BibitemShut
  {NoStop}%
\bibitem [{\citenamefont {Gustafson}\ \emph {et~al.}(2022)\citenamefont
  {Gustafson}, \citenamefont {Lamm}, \citenamefont {Lovelace},\ and\
  \citenamefont {Musk}}]{Gustafson:2022xdt}%
  \BibitemOpen
  \bibfield  {author} {\bibinfo {author} {\bibfnamefont {E.~J.}\ \bibnamefont
  {Gustafson}}, \bibinfo {author} {\bibfnamefont {H.}~\bibnamefont {Lamm}},
  \bibinfo {author} {\bibfnamefont {F.}~\bibnamefont {Lovelace}}, \ and\
  \bibinfo {author} {\bibfnamefont {D.}~\bibnamefont {Musk}},\ }\href@noop {}
  {\  (\bibinfo {year} {2022})},\ \Eprint {http://arxiv.org/abs/2208.12309}
  {arXiv:2208.12309 [quant-ph]} \BibitemShut {NoStop}%
\bibitem [{\citenamefont {Hackett}\ \emph {et~al.}(2019)\citenamefont
  {Hackett}, \citenamefont {Howe}, \citenamefont {Hughes}, \citenamefont {Jay},
  \citenamefont {Neil},\ and\ \citenamefont {Simone}}]{PhysRevA.99.062341}%
  \BibitemOpen
  \bibfield  {author} {\bibinfo {author} {\bibfnamefont {D.~C.}\ \bibnamefont
  {Hackett}}, \bibinfo {author} {\bibfnamefont {K.}~\bibnamefont {Howe}},
  \bibinfo {author} {\bibfnamefont {C.}~\bibnamefont {Hughes}}, \bibinfo
  {author} {\bibfnamefont {W.}~\bibnamefont {Jay}}, \bibinfo {author}
  {\bibfnamefont {E.~T.}\ \bibnamefont {Neil}}, \ and\ \bibinfo {author}
  {\bibfnamefont {J.~N.}\ \bibnamefont {Simone}},\ }\href {\doibase
  10.1103/PhysRevA.99.062341} {\bibfield  {journal} {\bibinfo  {journal} {Phys.
  Rev. A}\ }\textbf {\bibinfo {volume} {99}},\ \bibinfo {pages} {062341}
  (\bibinfo {year} {2019})}\BibitemShut {NoStop}%
\bibitem [{\citenamefont {Kreshchuk}\ \emph {et~al.}(2022)\citenamefont
  {Kreshchuk}, \citenamefont {Kirby}, \citenamefont {Goldstein}, \citenamefont
  {Beauchemin},\ and\ \citenamefont {Love}}]{Kreshchuk_2022}%
  \BibitemOpen
  \bibfield  {author} {\bibinfo {author} {\bibfnamefont {M.}~\bibnamefont
  {Kreshchuk}}, \bibinfo {author} {\bibfnamefont {W.~M.}\ \bibnamefont
  {Kirby}}, \bibinfo {author} {\bibfnamefont {G.}~\bibnamefont {Goldstein}},
  \bibinfo {author} {\bibfnamefont {H.}~\bibnamefont {Beauchemin}}, \ and\
  \bibinfo {author} {\bibfnamefont {P.~J.}\ \bibnamefont {Love}},\ }\href
  {\doibase 10.1103/physreva.105.032418} {\bibfield  {journal} {\bibinfo
  {journal} {Physical Review A}\ }\textbf {\bibinfo {volume} {105}} (\bibinfo
  {year} {2022}),\ 10.1103/physreva.105.032418}\BibitemShut {NoStop}%
\bibitem [{\citenamefont {Kreshchuk}\ \emph {et~al.}(2021)\citenamefont
  {Kreshchuk}, \citenamefont {Jia}, \citenamefont {Kirby}, \citenamefont
  {Goldstein}, \citenamefont {Vary},\ and\ \citenamefont {Love}}]{e23050597}%
  \BibitemOpen
  \bibfield  {author} {\bibinfo {author} {\bibfnamefont {M.}~\bibnamefont
  {Kreshchuk}}, \bibinfo {author} {\bibfnamefont {S.}~\bibnamefont {Jia}},
  \bibinfo {author} {\bibfnamefont {W.~M.}\ \bibnamefont {Kirby}}, \bibinfo
  {author} {\bibfnamefont {G.}~\bibnamefont {Goldstein}}, \bibinfo {author}
  {\bibfnamefont {J.~P.}\ \bibnamefont {Vary}}, \ and\ \bibinfo {author}
  {\bibfnamefont {P.~J.}\ \bibnamefont {Love}},\ }\href {\doibase
  10.3390/e23050597} {\bibfield  {journal} {\bibinfo  {journal} {Entropy}\
  }\textbf {\bibinfo {volume} {23}} (\bibinfo {year} {2021}),\
  10.3390/e23050597}\BibitemShut {NoStop}%
\bibitem [{\citenamefont {Kaplan}\ \emph {et~al.}(2003)\citenamefont {Kaplan},
  \citenamefont {Katz},\ and\ \citenamefont {Ünsal}}]{Kaplan_2003}%
  \BibitemOpen
  \bibfield  {author} {\bibinfo {author} {\bibfnamefont {D.~B.}\ \bibnamefont
  {Kaplan}}, \bibinfo {author} {\bibfnamefont {E.}~\bibnamefont {Katz}}, \ and\
  \bibinfo {author} {\bibfnamefont {M.}~\bibnamefont {Ünsal}},\ }\href
  {\doibase 10.1088/1126-6708/2003/05/037} {\bibfield  {journal} {\bibinfo
  {journal} {Journal of High Energy Physics}\ }\textbf {\bibinfo {volume}
  {2003}},\ \bibinfo {pages} {037} (\bibinfo {year} {2003})}\BibitemShut
  {NoStop}%
\bibitem [{\citenamefont {Buser}\ \emph {et~al.}(2021)\citenamefont {Buser},
  \citenamefont {Gharibyan}, \citenamefont {Hanada}, \citenamefont {Honda},\
  and\ \citenamefont {Liu}}]{Buser_2021}%
  \BibitemOpen
  \bibfield  {author} {\bibinfo {author} {\bibfnamefont {A.~J.}\ \bibnamefont
  {Buser}}, \bibinfo {author} {\bibfnamefont {H.}~\bibnamefont {Gharibyan}},
  \bibinfo {author} {\bibfnamefont {M.}~\bibnamefont {Hanada}}, \bibinfo
  {author} {\bibfnamefont {M.}~\bibnamefont {Honda}}, \ and\ \bibinfo {author}
  {\bibfnamefont {J.}~\bibnamefont {Liu}},\ }\href {\doibase
  10.1007/jhep09(2021)034} {\bibfield  {journal} {\bibinfo  {journal} {Journal
  of High Energy Physics}\ }\textbf {\bibinfo {volume} {2021}} (\bibinfo {year}
  {2021}),\ 10.1007/jhep09(2021)034}\BibitemShut {NoStop}%
\bibitem [{\citenamefont {Bender}\ and\ \citenamefont
  {Zohar}(2020)}]{Bender_2020}%
  \BibitemOpen
  \bibfield  {author} {\bibinfo {author} {\bibfnamefont {J.}~\bibnamefont
  {Bender}}\ and\ \bibinfo {author} {\bibfnamefont {E.}~\bibnamefont {Zohar}},\
  }\href {\doibase 10.1103/physrevd.102.114517} {\bibfield  {journal} {\bibinfo
   {journal} {Physical Review D}\ }\textbf {\bibinfo {volume} {102}} (\bibinfo
  {year} {2020}),\ 10.1103/physrevd.102.114517}\BibitemShut {NoStop}%
\bibitem [{\citenamefont {Unmuth-Yockey}(2019)}]{Unmuth-Yockey:2018xak}%
  \BibitemOpen
  \bibfield  {author} {\bibinfo {author} {\bibfnamefont {J.~F.}\ \bibnamefont
  {Unmuth-Yockey}},\ }\href {\doibase 10.1103/PhysRevD.99.074502} {\bibfield
  {journal} {\bibinfo  {journal} {Phys. Rev. D}\ }\textbf {\bibinfo {volume}
  {99}},\ \bibinfo {pages} {074502} (\bibinfo {year} {2019})},\ \Eprint
  {http://arxiv.org/abs/1811.05884} {arXiv:1811.05884 [hep-lat]} \BibitemShut
  {NoStop}%
\bibitem [{\citenamefont {Muschik}\ \emph {et~al.}(2017)\citenamefont
  {Muschik}, \citenamefont {Heyl}, \citenamefont {Martinez}, \citenamefont
  {Monz}, \citenamefont {Schindler}, \citenamefont {Vogell}, \citenamefont
  {Dalmonte}, \citenamefont {Hauke}, \citenamefont {Blatt},\ and\ \citenamefont
  {Zoller}}]{Muschik_2017}%
  \BibitemOpen
  \bibfield  {author} {\bibinfo {author} {\bibfnamefont {C.}~\bibnamefont
  {Muschik}}, \bibinfo {author} {\bibfnamefont {M.}~\bibnamefont {Heyl}},
  \bibinfo {author} {\bibfnamefont {E.}~\bibnamefont {Martinez}}, \bibinfo
  {author} {\bibfnamefont {T.}~\bibnamefont {Monz}}, \bibinfo {author}
  {\bibfnamefont {P.}~\bibnamefont {Schindler}}, \bibinfo {author}
  {\bibfnamefont {B.}~\bibnamefont {Vogell}}, \bibinfo {author} {\bibfnamefont
  {M.}~\bibnamefont {Dalmonte}}, \bibinfo {author} {\bibfnamefont
  {P.}~\bibnamefont {Hauke}}, \bibinfo {author} {\bibfnamefont
  {R.}~\bibnamefont {Blatt}}, \ and\ \bibinfo {author} {\bibfnamefont
  {P.}~\bibnamefont {Zoller}},\ }\href {\doibase 10.1088/1367-2630/aa89ab}
  {\bibfield  {journal} {\bibinfo  {journal} {New Journal of Physics}\ }\textbf
  {\bibinfo {volume} {19}},\ \bibinfo {pages} {103020} (\bibinfo {year}
  {2017})}\BibitemShut {NoStop}%
\bibitem [{\citenamefont {Schweizer}\ \emph {et~al.}(2019)\citenamefont
  {Schweizer}, \citenamefont {Grusdt}, \citenamefont {Berngruber},
  \citenamefont {Barbiero}, \citenamefont {Demler}, \citenamefont {Goldman},
  \citenamefont {Bloch},\ and\ \citenamefont {Aidelsburger}}]{Schweizer_2019}%
  \BibitemOpen
  \bibfield  {author} {\bibinfo {author} {\bibfnamefont {C.}~\bibnamefont
  {Schweizer}}, \bibinfo {author} {\bibfnamefont {F.}~\bibnamefont {Grusdt}},
  \bibinfo {author} {\bibfnamefont {M.}~\bibnamefont {Berngruber}}, \bibinfo
  {author} {\bibfnamefont {L.}~\bibnamefont {Barbiero}}, \bibinfo {author}
  {\bibfnamefont {E.}~\bibnamefont {Demler}}, \bibinfo {author} {\bibfnamefont
  {N.}~\bibnamefont {Goldman}}, \bibinfo {author} {\bibfnamefont
  {I.}~\bibnamefont {Bloch}}, \ and\ \bibinfo {author} {\bibfnamefont
  {M.}~\bibnamefont {Aidelsburger}},\ }\href {\doibase
  10.1038/s41567-019-0649-7} {\bibfield  {journal} {\bibinfo  {journal} {Nature
  Physics}\ }\textbf {\bibinfo {volume} {15}},\ \bibinfo {pages} {1168}
  (\bibinfo {year} {2019})}\BibitemShut {NoStop}%
\bibitem [{\citenamefont {Kokail}\ \emph {et~al.}(2019)\citenamefont {Kokail},
  \citenamefont {Maier}, \citenamefont {van Bijnen}, \citenamefont {Brydges},
  \citenamefont {Joshi}, \citenamefont {Jurcevic}, \citenamefont {Muschik},
  \citenamefont {Silvi}, \citenamefont {Blatt}, \citenamefont {Roos},\ and\
  \citenamefont {Zoller}}]{Kokail_2019}%
  \BibitemOpen
  \bibfield  {author} {\bibinfo {author} {\bibfnamefont {C.}~\bibnamefont
  {Kokail}}, \bibinfo {author} {\bibfnamefont {C.}~\bibnamefont {Maier}},
  \bibinfo {author} {\bibfnamefont {R.}~\bibnamefont {van Bijnen}}, \bibinfo
  {author} {\bibfnamefont {T.}~\bibnamefont {Brydges}}, \bibinfo {author}
  {\bibfnamefont {M.~K.}\ \bibnamefont {Joshi}}, \bibinfo {author}
  {\bibfnamefont {P.}~\bibnamefont {Jurcevic}}, \bibinfo {author}
  {\bibfnamefont {C.~A.}\ \bibnamefont {Muschik}}, \bibinfo {author}
  {\bibfnamefont {P.}~\bibnamefont {Silvi}}, \bibinfo {author} {\bibfnamefont
  {R.}~\bibnamefont {Blatt}}, \bibinfo {author} {\bibfnamefont {C.~F.}\
  \bibnamefont {Roos}}, \ and\ \bibinfo {author} {\bibfnamefont
  {P.}~\bibnamefont {Zoller}},\ }\href {\doibase 10.1038/s41586-019-1177-4}
  {\bibfield  {journal} {\bibinfo  {journal} {Nature}\ }\textbf {\bibinfo
  {volume} {569}},\ \bibinfo {pages} {355} (\bibinfo {year}
  {2019})}\BibitemShut {NoStop}%
\bibitem [{\citenamefont {Yang}\ \emph {et~al.}(2020)\citenamefont {Yang},
  \citenamefont {Sun}, \citenamefont {Ott}, \citenamefont {Wang}, \citenamefont
  {Zache}, \citenamefont {Halimeh}, \citenamefont {Yuan}, \citenamefont
  {Hauke},\ and\ \citenamefont {Pan}}]{Yang_2020}%
  \BibitemOpen
  \bibfield  {author} {\bibinfo {author} {\bibfnamefont {B.}~\bibnamefont
  {Yang}}, \bibinfo {author} {\bibfnamefont {H.}~\bibnamefont {Sun}}, \bibinfo
  {author} {\bibfnamefont {R.}~\bibnamefont {Ott}}, \bibinfo {author}
  {\bibfnamefont {H.-Y.}\ \bibnamefont {Wang}}, \bibinfo {author}
  {\bibfnamefont {T.~V.}\ \bibnamefont {Zache}}, \bibinfo {author}
  {\bibfnamefont {J.~C.}\ \bibnamefont {Halimeh}}, \bibinfo {author}
  {\bibfnamefont {Z.-S.}\ \bibnamefont {Yuan}}, \bibinfo {author}
  {\bibfnamefont {P.}~\bibnamefont {Hauke}}, \ and\ \bibinfo {author}
  {\bibfnamefont {J.-W.}\ \bibnamefont {Pan}},\ }\href {\doibase
  10.1038/s41586-020-2910-8} {\bibfield  {journal} {\bibinfo  {journal}
  {Nature}\ }\textbf {\bibinfo {volume} {587}},\ \bibinfo {pages} {392}
  (\bibinfo {year} {2020})}\BibitemShut {NoStop}%
\bibitem [{\citenamefont {Mil}\ \emph {et~al.}(2020)\citenamefont {Mil},
  \citenamefont {Zache}, \citenamefont {Hegde}, \citenamefont {Xia},
  \citenamefont {Bhatt}, \citenamefont {Oberthaler}, \citenamefont {Hauke},
  \citenamefont {Berges},\ and\ \citenamefont
  {Jendrzejewski}}]{doi:10.1126/science.aaz5312}%
  \BibitemOpen
  \bibfield  {author} {\bibinfo {author} {\bibfnamefont {A.}~\bibnamefont
  {Mil}}, \bibinfo {author} {\bibfnamefont {T.~V.}\ \bibnamefont {Zache}},
  \bibinfo {author} {\bibfnamefont {A.}~\bibnamefont {Hegde}}, \bibinfo
  {author} {\bibfnamefont {A.}~\bibnamefont {Xia}}, \bibinfo {author}
  {\bibfnamefont {R.~P.}\ \bibnamefont {Bhatt}}, \bibinfo {author}
  {\bibfnamefont {M.~K.}\ \bibnamefont {Oberthaler}}, \bibinfo {author}
  {\bibfnamefont {P.}~\bibnamefont {Hauke}}, \bibinfo {author} {\bibfnamefont
  {J.}~\bibnamefont {Berges}}, \ and\ \bibinfo {author} {\bibfnamefont
  {F.}~\bibnamefont {Jendrzejewski}},\ }\href {\doibase
  10.1126/science.aaz5312} {\bibfield  {journal} {\bibinfo  {journal}
  {Science}\ }\textbf {\bibinfo {volume} {367}},\ \bibinfo {pages} {1128}
  (\bibinfo {year} {2020})},\ \Eprint
  {http://arxiv.org/abs/https://www.science.org/doi/pdf/10.1126/science.aaz5312}
  {https://www.science.org/doi/pdf/10.1126/science.aaz5312} \BibitemShut
  {NoStop}%
\bibitem [{\citenamefont {Riechert}\ \emph {et~al.}(2022)\citenamefont
  {Riechert}, \citenamefont {Halimeh}, \citenamefont {Kasper}, \citenamefont
  {Bretheau}, \citenamefont {Zohar}, \citenamefont {Hauke},\ and\ \citenamefont
  {Jendrzejewski}}]{Riechert_2022}%
  \BibitemOpen
  \bibfield  {author} {\bibinfo {author} {\bibfnamefont {H.}~\bibnamefont
  {Riechert}}, \bibinfo {author} {\bibfnamefont {J.~C.}\ \bibnamefont
  {Halimeh}}, \bibinfo {author} {\bibfnamefont {V.}~\bibnamefont {Kasper}},
  \bibinfo {author} {\bibfnamefont {L.}~\bibnamefont {Bretheau}}, \bibinfo
  {author} {\bibfnamefont {E.}~\bibnamefont {Zohar}}, \bibinfo {author}
  {\bibfnamefont {P.}~\bibnamefont {Hauke}}, \ and\ \bibinfo {author}
  {\bibfnamefont {F.}~\bibnamefont {Jendrzejewski}},\ }\href {\doibase
  10.1103/physrevb.105.205141} {\bibfield  {journal} {\bibinfo  {journal}
  {Physical Review B}\ }\textbf {\bibinfo {volume} {105}} (\bibinfo {year}
  {2022}),\ 10.1103/physrevb.105.205141}\BibitemShut {NoStop}%
\bibitem [{\citenamefont {Childs}\ \emph {et~al.}(2021)\citenamefont {Childs},
  \citenamefont {Su}, \citenamefont {Tran}, \citenamefont {Wiebe},\ and\
  \citenamefont {Zhu}}]{Childs_2021}%
  \BibitemOpen
  \bibfield  {author} {\bibinfo {author} {\bibfnamefont {A.~M.}\ \bibnamefont
  {Childs}}, \bibinfo {author} {\bibfnamefont {Y.}~\bibnamefont {Su}}, \bibinfo
  {author} {\bibfnamefont {M.~C.}\ \bibnamefont {Tran}}, \bibinfo {author}
  {\bibfnamefont {N.}~\bibnamefont {Wiebe}}, \ and\ \bibinfo {author}
  {\bibfnamefont {S.}~\bibnamefont {Zhu}},\ }\href {\doibase
  10.1103/physrevx.11.011020} {\bibfield  {journal} {\bibinfo  {journal}
  {Physical Review X}\ }\textbf {\bibinfo {volume} {11}} (\bibinfo {year}
  {2021}),\ 10.1103/physrevx.11.011020}\BibitemShut {NoStop}%
\bibitem [{\citenamefont {Campbell}(2019)}]{Campbell_2019}%
  \BibitemOpen
  \bibfield  {author} {\bibinfo {author} {\bibfnamefont {E.}~\bibnamefont
  {Campbell}},\ }\href {\doibase 10.1103/physrevlett.123.070503} {\bibfield
  {journal} {\bibinfo  {journal} {Physical Review Letters}\ }\textbf {\bibinfo
  {volume} {123}} (\bibinfo {year} {2019}),\
  10.1103/physrevlett.123.070503}\BibitemShut {NoStop}%
\bibitem [{\citenamefont {Low}\ and\ \citenamefont {Chuang}(2017)}]{Low_2017}%
  \BibitemOpen
  \bibfield  {author} {\bibinfo {author} {\bibfnamefont {G.~H.}\ \bibnamefont
  {Low}}\ and\ \bibinfo {author} {\bibfnamefont {I.~L.}\ \bibnamefont
  {Chuang}},\ }\href {\doibase 10.1103/physrevlett.118.010501} {\bibfield
  {journal} {\bibinfo  {journal} {Physical Review Letters}\ }\textbf {\bibinfo
  {volume} {118}} (\bibinfo {year} {2017}),\
  10.1103/physrevlett.118.010501}\BibitemShut {NoStop}%
\bibitem [{\citenamefont {Low}\ and\ \citenamefont {Chuang}(2019)}]{Low_2019}%
  \BibitemOpen
  \bibfield  {author} {\bibinfo {author} {\bibfnamefont {G.~H.}\ \bibnamefont
  {Low}}\ and\ \bibinfo {author} {\bibfnamefont {I.~L.}\ \bibnamefont
  {Chuang}},\ }\href {\doibase 10.22331/q-2019-07-12-163} {\bibfield  {journal}
  {\bibinfo  {journal} {Quantum}\ }\textbf {\bibinfo {volume} {3}},\ \bibinfo
  {pages} {163} (\bibinfo {year} {2019})}\BibitemShut {NoStop}%
\bibitem [{\citenamefont {Berry}\ \emph {et~al.}(2015)\citenamefont {Berry},
  \citenamefont {Childs}, \citenamefont {Cleve}, \citenamefont {Kothari},\ and\
  \citenamefont {Somma}}]{Berry_2015}%
  \BibitemOpen
  \bibfield  {author} {\bibinfo {author} {\bibfnamefont {D.~W.}\ \bibnamefont
  {Berry}}, \bibinfo {author} {\bibfnamefont {A.~M.}\ \bibnamefont {Childs}},
  \bibinfo {author} {\bibfnamefont {R.}~\bibnamefont {Cleve}}, \bibinfo
  {author} {\bibfnamefont {R.}~\bibnamefont {Kothari}}, \ and\ \bibinfo
  {author} {\bibfnamefont {R.~D.}\ \bibnamefont {Somma}},\ }\href {\doibase
  10.1103/physrevlett.114.090502} {\bibfield  {journal} {\bibinfo  {journal}
  {Physical Review Letters}\ }\textbf {\bibinfo {volume} {114}} (\bibinfo
  {year} {2015}),\ 10.1103/physrevlett.114.090502}\BibitemShut {NoStop}%
\bibitem [{\citenamefont {Berry}\ and\ \citenamefont
  {Childs}(2012)}]{10.5555/2231036.2231040}%
  \BibitemOpen
  \bibfield  {author} {\bibinfo {author} {\bibfnamefont {D.~W.}\ \bibnamefont
  {Berry}}\ and\ \bibinfo {author} {\bibfnamefont {A.~M.}\ \bibnamefont
  {Childs}},\ }\href@noop {} {\bibfield  {journal} {\bibinfo  {journal}
  {Quantum Info. Comput.}\ }\textbf {\bibinfo {volume} {12}},\ \bibinfo {pages}
  {29–62} (\bibinfo {year} {2012})}\BibitemShut {NoStop}%
\bibitem [{\citenamefont {Yao}\ \emph {et~al.}(2021)\citenamefont {Yao},
  \citenamefont {Gomes}, \citenamefont {Zhang}, \citenamefont {Wang},
  \citenamefont {Ho}, \citenamefont {Iadecola},\ and\ \citenamefont
  {Orth}}]{Yao_2021}%
  \BibitemOpen
  \bibfield  {author} {\bibinfo {author} {\bibfnamefont {Y.-X.}\ \bibnamefont
  {Yao}}, \bibinfo {author} {\bibfnamefont {N.}~\bibnamefont {Gomes}}, \bibinfo
  {author} {\bibfnamefont {F.}~\bibnamefont {Zhang}}, \bibinfo {author}
  {\bibfnamefont {C.-Z.}\ \bibnamefont {Wang}}, \bibinfo {author}
  {\bibfnamefont {K.-M.}\ \bibnamefont {Ho}}, \bibinfo {author} {\bibfnamefont
  {T.}~\bibnamefont {Iadecola}}, \ and\ \bibinfo {author} {\bibfnamefont
  {P.~P.}\ \bibnamefont {Orth}},\ }\href {\doibase 10.1103/prxquantum.2.030307}
  {\bibfield  {journal} {\bibinfo  {journal} {{PRX} Quantum}\ }\textbf
  {\bibinfo {volume} {2}} (\bibinfo {year} {2021}),\
  10.1103/prxquantum.2.030307}\BibitemShut {NoStop}%
\bibitem [{\citenamefont {Stetina}\ \emph {et~al.}(2022)\citenamefont
  {Stetina}, \citenamefont {Ciavarella}, \citenamefont {Li},\ and\
  \citenamefont {Wiebe}}]{Stetina:2020abi}%
  \BibitemOpen
  \bibfield  {author} {\bibinfo {author} {\bibfnamefont {T.~F.}\ \bibnamefont
  {Stetina}}, \bibinfo {author} {\bibfnamefont {A.}~\bibnamefont {Ciavarella}},
  \bibinfo {author} {\bibfnamefont {X.}~\bibnamefont {Li}}, \ and\ \bibinfo
  {author} {\bibfnamefont {N.}~\bibnamefont {Wiebe}},\ }\href {\doibase
  10.22331/q-2022-01-18-622} {\bibfield  {journal} {\bibinfo  {journal}
  {Quantum}\ }\textbf {\bibinfo {volume} {6}},\ \bibinfo {pages} {622}
  (\bibinfo {year} {2022})},\ \Eprint {http://arxiv.org/abs/2101.00111}
  {arXiv:2101.00111 [quant-ph]} \BibitemShut {NoStop}%
\bibitem [{\citenamefont {Nguyen}\ \emph {et~al.}(2022)\citenamefont {Nguyen},
  \citenamefont {Tran}, \citenamefont {Zhu}, \citenamefont {Green},
  \citenamefont {Alderete}, \citenamefont {Davoudi},\ and\ \citenamefont
  {Linke}}]{Nguyen:2021hyk}%
  \BibitemOpen
  \bibfield  {author} {\bibinfo {author} {\bibfnamefont {N.~H.}\ \bibnamefont
  {Nguyen}}, \bibinfo {author} {\bibfnamefont {M.~C.}\ \bibnamefont {Tran}},
  \bibinfo {author} {\bibfnamefont {Y.}~\bibnamefont {Zhu}}, \bibinfo {author}
  {\bibfnamefont {A.~M.}\ \bibnamefont {Green}}, \bibinfo {author}
  {\bibfnamefont {C.~H.}\ \bibnamefont {Alderete}}, \bibinfo {author}
  {\bibfnamefont {Z.}~\bibnamefont {Davoudi}}, \ and\ \bibinfo {author}
  {\bibfnamefont {N.~M.}\ \bibnamefont {Linke}},\ }\href {\doibase
  10.1103/PRXQuantum.3.020324} {\bibfield  {journal} {\bibinfo  {journal} {PRX
  Quantum}\ }\textbf {\bibinfo {volume} {3}},\ \bibinfo {pages} {020324}
  (\bibinfo {year} {2022})},\ \Eprint {http://arxiv.org/abs/2112.14262}
  {arXiv:2112.14262 [quant-ph]} \BibitemShut {NoStop}%
\bibitem [{\citenamefont {Shaw}\ \emph {et~al.}(2020)\citenamefont {Shaw},
  \citenamefont {Lougovski}, \citenamefont {Stryker},\ and\ \citenamefont
  {Wiebe}}]{Shaw_2020}%
  \BibitemOpen
  \bibfield  {author} {\bibinfo {author} {\bibfnamefont {A.~F.}\ \bibnamefont
  {Shaw}}, \bibinfo {author} {\bibfnamefont {P.}~\bibnamefont {Lougovski}},
  \bibinfo {author} {\bibfnamefont {J.~R.}\ \bibnamefont {Stryker}}, \ and\
  \bibinfo {author} {\bibfnamefont {N.}~\bibnamefont {Wiebe}},\ }\href
  {\doibase 10.22331/q-2020-08-10-306} {\bibfield  {journal} {\bibinfo
  {journal} {Quantum}\ }\textbf {\bibinfo {volume} {4}},\ \bibinfo {pages}
  {306} (\bibinfo {year} {2020})}\BibitemShut {NoStop}%
\bibitem [{\citenamefont {Paulson}\ \emph {et~al.}(2021)\citenamefont {Paulson}
  \emph {et~al.}}]{Paulson:2020zjd}%
  \BibitemOpen
  \bibfield  {author} {\bibinfo {author} {\bibfnamefont {D.}~\bibnamefont
  {Paulson}} \emph {et~al.},\ }\href {\doibase 10.1103/PRXQuantum.2.030334}
  {\bibfield  {journal} {\bibinfo  {journal} {PRX Quantum}\ }\textbf {\bibinfo
  {volume} {2}},\ \bibinfo {pages} {030334} (\bibinfo {year} {2021})},\ \Eprint
  {http://arxiv.org/abs/2008.09252} {arXiv:2008.09252 [quant-ph]} \BibitemShut
  {NoStop}%
\bibitem [{\citenamefont {Farrell}\ \emph
  {et~al.}(2022{\natexlab{a}})\citenamefont {Farrell}, \citenamefont
  {Chernyshev}, \citenamefont {Powell}, \citenamefont {Zemlevskiy},
  \citenamefont {Illa},\ and\ \citenamefont
  {Savage}}]{https://doi.org/10.48550/arxiv.2207.01731}%
  \BibitemOpen
  \bibfield  {author} {\bibinfo {author} {\bibfnamefont {R.~C.}\ \bibnamefont
  {Farrell}}, \bibinfo {author} {\bibfnamefont {I.~A.}\ \bibnamefont
  {Chernyshev}}, \bibinfo {author} {\bibfnamefont {S.~J.~M.}\ \bibnamefont
  {Powell}}, \bibinfo {author} {\bibfnamefont {N.~A.}\ \bibnamefont
  {Zemlevskiy}}, \bibinfo {author} {\bibfnamefont {M.}~\bibnamefont {Illa}}, \
  and\ \bibinfo {author} {\bibfnamefont {M.~J.}\ \bibnamefont {Savage}},\
  }\href {\doibase 10.48550/ARXIV.2207.01731} {\enquote {\bibinfo {title}
  {Preparations for quantum simulations of quantum chromodynamics in 1+1
  dimensions: (i) axial gauge},}\ } (\bibinfo {year}
  {2022}{\natexlab{a}})\BibitemShut {NoStop}%
\bibitem [{\citenamefont {Farrell}\ \emph
  {et~al.}(2022{\natexlab{b}})\citenamefont {Farrell}, \citenamefont
  {Chernyshev}, \citenamefont {Powell}, \citenamefont {Zemlevskiy},
  \citenamefont {Illa},\ and\ \citenamefont {Savage}}]{Farrell:2022vyh}%
  \BibitemOpen
  \bibfield  {author} {\bibinfo {author} {\bibfnamefont {R.~C.}\ \bibnamefont
  {Farrell}}, \bibinfo {author} {\bibfnamefont {I.~A.}\ \bibnamefont
  {Chernyshev}}, \bibinfo {author} {\bibfnamefont {S.~J.~M.}\ \bibnamefont
  {Powell}}, \bibinfo {author} {\bibfnamefont {N.~A.}\ \bibnamefont
  {Zemlevskiy}}, \bibinfo {author} {\bibfnamefont {M.}~\bibnamefont {Illa}}, \
  and\ \bibinfo {author} {\bibfnamefont {M.~J.}\ \bibnamefont {Savage}},\
  }\href@noop {} {\  (\bibinfo {year} {2022}{\natexlab{b}})},\ \Eprint
  {http://arxiv.org/abs/2209.10781} {arXiv:2209.10781 [quant-ph]} \BibitemShut
  {NoStop}%
\bibitem [{\citenamefont {Murairi}\ \emph {et~al.}(2022)\citenamefont
  {Murairi}, \citenamefont {Cervia}, \citenamefont {Kumar}, \citenamefont
  {Bedaque},\ and\ \citenamefont {Alexandru}}]{Murairi:2022zdg}%
  \BibitemOpen
  \bibfield  {author} {\bibinfo {author} {\bibfnamefont {E.~M.}\ \bibnamefont
  {Murairi}}, \bibinfo {author} {\bibfnamefont {M.~J.}\ \bibnamefont {Cervia}},
  \bibinfo {author} {\bibfnamefont {H.}~\bibnamefont {Kumar}}, \bibinfo
  {author} {\bibfnamefont {P.~F.}\ \bibnamefont {Bedaque}}, \ and\ \bibinfo
  {author} {\bibfnamefont {A.}~\bibnamefont {Alexandru}},\ }\href@noop {} {\
  (\bibinfo {year} {2022})},\ \Eprint {http://arxiv.org/abs/2208.11789}
  {arXiv:2208.11789 [hep-lat]} \BibitemShut {NoStop}%
\bibitem [{\citenamefont {Kogut}\ and\ \citenamefont
  {Susskind}(1975)}]{PhysRevD.11.395}%
  \BibitemOpen
  \bibfield  {author} {\bibinfo {author} {\bibfnamefont {J.}~\bibnamefont
  {Kogut}}\ and\ \bibinfo {author} {\bibfnamefont {L.}~\bibnamefont
  {Susskind}},\ }\href {\doibase 10.1103/PhysRevD.11.395} {\bibfield  {journal}
  {\bibinfo  {journal} {Phys. Rev. D}\ }\textbf {\bibinfo {volume} {11}},\
  \bibinfo {pages} {395} (\bibinfo {year} {1975})}\BibitemShut {NoStop}%
\bibitem [{\citenamefont {Kan}\ and\ \citenamefont
  {Nam}(2021)}]{https://doi.org/10.48550/arxiv.2107.12769}%
  \BibitemOpen
  \bibfield  {author} {\bibinfo {author} {\bibfnamefont {A.}~\bibnamefont
  {Kan}}\ and\ \bibinfo {author} {\bibfnamefont {Y.}~\bibnamefont {Nam}},\
  }\href {\doibase 10.48550/ARXIV.2107.12769} {\enquote {\bibinfo {title}
  {Lattice quantum chromodynamics and electrodynamics on a universal quantum
  computer},}\ } (\bibinfo {year} {2021})\BibitemShut {NoStop}%
\bibitem [{\citenamefont {Preskill}(2018)}]{Preskill_2018}%
  \BibitemOpen
  \bibfield  {author} {\bibinfo {author} {\bibfnamefont {J.}~\bibnamefont
  {Preskill}},\ }\href {\doibase 10.22331/q-2018-08-06-79} {\bibfield
  {journal} {\bibinfo  {journal} {Quantum}\ }\textbf {\bibinfo {volume} {2}},\
  \bibinfo {pages} {79} (\bibinfo {year} {2018})}\BibitemShut {NoStop}%
\bibitem [{\citenamefont {Nielsen}\ and\ \citenamefont
  {Chuang}(2010)}]{nielsen_chuang_2010}%
  \BibitemOpen
  \bibfield  {author} {\bibinfo {author} {\bibfnamefont {M.~A.}\ \bibnamefont
  {Nielsen}}\ and\ \bibinfo {author} {\bibfnamefont {I.~L.}\ \bibnamefont
  {Chuang}},\ }\href {\doibase 10.1017/CBO9780511976667} {\emph {\bibinfo
  {title} {Quantum Computation and Quantum Information: 10th Anniversary
  Edition}}}\ (\bibinfo  {publisher} {Cambridge University Press},\ \bibinfo
  {year} {2010})\BibitemShut {NoStop}%
\bibitem [{\citenamefont {Welch}\ \emph {et~al.}(2014)\citenamefont {Welch},
  \citenamefont {Greenbaum}, \citenamefont {Mostame},\ and\ \citenamefont
  {Aspuru-Guzik}}]{Welch_2014}%
  \BibitemOpen
  \bibfield  {author} {\bibinfo {author} {\bibfnamefont {J.}~\bibnamefont
  {Welch}}, \bibinfo {author} {\bibfnamefont {D.}~\bibnamefont {Greenbaum}},
  \bibinfo {author} {\bibfnamefont {S.}~\bibnamefont {Mostame}}, \ and\
  \bibinfo {author} {\bibfnamefont {A.}~\bibnamefont {Aspuru-Guzik}},\ }\href
  {\doibase 10.1088/1367-2630/16/3/033040} {\bibfield  {journal} {\bibinfo
  {journal} {New Journal of Physics}\ }\textbf {\bibinfo {volume} {16}},\
  \bibinfo {pages} {033040} (\bibinfo {year} {2014})}\BibitemShut {NoStop}%
\bibitem [{\citenamefont {Bocharov}\ \emph {et~al.}(2015)\citenamefont
  {Bocharov}, \citenamefont {Roetteler},\ and\ \citenamefont
  {Svore}}]{Bocharov_2015}%
  \BibitemOpen
  \bibfield  {author} {\bibinfo {author} {\bibfnamefont {A.}~\bibnamefont
  {Bocharov}}, \bibinfo {author} {\bibfnamefont {M.}~\bibnamefont {Roetteler}},
  \ and\ \bibinfo {author} {\bibfnamefont {K.~M.}\ \bibnamefont {Svore}},\
  }\href {\doibase 10.1103/physrevlett.114.080502} {\bibfield  {journal}
  {\bibinfo  {journal} {Physical Review Letters}\ }\textbf {\bibinfo {volume}
  {114}} (\bibinfo {year} {2015}),\ 10.1103/physrevlett.114.080502}\BibitemShut
  {NoStop}%
\bibitem [{\citenamefont {Grabowska}\ \emph {et~al.}(2022)\citenamefont
  {Grabowska}, \citenamefont {Kane}, \citenamefont {Nachman},\ and\
  \citenamefont {Bauer}}]{https://doi.org/10.48550/arxiv.2208.03333}%
  \BibitemOpen
  \bibfield  {author} {\bibinfo {author} {\bibfnamefont {D.~M.}\ \bibnamefont
  {Grabowska}}, \bibinfo {author} {\bibfnamefont {C.}~\bibnamefont {Kane}},
  \bibinfo {author} {\bibfnamefont {B.}~\bibnamefont {Nachman}}, \ and\
  \bibinfo {author} {\bibfnamefont {C.~W.}\ \bibnamefont {Bauer}},\ }\href
  {\doibase 10.48550/ARXIV.2208.03333} {\enquote {\bibinfo {title} {Overcoming
  exponential scaling with system size in trotter-suzuki implementations of
  constrained hamiltonians: 2+1 u(1) lattice gauge theories},}\ } (\bibinfo
  {year} {2022})\BibitemShut {NoStop}%
\bibitem [{\citenamefont {Aoki}\ \emph {et~al.}(2012)\citenamefont {Aoki},
  \citenamefont {Christ}, \citenamefont {Flynn}, \citenamefont {Izubuchi},
  \citenamefont {Lehner}, \citenamefont {Li}, \citenamefont {Peng},
  \citenamefont {Soni}, \citenamefont {de~Water},\ and\ \citenamefont
  {and}}]{Aoki_2012}%
  \BibitemOpen
  \bibfield  {author} {\bibinfo {author} {\bibfnamefont {Y.}~\bibnamefont
  {Aoki}}, \bibinfo {author} {\bibfnamefont {N.~H.}\ \bibnamefont {Christ}},
  \bibinfo {author} {\bibfnamefont {J.~M.}\ \bibnamefont {Flynn}}, \bibinfo
  {author} {\bibfnamefont {T.}~\bibnamefont {Izubuchi}}, \bibinfo {author}
  {\bibfnamefont {C.}~\bibnamefont {Lehner}}, \bibinfo {author} {\bibfnamefont
  {M.}~\bibnamefont {Li}}, \bibinfo {author} {\bibfnamefont {H.}~\bibnamefont
  {Peng}}, \bibinfo {author} {\bibfnamefont {A.}~\bibnamefont {Soni}}, \bibinfo
  {author} {\bibfnamefont {R.~S.~V.}\ \bibnamefont {de~Water}}, \ and\ \bibinfo
  {author} {\bibfnamefont {O.~W.}\ \bibnamefont {and}},\ }\href {\doibase
  10.1103/physrevd.86.116003} {\bibfield  {journal} {\bibinfo  {journal}
  {Physical Review D}\ }\textbf {\bibinfo {volume} {86}} (\bibinfo {year}
  {2012}),\ 10.1103/physrevd.86.116003}\BibitemShut {NoStop}%
\bibitem [{\citenamefont {El-Khadra}\ \emph {et~al.}(1997)\citenamefont
  {El-Khadra}, \citenamefont {Kronfeld},\ and\ \citenamefont
  {Mackenzie}}]{El_Khadra_1997}%
  \BibitemOpen
  \bibfield  {author} {\bibinfo {author} {\bibfnamefont {A.~X.}\ \bibnamefont
  {El-Khadra}}, \bibinfo {author} {\bibfnamefont {A.~S.}\ \bibnamefont
  {Kronfeld}}, \ and\ \bibinfo {author} {\bibfnamefont {P.~B.}\ \bibnamefont
  {Mackenzie}},\ }\href {\doibase 10.1103/physrevd.55.3933} {\bibfield
  {journal} {\bibinfo  {journal} {Physical Review D}\ }\textbf {\bibinfo
  {volume} {55}},\ \bibinfo {pages} {3933} (\bibinfo {year}
  {1997})}\BibitemShut {NoStop}%
\bibitem [{\citenamefont {Aoki}\ \emph {et~al.}(2003)\citenamefont {Aoki},
  \citenamefont {Kuramashi},\ and\ \citenamefont {i.~Tominaga}}]{Aoki_2003}%
  \BibitemOpen
  \bibfield  {author} {\bibinfo {author} {\bibfnamefont {S.}~\bibnamefont
  {Aoki}}, \bibinfo {author} {\bibfnamefont {Y.}~\bibnamefont {Kuramashi}}, \
  and\ \bibinfo {author} {\bibfnamefont {S.}~\bibnamefont {i.~Tominaga}},\
  }\href {\doibase 10.1143/ptp.109.383} {\bibfield  {journal} {\bibinfo
  {journal} {Progress of Theoretical Physics}\ }\textbf {\bibinfo {volume}
  {109}},\ \bibinfo {pages} {383} (\bibinfo {year} {2003})}\BibitemShut
  {NoStop}%
\bibitem [{\citenamefont {Bernard}\ \emph {et~al.}(2011)\citenamefont {Bernard}
  \emph {et~al.}}]{FermilabLattice:2010rur}%
  \BibitemOpen
  \bibfield  {author} {\bibinfo {author} {\bibfnamefont {C.}~\bibnamefont
  {Bernard}} \emph {et~al.} (\bibinfo {collaboration} {Fermilab Lattice,
  MILC}),\ }\href {\doibase 10.1103/PhysRevD.83.034503} {\bibfield  {journal}
  {\bibinfo  {journal} {Phys. Rev. D}\ }\textbf {\bibinfo {volume} {83}},\
  \bibinfo {pages} {034503} (\bibinfo {year} {2011})},\ \Eprint
  {http://arxiv.org/abs/1003.1937} {arXiv:1003.1937 [hep-lat]} \BibitemShut
  {NoStop}%
\bibitem [{\citenamefont {Lin}\ and\ \citenamefont {Christ}(2007)}]{Lin_2007}%
  \BibitemOpen
  \bibfield  {author} {\bibinfo {author} {\bibfnamefont {H.-W.}\ \bibnamefont
  {Lin}}\ and\ \bibinfo {author} {\bibfnamefont {N.}~\bibnamefont {Christ}},\
  }\href {\doibase 10.1103/physrevd.76.074506} {\bibfield  {journal} {\bibinfo
  {journal} {Physical Review D}\ }\textbf {\bibinfo {volume} {76}} (\bibinfo
  {year} {2007}),\ 10.1103/physrevd.76.074506}\BibitemShut {NoStop}%
\bibitem [{\citenamefont {Aoki}\ \emph {et~al.}(2004)\citenamefont {Aoki},
  \citenamefont {Kayaba},\ and\ \citenamefont {Kuramashi}}]{Aoki_2004}%
  \BibitemOpen
  \bibfield  {author} {\bibinfo {author} {\bibfnamefont {S.}~\bibnamefont
  {Aoki}}, \bibinfo {author} {\bibfnamefont {Y.}~\bibnamefont {Kayaba}}, \ and\
  \bibinfo {author} {\bibfnamefont {Y.}~\bibnamefont {Kuramashi}},\ }\href
  {\doibase 10.1016/j.nuclphysb.2004.07.017} {\bibfield  {journal} {\bibinfo
  {journal} {Nuclear Physics B}\ }\textbf {\bibinfo {volume} {697}},\ \bibinfo
  {pages} {271} (\bibinfo {year} {2004})}\BibitemShut {NoStop}%
\bibitem [{\citenamefont {Bullock}\ and\ \citenamefont
  {Markov}(2003)}]{Bullock_2003}%
  \BibitemOpen
  \bibfield  {author} {\bibinfo {author} {\bibfnamefont {S.~S.}\ \bibnamefont
  {Bullock}}\ and\ \bibinfo {author} {\bibfnamefont {I.~L.}\ \bibnamefont
  {Markov}},\ }\href {\doibase 10.48550/ARXIV.QUANT-PH/0303039} {\  (\bibinfo
  {year} {2003}),\ 10.48550/ARXIV.QUANT-PH/0303039}\BibitemShut {NoStop}%
\bibitem [{\citenamefont {R.~K. Rao~Yarlagadda}(1997)}]{Yarlagadda_Hershey}%
  \BibitemOpen
  \bibfield  {author} {\bibinfo {author} {\bibfnamefont {J.~E.~H.}\
  \bibnamefont {R.~K. Rao~Yarlagadda}},\ }\href {\doibase
  https://doi.org/10.1007/978-1-4615-6313-6} {\emph {\bibinfo {title} {Hadamard
  Matrix Analysis and Synthesis}}}\ (\bibinfo  {publisher} {Springer},\
  \bibinfo {year} {1997})\BibitemShut {NoStop}%
\bibitem [{\citenamefont {Klco}\ and\ \citenamefont
  {Savage}(2019)}]{PhysRevA.99.052335}%
  \BibitemOpen
  \bibfield  {author} {\bibinfo {author} {\bibfnamefont {N.}~\bibnamefont
  {Klco}}\ and\ \bibinfo {author} {\bibfnamefont {M.~J.}\ \bibnamefont
  {Savage}},\ }\href {\doibase 10.1103/PhysRevA.99.052335} {\bibfield
  {journal} {\bibinfo  {journal} {Phys. Rev. A}\ }\textbf {\bibinfo {volume}
  {99}},\ \bibinfo {pages} {052335} (\bibinfo {year} {2019})}\BibitemShut
  {NoStop}%
\bibitem [{\citenamefont {Yuen}(1975)}]{CK_Yuen}%
  \BibitemOpen
  \bibfield  {author} {\bibinfo {author} {\bibfnamefont {C.-K.}\ \bibnamefont
  {Yuen}},\ }\href {\doibase 10.1109/T-C.1975.224271} {\bibfield  {journal}
  {\bibinfo  {journal} {IEEE Transactions on Computers}\ }\textbf {\bibinfo
  {volume} {C-24}},\ \bibinfo {pages} {590} (\bibinfo {year}
  {1975})}\BibitemShut {NoStop}%
\bibitem [{\citenamefont {Kushilevitz}\ and\ \citenamefont
  {Mansour}(1993)}]{decision_tree}%
  \BibitemOpen
  \bibfield  {author} {\bibinfo {author} {\bibfnamefont {E.}~\bibnamefont
  {Kushilevitz}}\ and\ \bibinfo {author} {\bibfnamefont {Y.}~\bibnamefont
  {Mansour}},\ }\href {\doibase 10.1137/0222080} {\bibfield  {journal}
  {\bibinfo  {journal} {SIAM Journal on Computing}\ }\textbf {\bibinfo {volume}
  {22}},\ \bibinfo {pages} {1331} (\bibinfo {year} {1993})},\ \Eprint
  {http://arxiv.org/abs/https://doi.org/10.1137/0222080}
  {https://doi.org/10.1137/0222080} \BibitemShut {NoStop}%
\bibitem [{\citenamefont {Carena}\ \emph {et~al.}(2021)\citenamefont {Carena},
  \citenamefont {Lamm}, \citenamefont {Li},\ and\ \citenamefont
  {Liu}}]{Carena:2021ltu}%
  \BibitemOpen
  \bibfield  {author} {\bibinfo {author} {\bibfnamefont {M.}~\bibnamefont
  {Carena}}, \bibinfo {author} {\bibfnamefont {H.}~\bibnamefont {Lamm}},
  \bibinfo {author} {\bibfnamefont {Y.-Y.}\ \bibnamefont {Li}}, \ and\ \bibinfo
  {author} {\bibfnamefont {W.}~\bibnamefont {Liu}},\ }\href {\doibase
  10.1103/PhysRevD.104.094519} {\bibfield  {journal} {\bibinfo  {journal}
  {Phys. Rev. D}\ }\textbf {\bibinfo {volume} {104}},\ \bibinfo {pages}
  {094519} (\bibinfo {year} {2021})},\ \Eprint
  {http://arxiv.org/abs/2107.01166} {arXiv:2107.01166 [hep-lat]} \BibitemShut
  {NoStop}%
\bibitem [{\citenamefont {Carena}\ \emph
  {et~al.}(2022{\natexlab{b}})\citenamefont {Carena}, \citenamefont
  {Gustafson}, \citenamefont {Lamm}, \citenamefont {Li},\ and\ \citenamefont
  {Liu}}]{Carena:2022hpz}%
  \BibitemOpen
  \bibfield  {author} {\bibinfo {author} {\bibfnamefont {M.}~\bibnamefont
  {Carena}}, \bibinfo {author} {\bibfnamefont {E.~J.}\ \bibnamefont
  {Gustafson}}, \bibinfo {author} {\bibfnamefont {H.}~\bibnamefont {Lamm}},
  \bibinfo {author} {\bibfnamefont {Y.-Y.}\ \bibnamefont {Li}}, \ and\ \bibinfo
  {author} {\bibfnamefont {W.}~\bibnamefont {Liu}},\ }\href@noop {} {\
  (\bibinfo {year} {2022}{\natexlab{b}})},\ \Eprint
  {http://arxiv.org/abs/2208.10417} {arXiv:2208.10417 [hep-lat]} \BibitemShut
  {NoStop}%
\bibitem [{\citenamefont {Clemente}\ \emph {et~al.}(2022)\citenamefont
  {Clemente}, \citenamefont {Crippa},\ and\ \citenamefont
  {Jansen}}]{Clemente:2022cka}%
  \BibitemOpen
  \bibfield  {author} {\bibinfo {author} {\bibfnamefont {G.}~\bibnamefont
  {Clemente}}, \bibinfo {author} {\bibfnamefont {A.}~\bibnamefont {Crippa}}, \
  and\ \bibinfo {author} {\bibfnamefont {K.}~\bibnamefont {Jansen}},\
  }\href@noop {} {\  (\bibinfo {year} {2022})},\ \Eprint
  {http://arxiv.org/abs/2206.12454} {arXiv:2206.12454 [hep-lat]} \BibitemShut
  {NoStop}%
\bibitem [{\citenamefont {Macridin}\ \emph
  {et~al.}(2018{\natexlab{a}})\citenamefont {Macridin}, \citenamefont
  {Spentzouris}, \citenamefont {Amundson},\ and\ \citenamefont
  {Harnik}}]{Macridin:2018oli}%
  \BibitemOpen
  \bibfield  {author} {\bibinfo {author} {\bibfnamefont {A.}~\bibnamefont
  {Macridin}}, \bibinfo {author} {\bibfnamefont {P.}~\bibnamefont
  {Spentzouris}}, \bibinfo {author} {\bibfnamefont {J.}~\bibnamefont
  {Amundson}}, \ and\ \bibinfo {author} {\bibfnamefont {R.}~\bibnamefont
  {Harnik}},\ }\href {\doibase 10.1103/PhysRevA.98.042312} {\bibfield
  {journal} {\bibinfo  {journal} {Phys. Rev. A}\ }\textbf {\bibinfo {volume}
  {98}},\ \bibinfo {pages} {042312} (\bibinfo {year} {2018}{\natexlab{a}})},\
  \Eprint {http://arxiv.org/abs/1805.09928} {arXiv:1805.09928 [quant-ph]}
  \BibitemShut {NoStop}%
\bibitem [{\citenamefont {Macridin}\ \emph
  {et~al.}(2018{\natexlab{b}})\citenamefont {Macridin}, \citenamefont
  {Spentzouris}, \citenamefont {Amundson},\ and\ \citenamefont
  {Harnik}}]{Macridin:2018gdw}%
  \BibitemOpen
  \bibfield  {author} {\bibinfo {author} {\bibfnamefont {A.}~\bibnamefont
  {Macridin}}, \bibinfo {author} {\bibfnamefont {P.}~\bibnamefont
  {Spentzouris}}, \bibinfo {author} {\bibfnamefont {J.}~\bibnamefont
  {Amundson}}, \ and\ \bibinfo {author} {\bibfnamefont {R.}~\bibnamefont
  {Harnik}},\ }\href {\doibase 10.1103/PhysRevLett.121.110504} {\bibfield
  {journal} {\bibinfo  {journal} {Phys. Rev. Lett.}\ }\textbf {\bibinfo
  {volume} {121}},\ \bibinfo {pages} {110504} (\bibinfo {year}
  {2018}{\natexlab{b}})},\ \Eprint {http://arxiv.org/abs/1802.07347}
  {arXiv:1802.07347 [quant-ph]} \BibitemShut {NoStop}%
\bibitem [{\citenamefont {Aleksandrowicz}\ \emph {et~al.}(2019)\citenamefont
  {Aleksandrowicz} \emph {et~al.}}]{gadi_aleksandrowicz_2019_2562111}%
  \BibitemOpen
  \bibfield  {author} {\bibinfo {author} {\bibfnamefont {G.}~\bibnamefont
  {Aleksandrowicz}} \emph {et~al.},\ }\href {\doibase 10.5281/zenodo.2562111}
  {\enquote {\bibinfo {title} {{Qiskit: An Open-source Framework for Quantum
  Computing}},}\ } (\bibinfo {year} {2019})\BibitemShut {NoStop}%
\bibitem [{\citenamefont {Dumitrescu}\ \emph {et~al.}(2018)\citenamefont
  {Dumitrescu}, \citenamefont {McCaskey}, \citenamefont {Hagen}, \citenamefont
  {Jansen}, \citenamefont {Morris}, \citenamefont {Papenbrock}, \citenamefont
  {Pooser}, \citenamefont {Dean},\ and\ \citenamefont
  {Lougovski}}]{PhysRevLett.120.210501}%
  \BibitemOpen
  \bibfield  {author} {\bibinfo {author} {\bibfnamefont {E.~F.}\ \bibnamefont
  {Dumitrescu}}, \bibinfo {author} {\bibfnamefont {A.~J.}\ \bibnamefont
  {McCaskey}}, \bibinfo {author} {\bibfnamefont {G.}~\bibnamefont {Hagen}},
  \bibinfo {author} {\bibfnamefont {G.~R.}\ \bibnamefont {Jansen}}, \bibinfo
  {author} {\bibfnamefont {T.~D.}\ \bibnamefont {Morris}}, \bibinfo {author}
  {\bibfnamefont {T.}~\bibnamefont {Papenbrock}}, \bibinfo {author}
  {\bibfnamefont {R.~C.}\ \bibnamefont {Pooser}}, \bibinfo {author}
  {\bibfnamefont {D.~J.}\ \bibnamefont {Dean}}, \ and\ \bibinfo {author}
  {\bibfnamefont {P.}~\bibnamefont {Lougovski}},\ }\href {\doibase
  10.1103/PhysRevLett.120.210501} {\bibfield  {journal} {\bibinfo  {journal}
  {Phys. Rev. Lett.}\ }\textbf {\bibinfo {volume} {120}},\ \bibinfo {pages}
  {210501} (\bibinfo {year} {2018})}\BibitemShut {NoStop}%
\bibitem [{\citenamefont {Temme}\ \emph {et~al.}(2017)\citenamefont {Temme},
  \citenamefont {Bravyi},\ and\ \citenamefont
  {Gambetta}}]{PhysRevLett.119.180509}%
  \BibitemOpen
  \bibfield  {author} {\bibinfo {author} {\bibfnamefont {K.}~\bibnamefont
  {Temme}}, \bibinfo {author} {\bibfnamefont {S.}~\bibnamefont {Bravyi}}, \
  and\ \bibinfo {author} {\bibfnamefont {J.~M.}\ \bibnamefont {Gambetta}},\
  }\href {\doibase 10.1103/PhysRevLett.119.180509} {\bibfield  {journal}
  {\bibinfo  {journal} {Phys. Rev. Lett.}\ }\textbf {\bibinfo {volume} {119}},\
  \bibinfo {pages} {180509} (\bibinfo {year} {2017})}\BibitemShut {NoStop}%
\bibitem [{\citenamefont {He}\ \emph {et~al.}(2020)\citenamefont {He},
  \citenamefont {Nachman}, \citenamefont {de~Jong},\ and\ \citenamefont
  {Bauer}}]{He:2020udd}%
  \BibitemOpen
  \bibfield  {author} {\bibinfo {author} {\bibfnamefont {A.}~\bibnamefont
  {He}}, \bibinfo {author} {\bibfnamefont {B.}~\bibnamefont {Nachman}},
  \bibinfo {author} {\bibfnamefont {W.~A.}\ \bibnamefont {de~Jong}}, \ and\
  \bibinfo {author} {\bibfnamefont {C.~W.}\ \bibnamefont {Bauer}},\ }\href
  {\doibase 10.1103/PhysRevA.102.012426} {\bibfield  {journal} {\bibinfo
  {journal} {Phys. Rev. A}\ }\textbf {\bibinfo {volume} {102}},\ \bibinfo
  {pages} {012426} (\bibinfo {year} {2020})},\ \Eprint
  {http://arxiv.org/abs/2003.04941} {arXiv:2003.04941 [quant-ph]} \BibitemShut
  {NoStop}%
\bibitem [{\citenamefont {D'Agostini}(1995)}]{DAgostini:1994fjx}%
  \BibitemOpen
  \bibfield  {author} {\bibinfo {author} {\bibfnamefont {G.}~\bibnamefont
  {D'Agostini}},\ }\href {\doibase 10.1016/0168-9002(95)00274-X} {\bibfield
  {journal} {\bibinfo  {journal} {Nucl. Instrum. Meth. A}\ }\textbf {\bibinfo
  {volume} {362}},\ \bibinfo {pages} {487} (\bibinfo {year}
  {1995})}\BibitemShut {NoStop}%
\bibitem [{\citenamefont {{Lucy}}(1974)}]{1974AJ.....79..745L}%
  \BibitemOpen
  \bibfield  {author} {\bibinfo {author} {\bibfnamefont {L.~B.}\ \bibnamefont
  {{Lucy}}},\ }\href {\doibase 10.1086/111605} {\bibfield  {journal} {\bibinfo
  {journal} {Astron. J.}\ }\textbf {\bibinfo {volume} {79}},\ \bibinfo {pages}
  {745} (\bibinfo {year} {1974})}\BibitemShut {NoStop}%
\bibitem [{\citenamefont {Richardson}(1972)}]{Richardson:72}%
  \BibitemOpen
  \bibfield  {author} {\bibinfo {author} {\bibfnamefont {W.~H.}\ \bibnamefont
  {Richardson}},\ }\href {\doibase 10.1364/JOSA.62.000055} {\bibfield
  {journal} {\bibinfo  {journal} {J. Opt. Soc. Am.}\ }\textbf {\bibinfo
  {volume} {62}},\ \bibinfo {pages} {55} (\bibinfo {year} {1972})}\BibitemShut
  {NoStop}%
\bibitem [{\citenamefont {Nachman}\ \emph {et~al.}(2020)\citenamefont
  {Nachman}, \citenamefont {Urbanek}, \citenamefont {de~Jong},\ and\
  \citenamefont {Bauer}}]{1910.01969}%
  \BibitemOpen
  \bibfield  {author} {\bibinfo {author} {\bibfnamefont {B.}~\bibnamefont
  {Nachman}}, \bibinfo {author} {\bibfnamefont {M.}~\bibnamefont {Urbanek}},
  \bibinfo {author} {\bibfnamefont {W.}~\bibnamefont {de~Jong}}, \ and\
  \bibinfo {author} {\bibfnamefont {C.}~\bibnamefont {Bauer}},\ }\href
  {\doibase 10.1038/s41534-020-00309-7} {\bibfield  {journal} {\bibinfo
  {journal} {npj Quantum Information}\ }\textbf {\bibinfo {volume} {6}}
  (\bibinfo {year} {2020}),\ 10.1038/s41534-020-00309-7},\ \Eprint
  {http://arxiv.org/abs/1910.01969} {arXiv:1910.01969 [hep-ph]} \BibitemShut
  {NoStop}%
\bibitem [{\citenamefont {Urbanek}\ \emph {et~al.}(2021)\citenamefont
  {Urbanek}, \citenamefont {Nachman}, \citenamefont {Pascuzzi}, \citenamefont
  {He}, \citenamefont {Bauer},\ and\ \citenamefont
  {de~Jong}}]{PhysRevLett.127.270502}%
  \BibitemOpen
  \bibfield  {author} {\bibinfo {author} {\bibfnamefont {M.}~\bibnamefont
  {Urbanek}}, \bibinfo {author} {\bibfnamefont {B.}~\bibnamefont {Nachman}},
  \bibinfo {author} {\bibfnamefont {V.~R.}\ \bibnamefont {Pascuzzi}}, \bibinfo
  {author} {\bibfnamefont {A.}~\bibnamefont {He}}, \bibinfo {author}
  {\bibfnamefont {C.~W.}\ \bibnamefont {Bauer}}, \ and\ \bibinfo {author}
  {\bibfnamefont {W.~A.}\ \bibnamefont {de~Jong}},\ }\href {\doibase
  10.1103/PhysRevLett.127.270502} {\bibfield  {journal} {\bibinfo  {journal}
  {Phys. Rev. Lett.}\ }\textbf {\bibinfo {volume} {127}},\ \bibinfo {pages}
  {270502} (\bibinfo {year} {2021})}\BibitemShut {NoStop}%
\bibitem [{\citenamefont {Endo}\ \emph {et~al.}(2021)\citenamefont {Endo},
  \citenamefont {Cai}, \citenamefont {Benjamin},\ and\ \citenamefont
  {Yuan}}]{Endo_2021}%
  \BibitemOpen
  \bibfield  {author} {\bibinfo {author} {\bibfnamefont {S.}~\bibnamefont
  {Endo}}, \bibinfo {author} {\bibfnamefont {Z.}~\bibnamefont {Cai}}, \bibinfo
  {author} {\bibfnamefont {S.~C.}\ \bibnamefont {Benjamin}}, \ and\ \bibinfo
  {author} {\bibfnamefont {X.}~\bibnamefont {Yuan}},\ }\href {\doibase
  10.7566/jpsj.90.032001} {\bibfield  {journal} {\bibinfo  {journal} {Journal
  of the Physical Society of Japan}\ }\textbf {\bibinfo {volume} {90}},\
  \bibinfo {pages} {032001} (\bibinfo {year} {2021})}\BibitemShut {NoStop}%
\end{thebibliography}%

\clearpage
\onecolumngrid
\appendix

\section{Numerical study of $L_1$ norm of Walsh coefficients of maximally coupled term}
\label{sec:L1_norm}
\begin{figure*}[t]
\subfloat[\label{fig:L1_vs_Npnq_many_alpha}]{\includegraphics[width=0.48\textwidth]{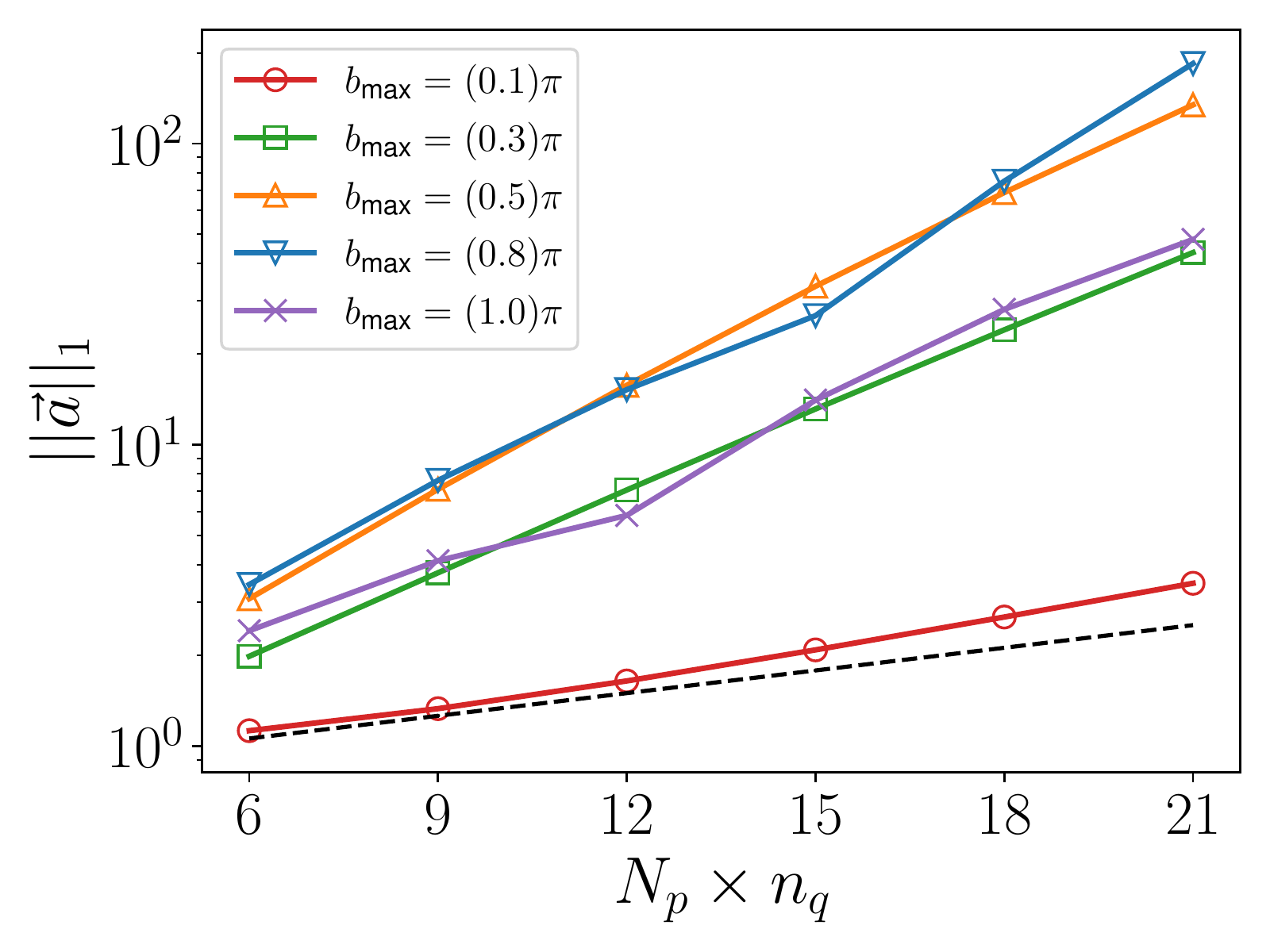}}
\hfill
\subfloat[\label{fig:L1_vs_Npnq_many_nq}]{\includegraphics[width=0.48\textwidth]{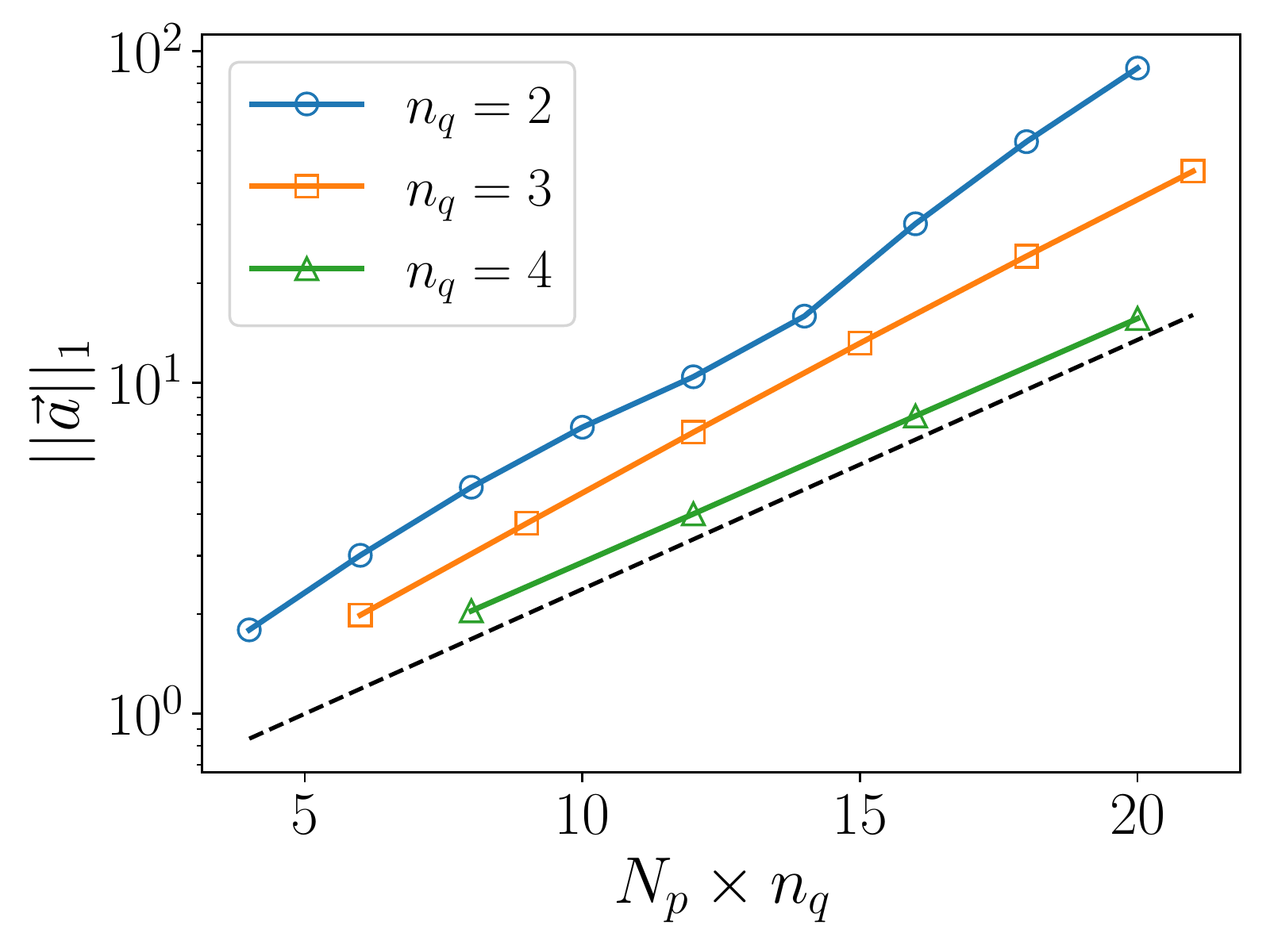}}
\caption{$L_1$ norm of the Walsh coefficients, $||\vec{a}||_1$, of the cosine of a sum of $N_p$ plaquettes as a function of $n_q N_p$. Fig.~\ref{fig:L1_vs_Npnq_many_alpha} shows $||\vec{a}||_1$ for fixed $n_q=3$. The different colored lines show different choices of $b_\text{max}$ chosen relative to the upper limit for $b_\text{max}$ given by $\pi$. The black dashed line shows the function $2^{(N_p n_q - 5)/(4n_q)}$ for reference. Fig.~\ref{fig:L1_vs_Npnq_many_nq} shows $||\vec{a}||_1$ for fixed $b_\text{max} = (0.5)\pi$. The different colored lines show different values of $n_q$. The black dashed line shows the function $2^{(N_p n_q-5)/4}$ for reference.}
\end{figure*}

In this section we present numerical studies of the $L_1$ norm of the Walsh coefficients of the maximally coupled term in the compact magnetic Hamiltonian. Specifically, the function we study is
\begin{align}
    \hat{f} = \cos \left( \sum_{i=1}^{N_p} \hat{B}_i \right).
\end{align}
Each lattice site is represented with $n_q$ qubits and so the dimension of the operator is $N \equiv 2^{N_p n_q}$. The Walsh coefficients $\vec{a} = (a_0, \dots, a_{N-1})$ are given by
\begin{align}
    a_j = \frac{1}{N} \Tr[ \hat{f} \hat{w}_j ].
\end{align}
The $L_1$ norm of these Walsh coefficients is given by
\begin{align}
    ||\vec{a}||_1 = \sum_{j=0}^{N-1} |a_j|.
\end{align}
Calculating all of the Walsh coefficients using the fast Walsh transform requires $\CO(n_q N_p 2^{n_q N_p})$ floating point operations. Due to limited computational resources, we limit our numerical study to $N_p n_q \leq 21$. Figure~\ref{fig:L1_vs_Npnq_many_alpha} shows $||\vec{a}||_1$ as a function of $N_p n_q$ for several choices of $b_\text{max}$ for fixed $n_q=3$. We see that for $6 \leq N_p n_q \leq 21$, the scaling is exponential, with each choice of $b_\text{max}$ growing faster than $2^{N_p (n_q-5)/(4n_q)}$. We find similar results for $n_q=2,4$. Figure~\ref{fig:L1_vs_Npnq_many_alpha} shows $||\vec{a}||_1$ for the choice of $b_\text{max} = (0.5)\pi$ for different values of $n_q$. For the entire range of $N_p n_q$ studied and for each $n_q$, the value of $||\vec{a}||_1$ grows at least as fast as $2^{(N_p n_q-5)/4}$. While we were unable to show that $||\vec{a}||_1$ will grow exponentially analytically, we expect the exponential growth to continue for values of $N_p n_q$ required for realistic calculations. Because the decision tree algorithm in Ref.~\cite{decision_tree} requires that $||\vec{a}||_1 \leq \text{poly}(N_p n_q)$, our numerical studies indicate that this method cannot be used to break the classical exponential scaling required to compute the exponentially many Walsh coefficients.

\section{Single cosine Walsh coefficients}
\begin{figure*}[t]
    \centering
    \includegraphics[width=0.45\textwidth]{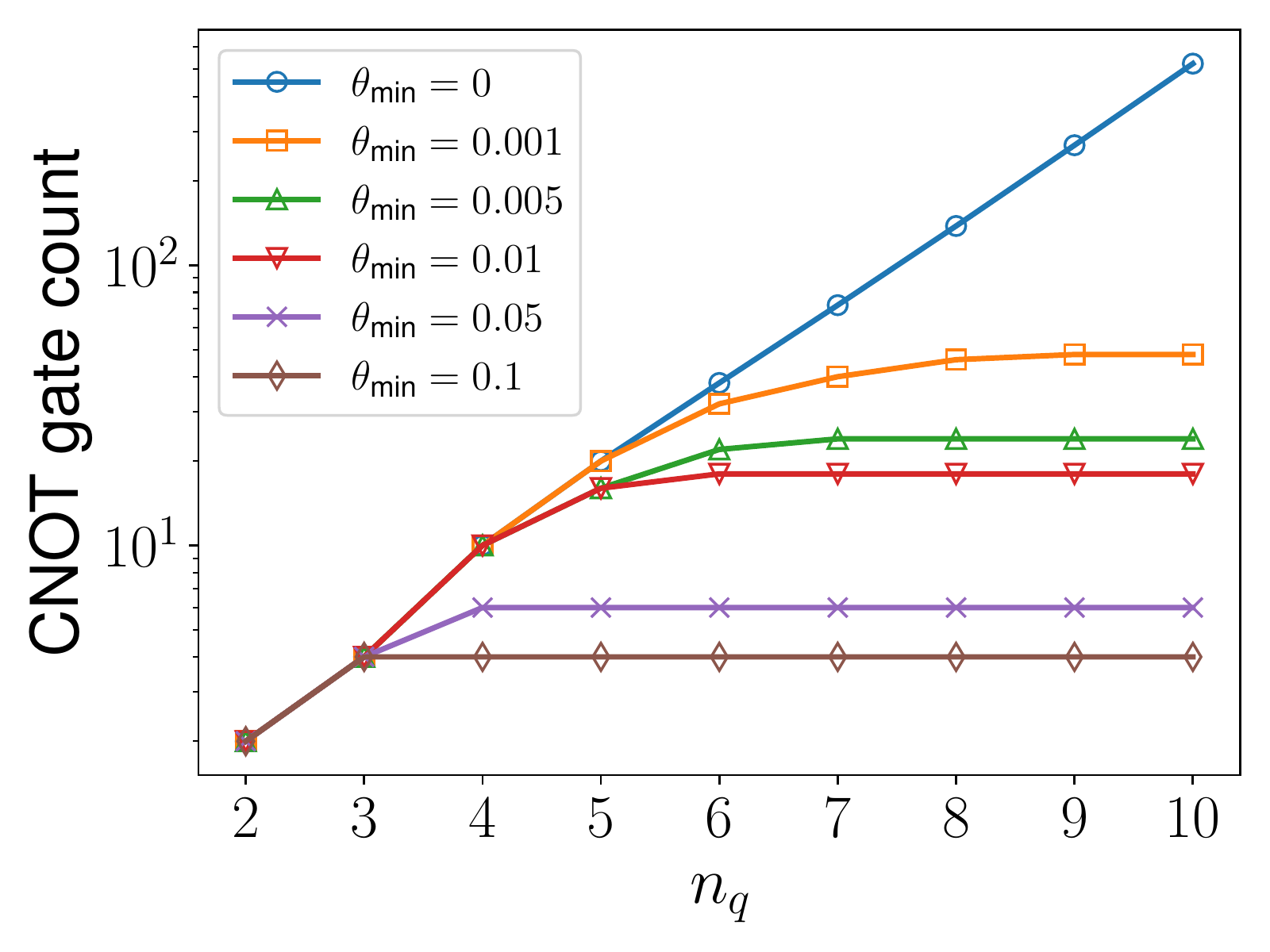}
    \caption{Number of CNOT gates required to implement $\exp(i \cos(\hat{B}))$ as a function of the number of qubits used to represent the operator $n_q$. The different colored lines indicate different values of the cutoff $\theta_\text{min}$.}
    \label{fig:gate_count_single_cosine}
\end{figure*}

Raising and lowering operators are ubiquitous throughout physics. Because Hamiltonian operators must be hermitian, sums of raising(lowering) operators are always accompanied by lowering(raising) operators. The sum of a raising and lowering operator (assuming periodic boundary conditions on the states) is diagonalized by a Fourier transformation. The resulting diagonal matrix is twice the cosine of evenly spaced entries on the diagonal. Understanding how the Walsh coefficients scale for a single cosine operator is therefore of interest. Figure \ref{fig:gate_count_single_cosine} shows the 
CNOT gate count required to implement such a term, in the notation of the main text $\exp(i \cos(\hat{B}))$, as a function of $n_q$. The construction of the $\hat{B}$ operator follows the same procedure as outlined in Sec.~\ref{ssec:originalBasisReview}. We notice that for a given value of $\theta_\text{min}$ the gate count levels off. This behavior occurs because as one increases the number of qubits used to represent the cosine, the additional Walsh coefficients are all below the cutoff $\theta_\text{min}/2$ and dropped from the circuit. Any unused qubits could be dropped from the calculation.

\section{Analytic gate count scaling of repeated products of functions}
In this part of the appendix we derive the scaling of implementing a diagonal unitary operator given by $\exp(i \bigotimes_{i=1}^{N} f(\hat{x}))$ as a function of $N$, where $f(\hat{x})$ is an arbitrary function of the diagonal operator $\hat{x}$ which has evenly spaced entries on the diagonal.

Each operator $f(\hat{x})$ is represented by $n$ qubits and can be decomposed into Walsh functions as
\begin{align}
	f(\hat{x}) = \sum_{j=0}^{2^{n}-1} a_j \hat{w}_j,
\end{align}
where $a_j$ and $\hat{w}_j$ are the $j^\text{th}$ Walsh coefficient and Walsh function. We denote the $n^\text{th}$ largest magnitude Walsh coefficient of the functin $f(\hat{x})$ as $A_n$. So $A_1$ and $A_2$ are the largest and second largest magnitude Walsh coefficients, respectively. The tensor product of two of these terms is given by
\begin{align}
\begin{split}
	f(\hat{x}) \otimes f(\hat{x}) &= \sum_{j,k=0}^{2^n-1} a_j a_k \hat{w}_j \otimes \hat{w_k}
	\\
	&\equiv \sum_{j,k=0}^{2^n-1} a_{jk} \hat{w}_{jk},
\end{split}
\end{align}
where we have introduced the notation $a_{jk} \equiv a_j a_k$ and $\hat{w}_{jk} \equiv \hat{w}_j \otimes \hat{w_k}$. For a product of $N$ terms we have
\begin{align}
	\bigotimes_{i=1}^{N} f(\hat{x}) = \sum_{j_1, j_2, \dots j_{N_p} = 0}^{2^{n}-1} a_{j_1 \dots j_{N_p}} \hat{w}_{j_1 \dots j_{N}}.
\end{align}
where, similarly, $a_{j_1 j_2 \dots j_{N}} \equiv a_{j_1} a_{j_2} \dots a_{j_N}$, and $\hat{w}_{j_1 j_2 \dots j_{N}} \equiv \hat{w}_{j_1} \otimes \hat{w}_{j_2} \otimes \dots \otimes \hat{w}_{j_N}$. Note that coefficients $a_{j_1 j_2 \dots j_{N}}$ with the same values of $j_i$ are equal regardless of the order of the indices $j_i$ due to the commutativity of multiplication of real numbers. The same is not true for the tensor product of different operators. We want to understand how many Walsh coefficients are larger than some threshold $\theta_\text{min}/2$ as a function of $\theta_\text{min}/2$ (we divide by 2 because the $R_z$ gate angle is twice the magnitude of the Walsh coefficient). In order to show this, we break up the sum according to the number of common indices each term has. For example, with $N=2$ we would write
\begin{align}
\begin{split}
	\bigotimes_{i=1}^{2} f(\hat{x}) &= \sum_{j_1, j_2 = 0}^{2^{n}-1} a_{j_1 j_2} \hat{w}_{j_1 j_2}
	\\
	&= \sum_{j=0}^{2^{n}-1} a_j^2 \hat{w}_{jj} + \sum_{k>l} a_k a_l \Big(\hat{w}_{kl} + \hat{w}_{lk} \Big)
	\\
	&= A_1^2 \Bigg[ \sum_{j=0}^{2^{n}-1} \left(\frac{a_{j}}{A_1}\right)^2 \hat{w}_{jj} + \frac{A_2}{A_1} \sum_{k>l} \frac{a_k a_l}{A_1 A_2} \Big(\hat{w}_{kl} + \hat{w}_{lk} \Big) \Bigg]
\end{split}
\end{align}
where going to the last line we have factored out $A_1^2$. Notice that because $A_1$ is the largest magnitude Walsh coefficient, for all $j$ we have $(a_j/A_1)^2 \leq 1$. Similar arguments can be used to show for all $k,l$ that $|(a_k a_l)/(A_1 A_2)|\leq 1$. This implies that, in general, each term in the sum over $k,l$ is suppressed by a factor of $A_2/A_1$ relative to terms in the sum over $j$. For $N=4$ we would write 
\begin{align}
\begin{split}
	\bigotimes_{i=1}^{4} f(\hat{x}) &= A_1^4 \Bigg[ \sum_{j=0}^{2^{n}-1} \left(\frac{a_{j}}{A_1}\right)^4 \hat{w}_{jjjj} + \frac{A_2}{A_1}\sum_{j>k} \left(\frac{a_j}{A_1}\right)^3 \frac{a_k}{A_2} \Big(\hat{w}_{jjjk} + \hat{w}_{jjkj} + \hat{w}_{jkjj} + \hat{w}_{kjjj} \Big)
	\\
	&\hspace{0.2in} + \left(\frac{A_2}{A_1}\right)^2\sum_{\substack{k,l=0 \\ j>k,l}} \left(\frac{a_j}{A_1}\right)^2 \frac{a_k}{A_2} \frac{a_l}{A_2} \Big(\hat{w}_{jjkl} + \hat{w}_{jkjl} + \hat{w}_{jklj} + \hat{w}_{kjjl} + \hat{w}_{kjlj} + \hat{w}_{kljj}\Big)
	\\
	&\hspace{0.2in} + \frac{A_2}{A_1} \frac{A_3}{A_1} \frac{A_4}{A_1} \sum_{j>k>l>m} \frac{a_j a_k a_l}{A_1 A_2 A_3 A_4} \Big(\hat{w}_{jklm} + \text{all other permutations}\Big) \Bigg].
\end{split}
\end{align}

Notice that the sum over $j,k,l$ in the second line is done such that $k$ can equal $l$, but $j$ can never equal $k$ or $l$. This organization means that there is always one pair of repeated indices, and when $k=l$ there are two pairs of repeated indices. To account for the scenario when $k=l$, we factor out two factors of $A_2$. This guarantees $|a_j^2 a_k a_l/(A_1^2 A_2^2)| \leq 1$ for all $j,k,l$ being summed over. The sum on the last line is done such that there are no repeated indices, and we therefore pull out one factor of $A_2$, $A_3$ and $A_4$. Making similar arguments as for the $N=2$ example, terms with three repeated indices are suppressed by a factor of $A_2/A_1$. Terms with at least one pair of repeated indices are suppressed by a factor of $(A_2/A_1)^2$. Terms with no repeated indices are more greatly suppressed, with a factor of $(A_2 A_3 A_4)/A_1^3$ suppression. This pattern holds for general $N$, as seen by
\begin{align}
\begin{split}
	\bigotimes_{i=1}^{N} f(\hat{x}) &= A_1^N \Bigg[ \overbrace{\sum_{j=0}^{2^{n}-1} \left(\frac{a_{j}}{A_1}\right)^N \hat{w}_{j \dots j}}^{2^{n_q} \text{ terms}}
	\\
	&\hspace{0.2in} + \frac{A_2}{A_1} \sum_{j_1 > j_2}^{2^n-1} \left(\frac{a_{j_1}}{A_1}\right)^{N-1} \frac{a_{j_2}}{A_2} \overbrace{\Big(\hat{w}_{j_1 j_1 \dots j_1 j_2} + \hat{w}_{j_1 j_1 \dots j_2 j_1} + \dots + \hat{w}_{j_2 j_1 \dots j_1}\Big)}^{N \text{ terms}}
	\\
	&\hspace{0.2in} + \left(\frac{A_2}{A_1}\right)^2\sum_{\substack{j_2,j_3=0 \\ j_1>j_2,j_3}}^{2^n-1} \left(\frac{a_{j_1}}{A_1}\right)^{N-2} \frac{a_{j_2}}{A_2} \frac{a_{j_3}}{A_2} \overbrace{ \Big(\hat{w}_{j_1 j_1 \dots j_1 j_2 j_3} + \text{all other permutations} \Big)}^{N (N-1)/2 \text{ terms}}
	\\
	&\hspace{0.2in} + \dots
	\\
	&\hspace{0.2in} + \frac{A_2 A_3 \dots A_N}{A_1^N}\sum_{j_1 > j_2 > \dots > j_{N}} \frac{a_{j_1} a_{j_2} \dots a_{j_N}}{(A_2)^N} \overbrace{\Big( \text{all permutations} \Big)}^{N! \text{ terms}} \Bigg].
\end{split}	
\end{align}

In other words, terms in a given sum with at least one set of $r$ repeated indices are suppressed by a factor of $(A_2/A_1)^r$. If there is a large gap between the first and second largest magnitude Walsh coefficients, \textit{i.e.} $|A_2/A_1| \ll 1$, the gap between the maximum coefficient in each sum will be large. This implies that choosing  $\theta_\text{min}/2 > A_1^{N-r} A_2^r$ ensures that every term in each sum with $\CO(N^{r+1})$ or greater number of terms will be dropped completely (except for when $r=0$, when terms are suppressed by a factor of $A_2 A_3 \dots A_N/A_1^N$). The resulting CNOT scaling for this choice of $\theta_\text{min}/2$ will then be $\CO(N^r)$. One complication is that the scaling is being studied as a function of $N$ and so each gate count will have a different $N$. This makes the specific choice of $N$ when choosing $\theta_\text{min}$ somewhat challenging.

As an example, consider the function
\begin{align}
    f(\hat{B}) = \cos ( \hat{B} )
\end{align}
where $\hat{B}$ is the magnetic field operator represented by $n_q$ qubits. The diagonal $\hat{B}$ operator is sampled using a $b_\text{max}$ value associated with the choice of coupling $g=0.1$. The product of $N_p$ of these terms is given by
\begin{align}
    \bigotimes_{i=1}^{N_p} f(\hat{B}_i) = \bigotimes_{i=1}^{N_p} \cos \left( \hat{B}_i \right ).
\end{align}
To understand the CNOT gate count scaling as a function of $N_p$, we first calculate the Walsh coefficients for a single $\cos(\hat{B})$ operator. The Paley-ordered Walsh coefficients for the operator $\cos(\hat{B})$ using $n_q=2$ are shown in Table~\ref{tab:cos_walsh_coeffs_nq2_nq3}.
\begin{table}[h]
    \centering
    \begin{tabular}{c|c|c|c|c}
    $n_q$ & $a_0$ & $a_1$ & $a_2$ & $a_3$ \\
    \hline \hline 
    2 & $9.83 \times 10^{-1} $ & $1.10 \times 10^{-2}$ & $-1.10 \times 10^{-2}$ & $5.48 \times 10^{-3}$
    \end{tabular}
    \caption{Paley-ordered Walsh coefficients for the operator $\cos(\hat{B})$ for $n_q=2$.}
    \label{tab:cos_walsh_coeffs_nq2_nq3}
\end{table}
The first and second largest magnitude Walsh coefficients are $A_1=9.83\times 10^{-1}, A_2=1.10\times 10^{-2}$. In the following discussion we will use $A_1=1$ for simplicity. Doing so avoids the problem of what value of $N_p$ to raise $A_1$ to. The cutoff where we expect $\CO(N_p^r)$ scaling to turn is then given by
\begin{align}
    \theta_\text{min}^{(r)} \approx 2 \left( A_{2}\right)^r.
    \label{eq:scaling_cutoff_prediction}
\end{align}
\begin{figure}
    \centering
    \includegraphics[width=0.6\textwidth]{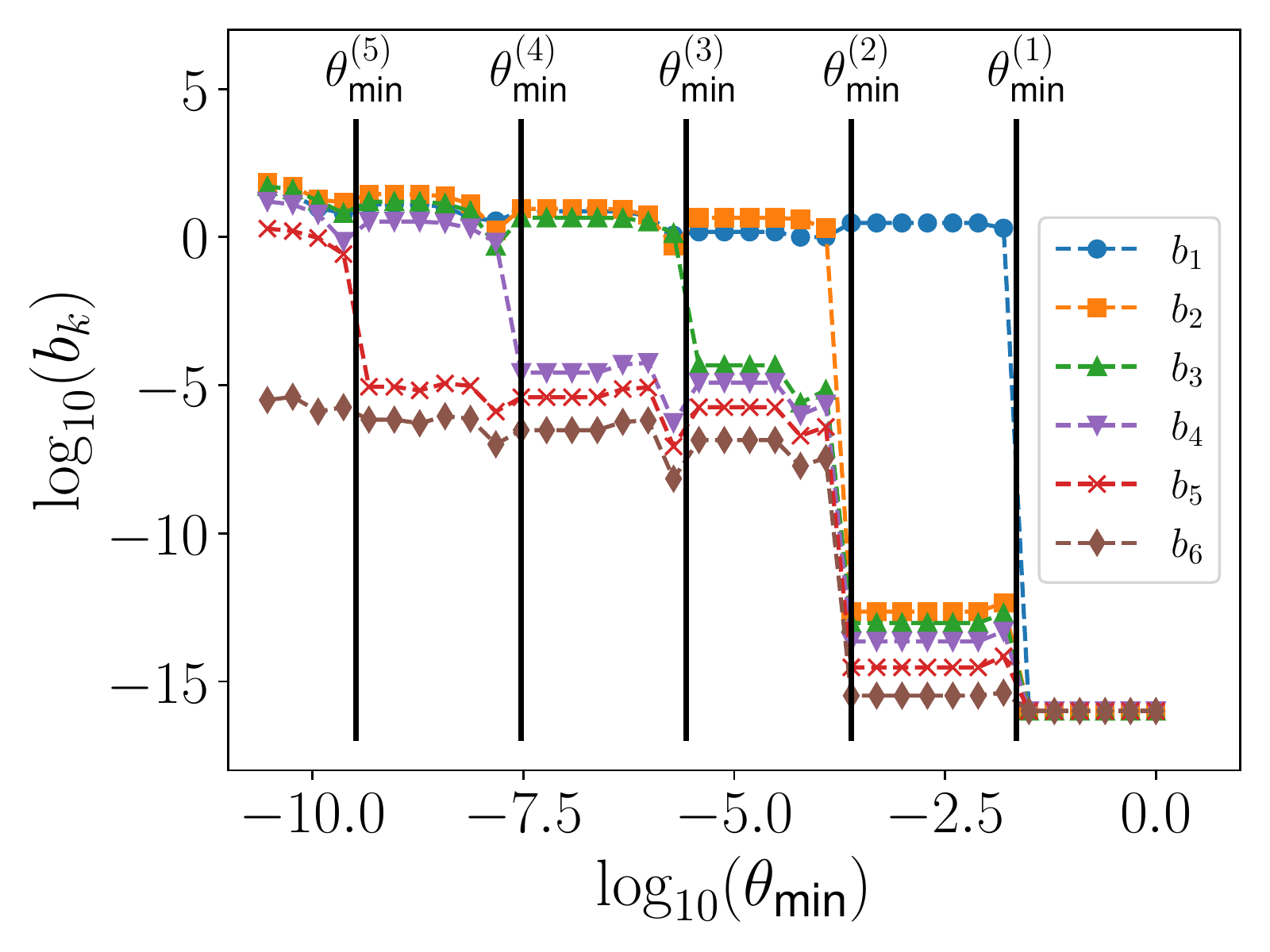}
    \caption{Fit parameters as a function of $\theta_\text{min}$. The fit parameters were found fitting the CNOT gate count required to implement $\exp(i \bigotimes_{j=1}^{N_p} \cos(\hat{B}_j))$ for $N_p=1, \dots,8$ to the fit form $\text{g}(N_p) = \sum_{k=0}^{7} b_k N_p^{k}$. The different colored points indicate different fit parameters $b_k$. The vertical black lines indicate the value of $\theta_\text{min}^{(r)}$ given by Eq.~\eqref{eq:scaling_cutoff_prediction} for $r=1,\dots, 5$.}
    \label{fig:fit_Cx_vs_Np_nq2}
\end{figure}
To test this, we calculate the number of gates to implement $\exp(i \bigotimes_{j=1}^{N_p} \cos(\hat{B}_j))$ for values $1 \leq N_p \leq 8$. We do so for all cutoff values between $2^{-36} \leq \theta_\text{min} \leq 1$ spaced by factors of $1/2$. Fits are then performed to the CNOT gate count assuming a polynomial fit form $\text{g}(N_p) = \sum_{k=0}^{7} b_k N_p^{k}$. This fit form was chosen because it has enough parameters to perfectly describe the data. Figure~\ref{fig:fit_Cx_vs_Np_nq2} shows the result of the fit parameters $b_k$ as a function of $\theta_\text{min}$. For the largest choice of $\theta_\text{min}$, all fit parameters are zero except for $b_1$, indicating the scaling is linear in $N_p$. As $\theta_\text{min}$ is decreased, more fit parameters become non-zero, indicating the scaling has changed. The black vertical lines show the values of $\theta_\text{min}^{(r)}$ for $r=1,2,3,4,5$ predicted using Eq.~\ref{eq:scaling_cutoff_prediction}. We see that the predicted values of $\theta_\text{min}^{(r)}$ line up with the actual values determined by the fit. The slight deviations are likely caused by the issue of choosing what power to raise $A_1$ to.

\section{Details of the Walsh Function Formalism}
\label{app:WalshImplementation}

In this section we provide a procedure for using the Walsh function formalism to construct diagonal unitary matrices. A more complete introduction can be found in Ref.~\cite{Welch_2014}.

The Walsh function formalism works as follows. Suppose we sample a function $f(x)$ at $N=2^{n}$ evenly spaced discrete points $x_k = k / N$ on the interval $x_k \in [0, 1-1/N]$. The Paley/dyadic-ordered Walsh functions are defined as
\begin{align}
    w_j(x_k) = (-1)^{\sum_{i=1}^N j_i k_i}
    \,,
    \label{eq:walsh_def}
\end{align}
where $j=0,1,..., N-1$, $j_i$ is the $i$'th bit in the \textit{binary expansion} 
\begin{align}
    j=\sum_{i=1}^n j_i 2^{(i-1)}
    \,,
\end{align}
and $k_i$ is the $i$'th bit in the \textit{dyadic binary} expansion
\begin{align}
    k=\sum_{i=1}^n k_i 2^{(n-i)}
    \,.
    \label{dyadic}
\end{align}
Notice that for any integer, the dyadic binary string is simply the bit-reversed binary string. We use the convention that all classes of binary strings are written as $(j_n j_{n-1} \dots j_1)$. As an example, consider the integer $13$. Using the definitions and notation above, the binary string is $(1101)$ and the dyadic binary string is $(1011)$. Using these strings, we find $w_{13}(x_{13}) = (-1)^2 = 1$.
Two other orderings of the Walsh functions are sequency ordered where $k_i$ is instead the $i^\text{th}$ bit of the \textit{Gray binary expansion} of $k$, and the Hadamard/natural ordered where $k_i$ is the $i^\text{th}$ bit of the \textit{binary expansion} of $k$. 

Analogous to the discrete Fourier series, the Walsh functions form an orthogonal basis of discretely sampled functions. They satisfy orthogonality relations
\begin{align}
    \sum_{i=1}^{N} w_j(x_i) w_k(x_i) = N \delta_{jk}
    \,.
\end{align}
The function $f$ can be written as 
\begin{align}
    f(x_k) = \sum_{j=1}^{N} a_j w_j(x_k)
    \,,
\end{align}
where the Walsh coefficients can be calculated by
\begin{align}
    a_j = \frac{1}{N} \sum_{i=1}^{N} f(x_i) \,  w_j(x_i)
    \,.
\end{align}
As pointed out in the main text, the Walsh coefficients can be calculated using the fast-Walsh-transform which requires $\CO(N \log_2 N)$ floating point operations \cite{Yarlagadda_Hershey}. 

The ability to represent discretely sampled functions makes Walsh functions useful in quantum computing where one must always sample operators at discrete points, i.e. digitize any continuous theory. On a quantum computer, Walsh operators are diagonal operators with the Walsh function values on the diagonal. The Paley-ordered Walsh operators for an $n$ qubit system are given by
\begin{align}
    \hat{w}_j = \bigotimes_{i=1}^{n} (\sigma^z_i)^{j_i}
    \,,
\end{align}
acting on a state representing a dyadic binary representation of the integer $k$ where, again, $j_i$ is the $i^{\text{th}}$ bit of the binary expansion of $j$, and $\sigma^z_i$ is the Pauli matrix acting on the $i^{\text{th}}$ qubit. These operators satisfy
\begin{align}
    \hat{w}_j \ket{k} = w_j(x_k) \ket{k}\,,
\end{align}
where the function $w_j(x_k)$ is the Walsh function defined in~\eqref{eq:walsh_def}. The set of Walsh functions form a basis for real diagonal operators. 

Walsh operators can be exponentiated using the identity
\begin{align}
    \exp(i a_j \hat{w}_j) = \cos(a_j) \mathds{1} + i \sin(a_j) \hat{w}_j
    \,.
\end{align}
Thus, this operator applies the phase $e^{i a_j}$ if the number of 1's in the dyadic binary representation of $k$ is even and $e^{-i a_j}$ if it is odd, which can be implemented using a single rotation gate $R_z(-2 a_j)$ and CNOT gates applied in the following way:
\begin{enumerate}
    \item Write the binary string $(j_n j_{n-1} \dots j_1)$ such that $j=\sum_{i=1}^{n} j_i 2^{i-1}$
    \item Identify the index $i_\text{msb}$ of the most significant bit (msb) with value 1
    \item Place an $R_z(\theta)$ gate on qubit $i_\text{msb}$ with rotation angle $\theta = -2a_j$
    \item For all other values of 1 in the binary string, place two CNOT gates, one on either side of the $R_z$ gate, acting on qubit $i_\text{msb}$ controlled the index of the 1-valued entry
\end{enumerate}
As an example we show the four qubit circuit for $j = 13$, which in binary notation is $j = 1101$. We therefore find $i_{\rm msb}$ = 4, leading to the circuit for $\exp(i a_{13} w_{13})$ shown in Fig.~\ref{fig:exp_of_walsh_example}. The diagonal unitary operator with the phases on the diagonal given by $\hat{H} = \sum_{i=1}^n a_i \hat{w}_i$ is implemented as a product of exponentials of Walsh operators, $\hat{U} = e^{i \hat{H}} = \prod_{j=1}^{N} e^{i a_j \hat{w}_j}$.

Note that because the Walsh operators are diagonal, the terms in the product commute and one is free to choose an order that minimizes the number of CNOT gates in the circuit. 
To find the optimal ordering, we first recall that the index $i_\text{msb}$ of a Walsh operator indicates on which qubit the CNOT gate targets. 
One can therefore simplify the CNOT gates between adjacent Walsh operators with the same $i_\text{msb}$ using the property that CNOT gates targeted on the same qubit commute. 
The controls of the simplified set of CNOT gates between the circuits for $e^{i a_j \hat{w}_j}$ and $e^{i a_k \hat{w}_k}$ are therefore given by the locations of the 1-valued bits in the bit-wise exclusive or (XOR) of the binary expansion of $j$ and $k$. Therefore, to minimize the CNOT gate count we need to minimize the number of binary transitions between indices of adjacent Walsh operators. 
This is achieved by ordering the Walsh operators according to their Gray binary ordering, i.e. using sequency ordered Walsh operators. This procedure works for subsets of Walsh operators with the same $i_\text{msb}$. Before placing the first gate for a given $i_\text{msb}$, a single CNOT gate is placed targeted on qubit $i_\text{msb}$ and controlled on qubit $i_\text{msb}-1$. The CNOT gates to the left of $R_{z,3}$ and $R_{z,6}$ in Fig.~\ref{fig:circ_simplification} are examples of CNOT gates that are placed between sets of different $i_\text{msb}$. 
For an $n$-qubit circuit, the optimal ordering requires $(2^{n}-1)$ $R_z$ gates and $(2^{n}-2)$ CNOT gates for a total of $(2^{n+1}-3)$ gates, which, as stated previously, was shown to be asymptotically optimal in Ref.~\cite{Bullock_2003}. 
As already mentioned, a nice worked-through example for constructing the optimal three qubit circuit can be found in Ref.~\cite{Welch_2014}.

We now review the procedure given in Ref.~\cite{Welch_2014} for constructing the circuit of a general approximate time evolution operator $U_\epsilon(t)$ using a cutoff $\theta_\text{min}$. As explained in Sec.~\ref{ssec:removingSmallWalsh}, Walsh operators are placed in sequency order, but because some have been dropped, the number of binary transitions between the adjacent Walsh operators can be in general more than one. As before, the control qubit of the CNOT gates placed between two adjacent Walsh operators are the locations of the bits equal to one in the XOR of the Walsh operator indices. Similarly to the exact circuit construction, this procedure works for subsets of Walsh operators with the same $i_\text{msb}$. A single CNOT gate is still placed targeted on qubit $i_\text{msb}$ and controlled on qubit $i_\text{msb}-1$ before placing the first gate for a given $i_\text{msb}$. The CNOT gate count for circuits constructed in this way can likely be reduced using circuit identities. The $R_z$ gates on the other hand are fixed for a certain choice of $\theta_\text{min}$.

\section{CNOT gate count for the original basis}
\label{app:truncationOriginalBasis}
\begin{figure*}[t]
\subfloat[\label{fig:Cx_vs_Np_unweaved_dt}Walsh function truncation scale, $\theta_\text{min}$, chosen relative to $\delta t$. ]{\includegraphics[width=0.48\textwidth]{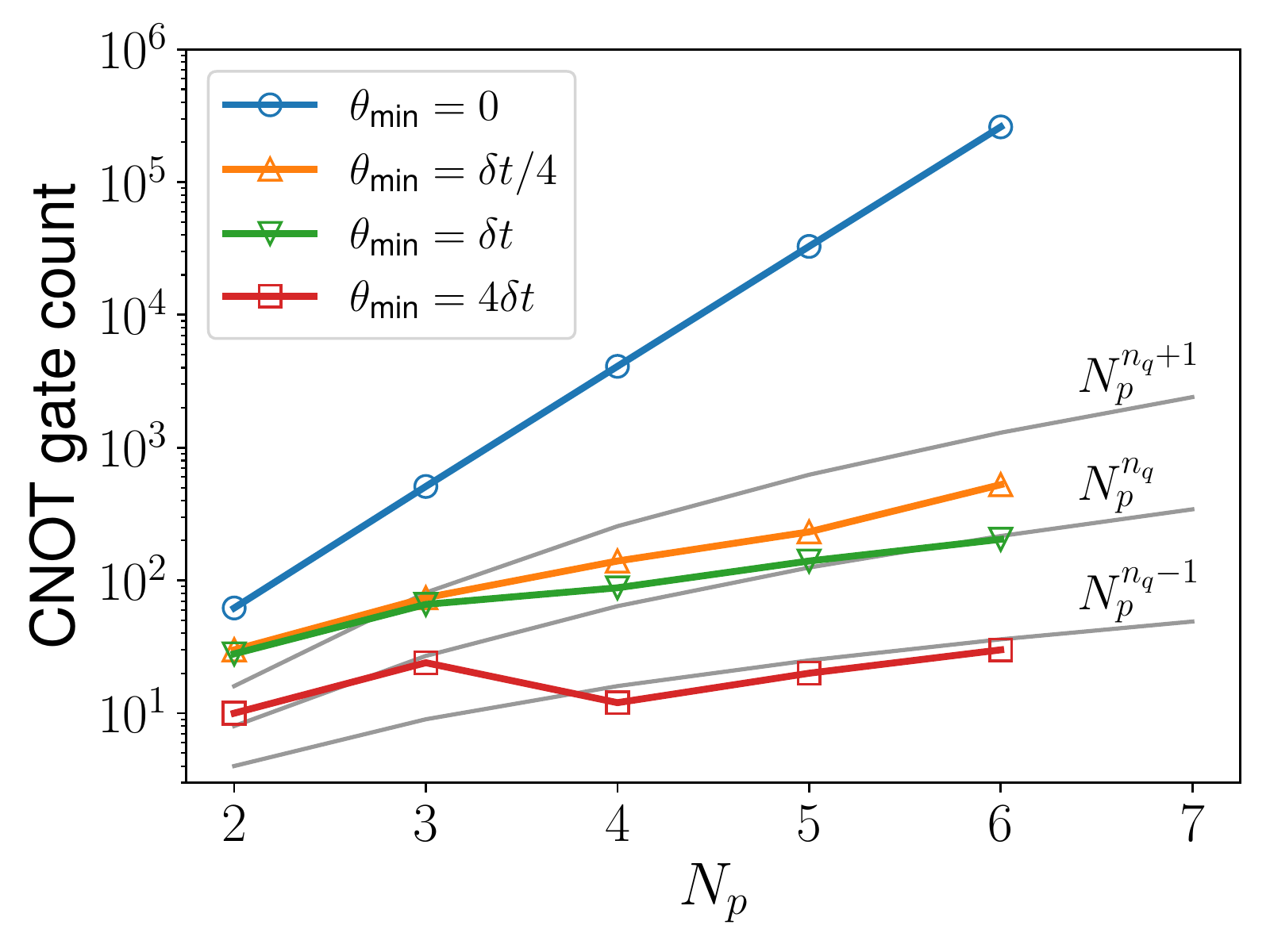}}
\hfill
\subfloat[\label{fig:Cx_vs_Np_unweaved_dtsq} Walsh function truncation scale, $\theta_\text{min}$, chosen relative to $\delta t^2$ with $\delta t = 1/4$.]{\includegraphics[width=0.48\textwidth]{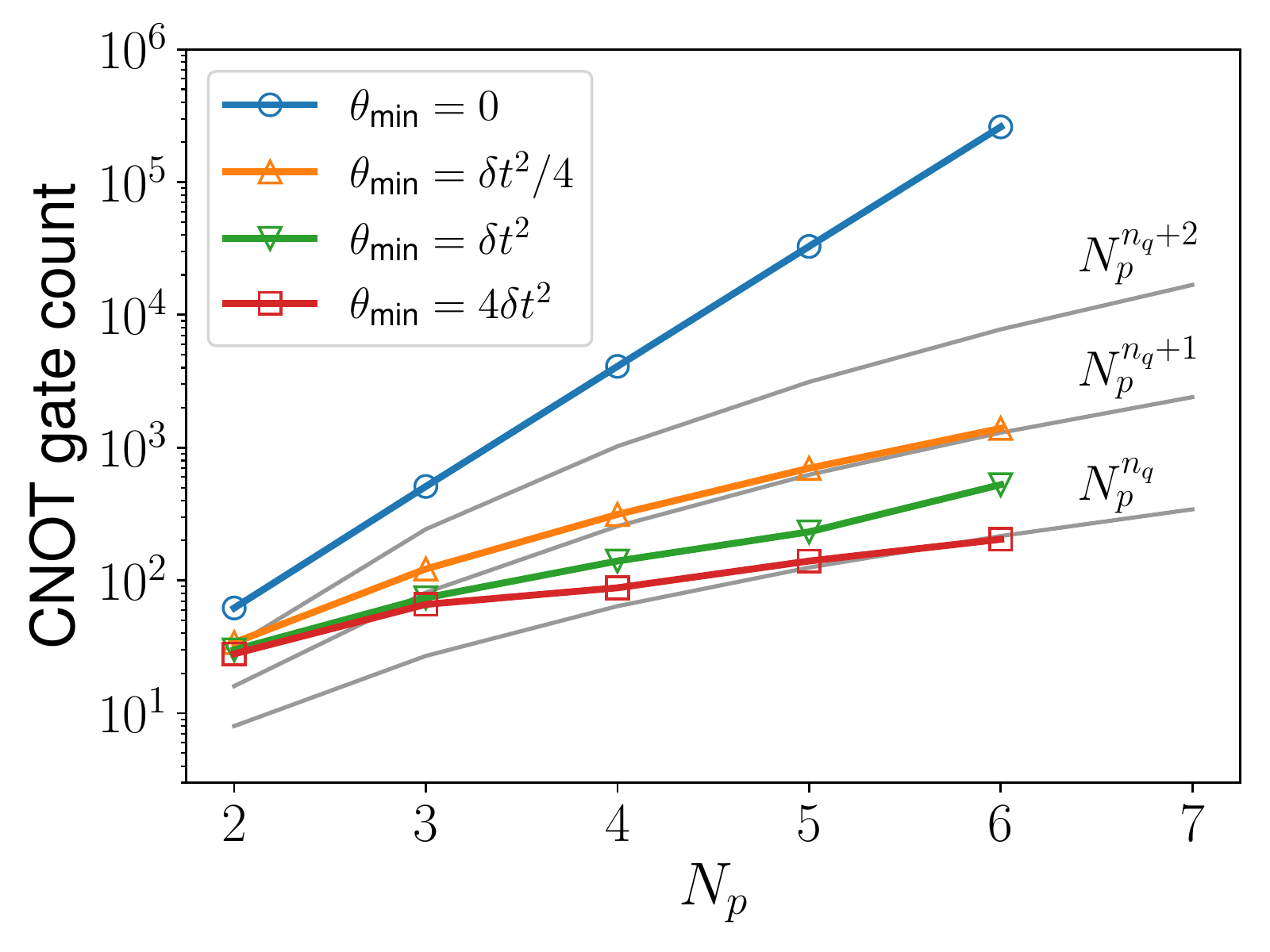}}
\hfill
\caption{CNOT gate count to implement the complex exponential of the the maximally coupled operator in the compact magnetic Hamiltonian in the original basis as a function of the number of plaquettes $N_p$ using $g = 0.1$ and $n_q=3$. Figure~\ref{fig:Cx_vs_Np_unweaved_dt} show the gate count if $\theta_\text{min}$ is chosen relative to $\delta t$; gray lines show the functions $N_p^{n_q-1}, N_p^{n_q}$ and $N_p^{n_q+1}$ for reference. Figure~\ref{fig:Cx_vs_Np_unweaved_dtsq} shows gate count if $\theta_\text{min}$ is chosen relative to $\delta t^2$. Gray lines show the functions $N_p^{n_q}, N_p^{n_q+1}$ and $N_p^{n_q+2}$ for reference.}
\end{figure*}

In order to test how well this truncation method can improve upon the exponential gate count scaling in the original basis, we calculate the CNOT gate count needed to implement the cosine of a sum of $N_p$ magnetic field operators for various values of $\theta_\text{min}$ assuming both a 1$^\text{st}$ order and 2$^{\text{nd}}$ order Suzuki-Trotter implementation. Figure~\ref{fig:Cx_vs_Np_unweaved_dt} shows the gate count choosing $\theta_\text{min}$ relative to $\delta t$, and Figure~\ref{fig:Cx_vs_Np_unweaved_dtsq} shows the gate count choosing $\theta_\text{min}$ relative to $\delta t^2$ for a modest choice of $\delta t=1/4$. Both plots were made using $g=0.1$ and $n_q = 3$. 
We see that in both cases, imposing a cutoff does break the exponential scaling. Looking first at Fig.~\ref{fig:Cx_vs_Np_unweaved_dt}, for values of $1/8 \leq \theta_\text{min}/\delta t < 1$ the scaling appears to roughly follow  between $\CO(N_p^{n_q})$ and $\CO(N_p^{n_q+1})$. From Fig.~\ref{fig:Cx_vs_Np_unweaved_dtsq}, we see that for values of $1/8 \leq \theta_\text{min}/(\delta t)^2 < 1$ the scaling appears to roughly follow  between $\CO(N_p^{n_q})$ and $\CO(N_p^{n_q+2})$. We observed similar gate count scaling for $n_q=2,4$. 

\begin{figure*}[t]
\subfloat[\label{fig:Cx_vs_nq_unweaved_dt} Walsh function truncation scale $\theta_\text{min}$, chosen relative to $\delta t$.]{\includegraphics[width=0.48\textwidth]{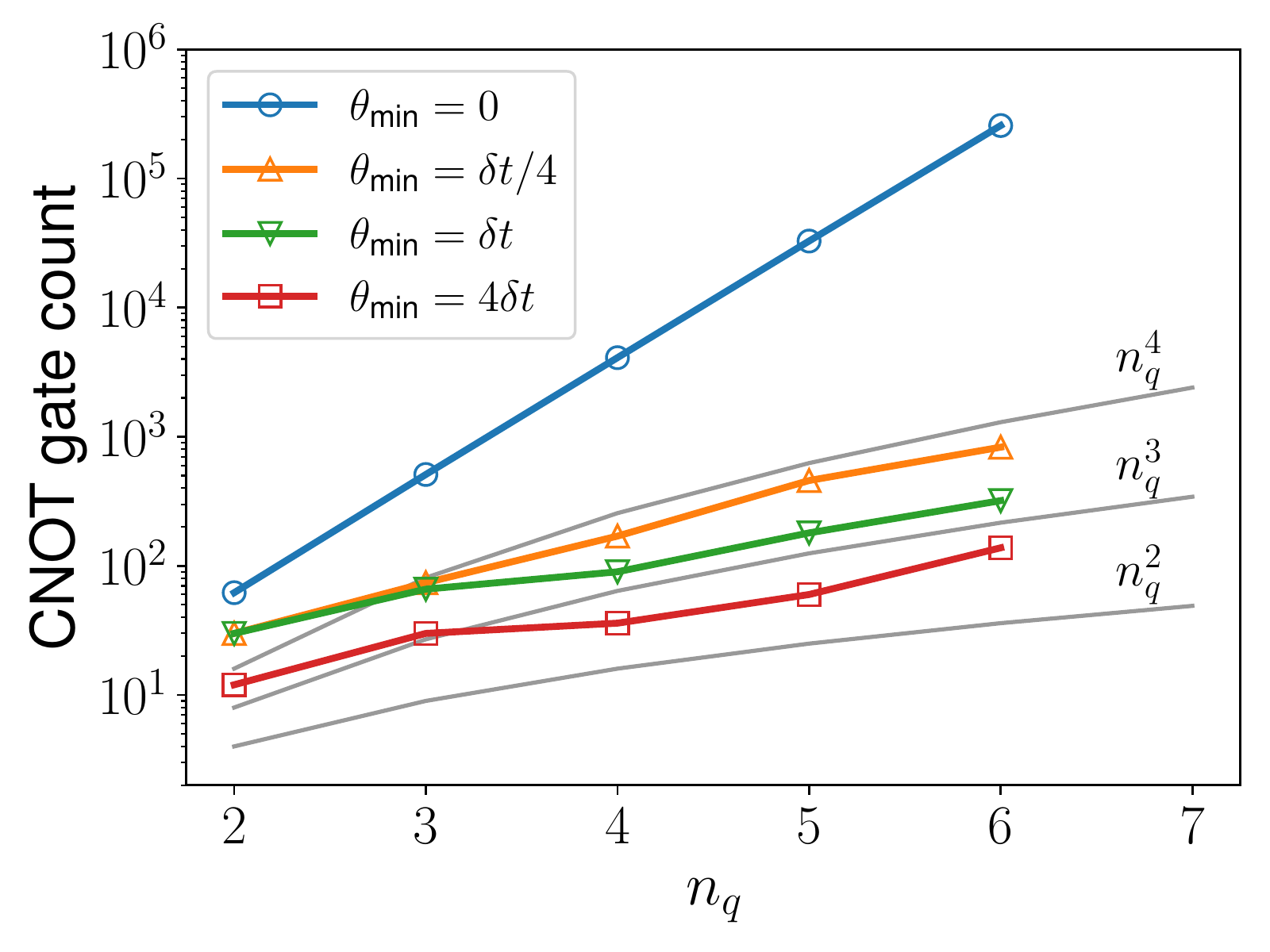}}
\hfill
\subfloat[\label{fig:Cx_vs_nq_weaved_dt} Walsh function truncation scale $\theta_\text{min}$, chosen relative to $\delta t^2$, with $\delta t = 1/4$.]{\includegraphics[width=0.48\textwidth]{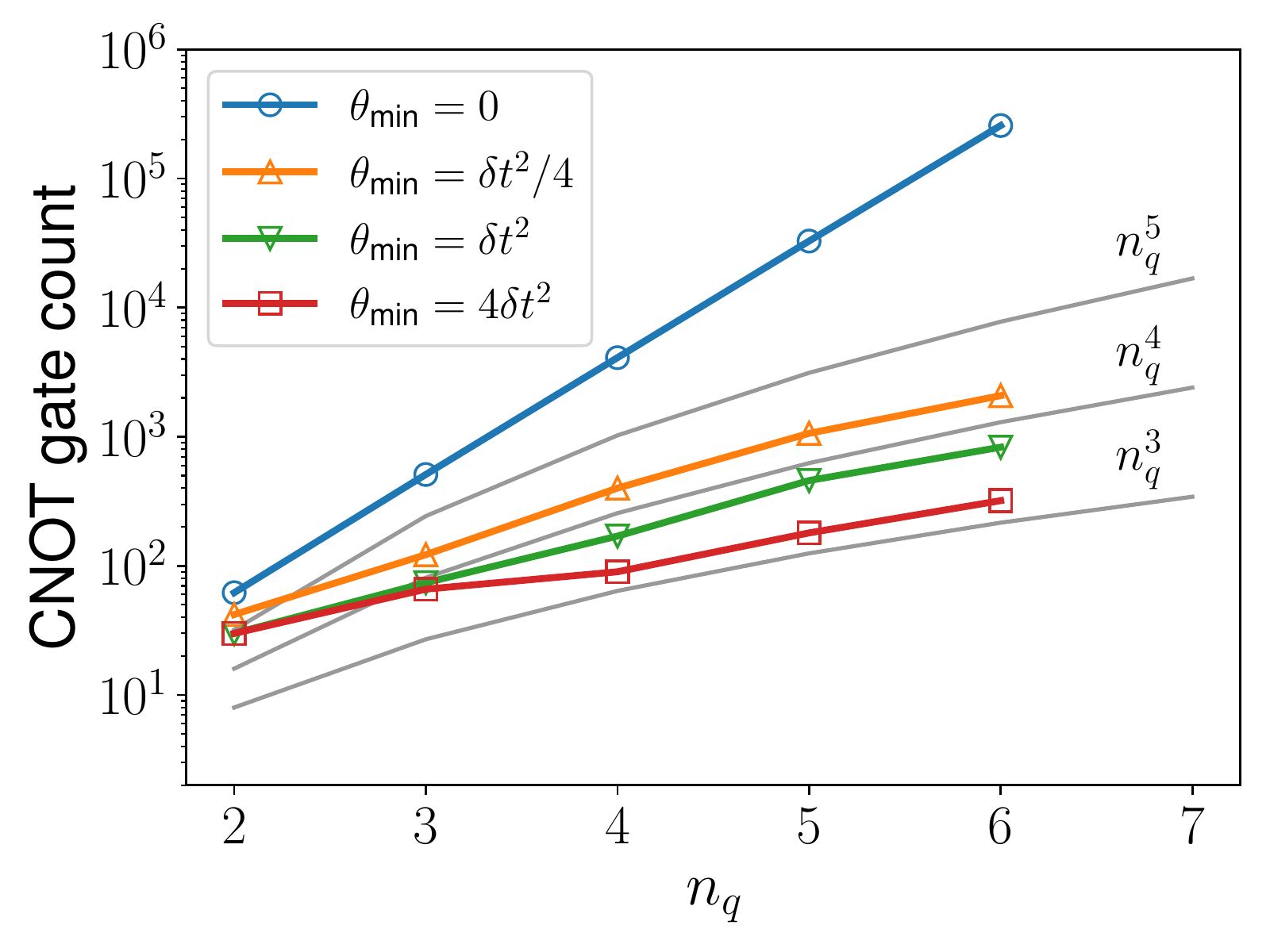}}
\caption{CNOT gate count to implement the complex exponential of the maximally coupled operator in the compact magnetic Hamiltonian in the original basis as a function of the number of qubits per plaquette $n_q$ for fixed $N_p=3$ with $g = 0.1$. The values of $b_\text{max}$ were chosen using the procedure in Sec.~\ref{ssec:weavedBasisDigitization}. In the left plot, the gray lines show the functions $n_q^2, n_q^3$, and $n_q^4$ for reference. In the right plot, the gray lines show the functions $n_q^3, n_q^4$, and $n_q^5$ for reference.}
\end{figure*}

\section{Upper bound on Walsh series truncation error (with non-zero $\theta_\text{min}$)}
\label{app:errorUpperBound}
In this appendix we derive an upper bound on the error introduced by truncating the Walsh series using a cutoff $\theta_\text{min}$ in terms of the Suzuki-Trotter step-size $\delta t$.

We define the difference between two $n$-qubit operators as $E(U_1, U_2) \equiv || U_1 - U_2 ||$, where $||A|| \equiv \text{max}_{\ket{\psi}} |A \ket{\psi}|$ is the spectral norm of the operator $A$ and $\ket{\psi}$ is any normalized vector. The error between the exact time evolution operator for a single Suzuki-Trotter step $U(\delta t) = e^{i H t}$ and the approximate time evolution operator $U_\epsilon(\delta t) = e^{i H_\epsilon \delta t}$ is $E(U(\delta t), U_\epsilon(\delta t)) \leq \epsilon$. The approximate operator $\delta t H_{\epsilon}$ is calculated by dropping all Walsh operators with Walsh coefficients $|a_j| < \theta_\text{min}/2$. This approximation is equivalent to rounding some fixed point accuracy \cite{CK_Yuen}, and the upper bound on the error assumes that the error from dropping each individual Walsh operator adds coherently. For a general operator $H$ the upper bound on the error is
\begin{equation}
    E(H, H_\epsilon) \leq N_\text{drop}(\theta_\text{min}) \theta_\text{min},
\end{equation}
where $N_\text{drop}(\theta_\text{min})$ is the number of Walsh operators dropped from the Walsh series used to calculate $H_\epsilon$. Note that $N_\text{drop}(\theta_\text{min})$ is a function of $\theta_\text{min}$ (note that this upper bound is likely overly pessimistic, because errors from dropping different Walsh operators will in general not always add coherently \cite{CK_Yuen}).

Now suppose that one uses a cutoff $\theta_\text{min} = \beta (\delta t)^m$, where $\beta$ is a constant and $m$ is the order of the Suzuki-Trotter scheme being used. The upper bound on the error is now
\begin{equation}
    E(H, H_\epsilon) \leq \beta N_\text{drop}(\delta t) (\delta t)^m.
\end{equation}
Even though $N_\text{drop}(\delta t)$ is a complicated function of $\delta t$, we can, however, make some general arguments on how $N_\text{drop}(\delta t)$ changes with $\delta t$. To do so, first recall that the Walsh coefficients are given by
\begin{equation}
    a_j = \frac{\delta t}{2^n} \Tr \left[\hat{H} \hat{w}_j\right].
\end{equation}
A Walsh operator $\hat{w}_j$ is dropped if $|a_j| < \beta (\delta t)^m/2$. Plugging in the expression for $a_j$, we see that an equivalent condition for dropping $a_j$ is if
\begin{equation}
    \frac{1}{2^n} \Big|\Tr \left[ H w_j \right] \Big| < \frac{\beta}{2} (\delta t)^{m-1}.
\end{equation}
Because the left hand side of the above inequality is independent of $\delta t$, the number of Walsh operators dropped can either stay the same or increase as $\delta t$ is increased, but never decrease. Similarly, as $\delta t$ is decreased, the number of Walsh operators dropped can either stay the same or decrease, but never increase. This implies that $N_\text{drop}(\delta t)$ is a monotonically non-decreasing function of $\delta t$. The error can therefore be bounded by
\begin{equation}
    E(H, H_\epsilon) \leq c\, (\delta t^m)
\end{equation}
where $c$ is some constant (we have absorbed $\beta$ into $c$). If we use an $m$ order Suzuki-Trotter scheme, the error between the exact and approximate time evolution operator is given by
\begin{equation}
    E(U(t), U_\epsilon(t)) \leq \alpha_m (\delta t)^m t + c\, (\delta t)^m t,
\end{equation}
where $\alpha_m$ is the pre-factor of the Suzuki-Trotter error. 
\end{document}